\documentclass[12pt]{article}
\usepackage{enumitem} 
\usepackage{rotfloat} 
\usepackage{graphicx,graphics}
\usepackage{epsfig,wrapfig,color,lscape}
\usepackage{amssymb, amsmath}
\usepackage{verbatim}
\usepackage{natbib}
\usepackage{authblk}
\usepackage{kotex}
\usepackage{multirow}
\usepackage{lastpage}
\usepackage[normalem]{ulem}
\usepackage[ruled,vlined]{algorithm2e}
\usepackage{algpseudocode}
\usepackage{bm}
\usepackage[title]{appendix}

\usepackage{siunitx}
\usepackage{url} 

\usepackage{longtable} 
\usepackage{arydshln}
\usepackage[usenames,dvipsnames,svgnames,table]{xcolor} 
\usepackage[colorlinks]{hyperref} 
\usepackage{subcaption} 
\usepackage{array}

\definecolor{Gray}{gray}{0.9}
\usepackage{booktabs}
\newcommand{\hide}[1]{}

\makeatletter

\newtheorem{theorem}{Theorem}[section]
\newtheorem{lemma}[theorem]{Lemma}
\newtheorem{proposition}[theorem]{Proposition}

\newenvironment{proof}[1][Proof]{\begin{trivlist}
\item[\hskip \labelsep {\bfseries #1}]}{\end{trivlist}}

\newcommand{\bbR}{\mathbb{R}}

\newcommand{\bbN}{\mathbb{N}}
\newcommand{\bbX}{\mathbb{X}}

\newcommand{\bfS}{\mathbf{S}}
\newcommand{\bfLambda}{\mathbf{\Lambda}}

\newcommand{\sfD}{\mathsf{D}} 
\newcommand{\sfC}{\mathsf{C}}   
\newcommand{\sfVR}{\mathsf{VR}} 

\newcommand{\Frechet}{Fr\'{e}chet }

\newcommand{\Erdos}{Erd\H{o}s}
\newcommand{\Renyi}{R\'{e}nyi }

\newcommand{\rmE}{\textrm{E}}

\begin{document}

\def\spacingset#1{\renewcommand{\baselinestretch}%
{#1}\small\normalsize} \spacingset{1}


\title{Comparing multiple latent space embeddings using topological analysis}
\author[1]{Kisung You}
\author[2,3]{Ilmun Kim}
\author[2,3]{Ick Hoon Jin}
\author[4]{Minjeong Jeon}
\author[1]{Dennis Shung}
\affil[1]{Department of Internal Medicine, Yale University}
\affil[2]{Department of Applied Statistics, Yonsei University}
\affil[3]{Department of Statistics and Data Science, Yonsei University}
\affil[4]{Graduate School of Education \& Information Studies, University of California, Los Angeles}
\date{}

\maketitle

\bigskip
\begin{abstract}
The latent space model is one of the well-known methods for statistical inference of network data. While the model has been much studied for a single network, it has not attracted much attention to analyze collectively when multiple networks and their latent embeddings are present. We adopt a topology-based representation of latent space embeddings to learn over a population of network model fits, which allows us to compare networks of potentially varying sizes in an invariant manner to label permutation and rigid motion. This approach enables us to propose algorithms for clustering and multi-sample hypothesis tests by adopting well-established theories for Hilbert space-valued analysis. After the proposed method is validated via simulated examples, we apply the framework to analyze educational survey data from Korean innovative school reform. 
\end{abstract}

\noindent%
\vfill

\newpage
\spacingset{1.9} 
\section{Introduction}\label{sec:introduction}

The network is a prevalent form of data containing information about the relationship among objects and has attracted much interest from a number of disciplines, including neuroscience, sociology, political science, linguistics, and so on \citep{newman_networks_2010, barabasi_network_2016}. Its universality has also driven methodological developments in various fields with specific designs to capture different characteristics of the data \citep{goldenberg_survey_2009}. The latent space model (LSM) was proposed in a seminal paper of \cite{hoff_latent_2002-1}, which popularized model-based statistical analysis of network-valued data. Given an observed binary, an undirected network of $n$ nodes represented by an adjacency matrix $A$, the original distance model of \cite{hoff_latent_2002-1} assumes that 
\begin{equation}\label{eq:lsm_binary}
\begin{gathered}
A_{ij} = A_{ji} \overset{ind}{\sim} \text{Bernoulli}(p_{ij}),\quad  i,j=1,\ldots,n, \\
\textrm{with }  \text{logit} (p_{ij}) = \log \left( \frac{p_{ij}}{1-p_{ij}}\right) = \alpha  - \delta (z_i, z_j),
\end{gathered}
\end{equation}
for an intercept parameter $\alpha \in \bbR$ and latent vectors $\lbrace z_i \rbrace_{i=1}^n$ that correspond to positions of the nodes in some low-dimensional Euclidean space $\bbR^p$ endowed with standard $L_2$ norm or a general metric space $(\bbX, \delta)$ \citep{smith_geometry_2019}. 
The model postulates that when two nodes are close in an underlying latent space, it is more likely that there exists an edge between two nodes. As shown in Equation \eqref{eq:lsm_binary}, LSM is closely connected to generalized linear models \citep{nelder_generalized_1972}; hence it provides a flexible framework for broader classes of networks via a choice of link functions, higher degree of interpretability, and statistical treatment of uncertainty in network analysis.

An interesting direction for network-valued data analysis is to study a collection of networks  \citep{tantardini_comparing_2019}. One notable example is a group-level analysis of human brains where each individual's functional or structural network is constructed from non-invasive measurements such as magnetic resonance imaging, and a collection of attained networks is analyzed to find differences of network patterns between normal and control samples \citep{park_structural_2013-1}. The topic of inference for a population of networks has attained much attention recently with a variety of tasks such as hypothesis testing \citep{10.1214/16-AOAS1015,9233364}, clustering \citep{10.5555/3295222.3295450}, classification \citep{10.1214/19-AOAS1252} to name a few.


Moving forward, our main interest is to perform common statistical tasks such as clustering, visualization, or hypothesis testing based on the joint estimates of multiple latent embeddings. This means that the unit of analysis is point sets or empirical measures. Studying point sets has long been a central theme in statistical shape analysis \citep{dryden_statistical_1998} where a shape is represented as a set of points called landmarks and analysis of multiple shapes is performed using the language of statistics on Riemannian manifolds \citep{bhattacharya_nonparametric_2015}. The latter view of latent embeddings as empirical measures is closely related to the theory of optimal transport and has recently gained popularity due to accompanying progress in computational apparatus \citep{villani_optimal_2009}.

Both approaches, however, may not be suitable for learning with multiple latent space embeddings. The distance-based model is identifiable only up to isometry according to the formulation in Equation \eqref{eq:lsm_binary}. For example, a rigid transformation $T(x) = Rx + t$ preserves the distance between two points in Euclidean space, i.e., $d(z_i, z_j) = d(T(z_i), T(z_j))$. This prohibits a direct application of the aforementioned approaches for inference since two sets of latent vectors $\lbrace z_i \rbrace_{i=1}^n$ and $\lbrace T(z_i) \rbrace_{i=1}^n$ carry identical information regarding the shape of a network. The approach based on shape analysis has further drawbacks in a realistic setting. First, aligning multiple shapes requires all point sets with equal cardinalities to apply Procrustes analysis after normalization. Second, shape representations require explicit correspondence and orderings of points in general. These issues make learning with multiple networks problematic if we have a number of latent embeddings whose sizes are not identical or labeling each node to construct correspondence across multiple network embeddings is not available.

We propose a novel framework to analyze multiple latent embeddings for a collection of networks when they are potentially of different sizes and no correspondence is available using the language of topological data analysis (TDA). For each latent embedding, persistent homology, which quantifies topological characteristics of a given point set, is recovered by choice of the simplicial complex and topological information is fully encoded as a multiset of points called persistence diagram. The next step converts each persistence diagram into an informative representation called persistence landscape. This transformation allows each network to be represented as a functional object in Banach or Hilbert space, the theory thereof has been long established in the branch of functional data analysis. With multiple persistence landscapes derived from latent embeddings, we focus on hypothesis testing and cluster analysis among many learning tasks. We first introduce two hypothesis testing procedures for the equality of multiple distributions based on the theory of energy statistics \citep{szekely_e-statistics_2002, szekely_energy_2017}. We also present $k$-medoids \citep{kaufman_partitioning_1990}, $k$-groups \citep{li_k-groups_2017}, and spectral clustering \citep{von_luxburg_tutorial_2007-1} algorithms for cluster analysis.

The rest of the paper is organized as follows. Section \ref{sec:background} introduces the educational panel data that motivates this study and provides a concise introduction to TDA with minimal exposure of relevant concepts. Methods for multi-sample hypothesis testing and cluster analysis are described in detail in Section \ref{sec:main} along with theoretical validation for the application of energy statistics to the space of persistence landscapes. Section \ref{sec:experiment} demonstrates the effectiveness of our framework on two simulated settings where networks of varying sizes are generated from different network models that incorporate heterogeneous topological properties.  We also apply our framework to analysis of the aforementioned panel data in the context of latent space modeling of the item response data. We conclude in Section \ref{sec:conclusion} by highlighting the unique advantages of our framework and discussing potential directions for future work.

\section{Background}\label{sec:background}
\subsection{Motivation}

A motivating data is taken from  \cite{jin_hierarchical_2020} that investigated the impact of  the ``innovation school program'' of South  Korea. 
The innovation school program was initiated in 2009 as a response to large criticism over the public K-12 education system of South Korea for its excessively competitive environment. This educational reform program aimed to endow schools with substantial degree of autonomy and foster self-directed learning and creative environment  \citep{gyeonggi}. To evaluate the impact of the program using student responses to item-level questions,  \cite{jin_hierarchical_2020} used the Gyeonggi Education Panel Study (GEPS) data, a large-scale panel survey on representative samples of K-12 students in Gyeonggi province which is one of the first provinces that adopted the innovation school program. The GEPS data contain a rich set of student- and school-level variables across three school levels - elementary, middle, and high schools. The data also contains student-level measures on psychological/attitude attributes perceived by students, such as mental well-being, self-efficacy, academic stress, relationship with friends, etc. From stakeholders' perspectives, a significant question to be answered would be whether the newly adopted program was able to make intended differences in the non-cognitive outcomes of schooling. 

\cite{jin_doubly_2019} applied a special type of latent space approach that  models items and individuals simultaneously in a common latent space, so-called the network item response model (NIRM). To briefly explain,  the joint modeling framework  involves two intercept-embedding pairs $(\beta, W)$ and $(\theta, Z)$ for items and individuals respectively at each school and multiple schools are modeled hierarchically. NIRM was estimated by using fully Bayesian approach with MCMC. We refer to \cite{jin_doubly_2019} for additional details of the model specification and estimation. 

Although NIRM can provide intuitive explanations of the differences in the dependence structures among items and individuals between the school types based on visualized dependences in the latent space, the rigorous analysis of the latent embeddings at the population level is challenging due to the varying within-school sample sizes across schools  as well as invariance under the rigid motion such as translation and rotation of the model estimates. The last column in Table \ref{table:real_exploratory} shows that the range of the school size, number of students within school, is fairly large at the all three school levels (elementary, middle, and high schools).

\begin{table}[htbp]
	\begin{center}
		\begin{tabular}{l|c|c|c|c|c}
			\hline
			\multirow{2}{*}{level} & \multicolumn{3}{c|}{\# schools} & \multirow{2}{*}{\# items} & \multirow{2}{*}{school size} \\ \cline{2-4}
			& innovative     & regular     & total    &  &  \\ \hline
			elementary & 17 & 54 & 71 & 60 & [21,63] \\
			middle & 21 & 42 & 63 & 70 & [29,78] \\ 
			high & 16 & 46 & 62 & 72 & [37,81] \\ \hline
		\end{tabular}
	\end{center}
	\caption{Number of schools, items, and range of the number of individuals per site (school size) for three school levels.}
	\label{table:real_exploratory}
\end{table}

\subsection{Brief Introduction to Topological Data Analysis}
We now introduce some basic concepts in TDA at the minimal level to suffice for what follows. We refer interested readers to \cite{Zomorodian_2005, edelsbrunner_computational_2010, Wasserman_2018} for formal introduction and details of the topic.

\begin{figure}[h]
	\centering
	\includegraphics[width=1.0\textwidth]{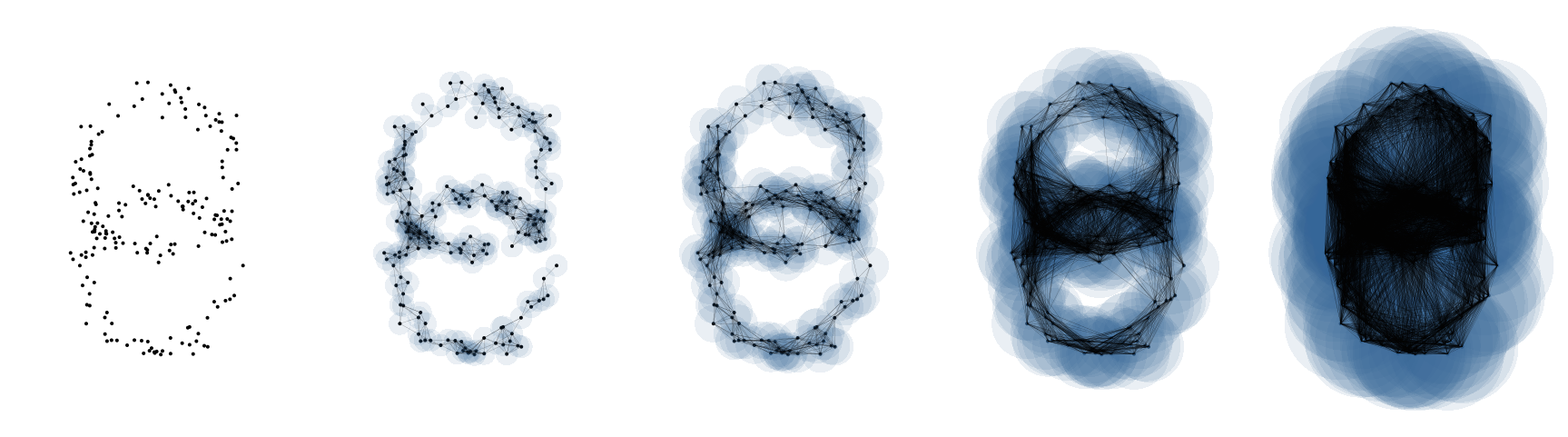}
	\caption{A randomly generated sample $X$ from two intersecting circles at the left and 1-skeletons of the \v{C}ech complexes from $X_r$ for $r = \lbrace0.05, 0.1, 0.25, 0.50\rbrace\cdot r_\text{max}$ where $r_\text{max} = \text{max} \lbrace d(x_i, x_j)~\vert~ x_i, x_j \in X \rbrace $.}
	\label{fig:filtration}
\end{figure}

Let $X=\lbrace x_1, x_2, \ldots, x_m \rbrace$  be a set of points in a metric space $(\bbX, \delta)$, $B_r (x_i) = \lbrace y \in \bbX ~\vert~ \delta(x_i, y) \leq r \rbrace$ a ball of radius $r$ centered at $x_i$, and $X_r = \bigcup_{i=1}^m  B_r (x_i)$ a union of balls across all points in $X$. As shown in Figure \ref{fig:filtration}, we first observe that when $r=0$, $X_r$ has $m$ connected components that consist of singletons. As $r \rightarrow \infty$, some balls in $X_r$ coalesce over the course of radius, generating a smaller number of larger connected components and $X_r$ eventually merges into one big component for sufficiently large value of $r$. In algebraic topology, the numbers $\beta_0, \beta_1, \beta_2, \ldots$ are known as Betti numbers where $\beta_k$ is the rank of the $k$-th homology group. Homology groups of $0$-th, $1$-st, and $2$-nd orders/dimensions characterize connected components, loops, and voids, respectively. For example, a filled circle has $\beta_0 = 1, \beta_1 = 0$ since it is connected and has no one-dimensional hole. 

The previous observation of varying topological structures in $X_r$ over a range of the radius necessitates to take a multiscale perspective, which is examined by persistent homology. As $r$ varies, topological features such as connected components and holes may appear, alter, or disappear. Based on the set of balls $X_r$, the \v{C}ech complex $\sfC_r (X)$ is the simplicial complex with vertices $\lbrace x_i\rbrace $ and $k$-simplices correspond to $k+1$ balls whose intersections are non-empty. For example, for a fixed $r$, $\sfC_r (X)$ contains all singletons $\lbrace x_i\rbrace$ which are $0$-dimensional simplices. For $1$-dimensional simplices, all pairs $(i,j)$ such that $ d(x_i, x_j) \leq 2r$ are included in $\sfC_r (X)$. Likewise, any triplets $(i,j,k)$ where the intersection of $B_r (x_i)$, $B_r (x_j)$, and $B_r (x_k)$ is non-empty are included as $2$-dimensional simplices. The benefit of \v{C}ech complex is that $\sfC_r (X)$ is homotopy equivalent to $X_r$ if the amibient space is Euclidean, i.e. $\bbX = \mathbb{R}^p$. Roughly speaking, the homotopy equivalence means that one can be continuously transformed into another. A collection of \v{C}ech complexes $\lbrace \sfC_r (X)\rbrace_{r \geq 0}$ forms a filtered simplicial complex and the persistent homology is obtained thereof. Even though the homology of the \v{C}ech complex can be computed using elementary matrix operations \citep{edelsbrunner_computational_2010}, it is still computationally expensive for large input matrices. A popular alternative to the \v{C}ech complex is the Vietoris-Rips (VR) complex $\sfVR_r (X)$ whose persistent homology approximates that defined by the \v{C}ech complex \citep{de_silva_coverage_2007}. In VR complex, $k$-simplices are included only if all pairwise intersections are non-empty. 

\begin{figure}
\begin{subfigure}{.5\textwidth}
  \centering
  \caption{}
  \includegraphics[width=6cm]{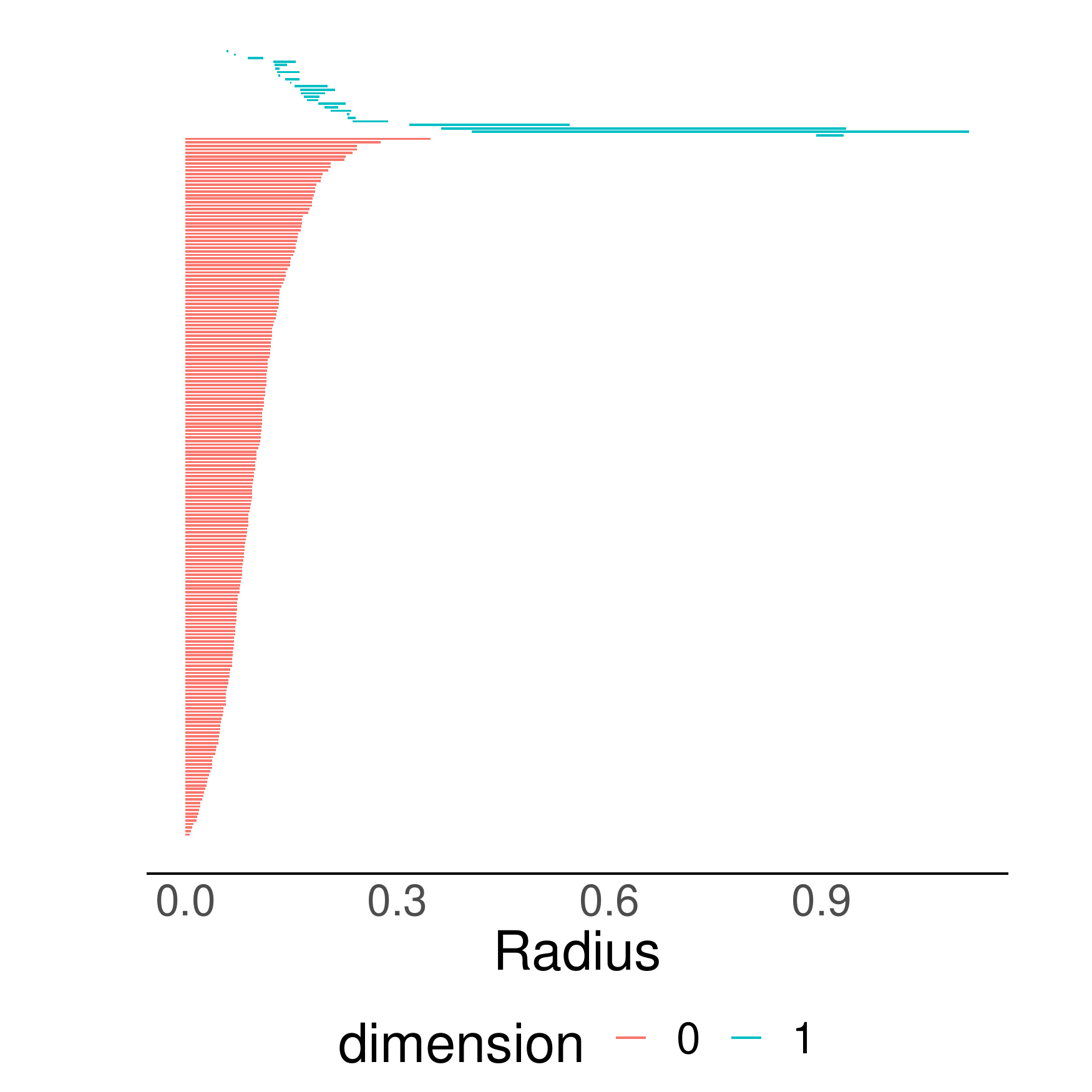}
\end{subfigure}%
\begin{subfigure}{.5\textwidth}
  \centering
  \caption{}
  \includegraphics[width=6cm]{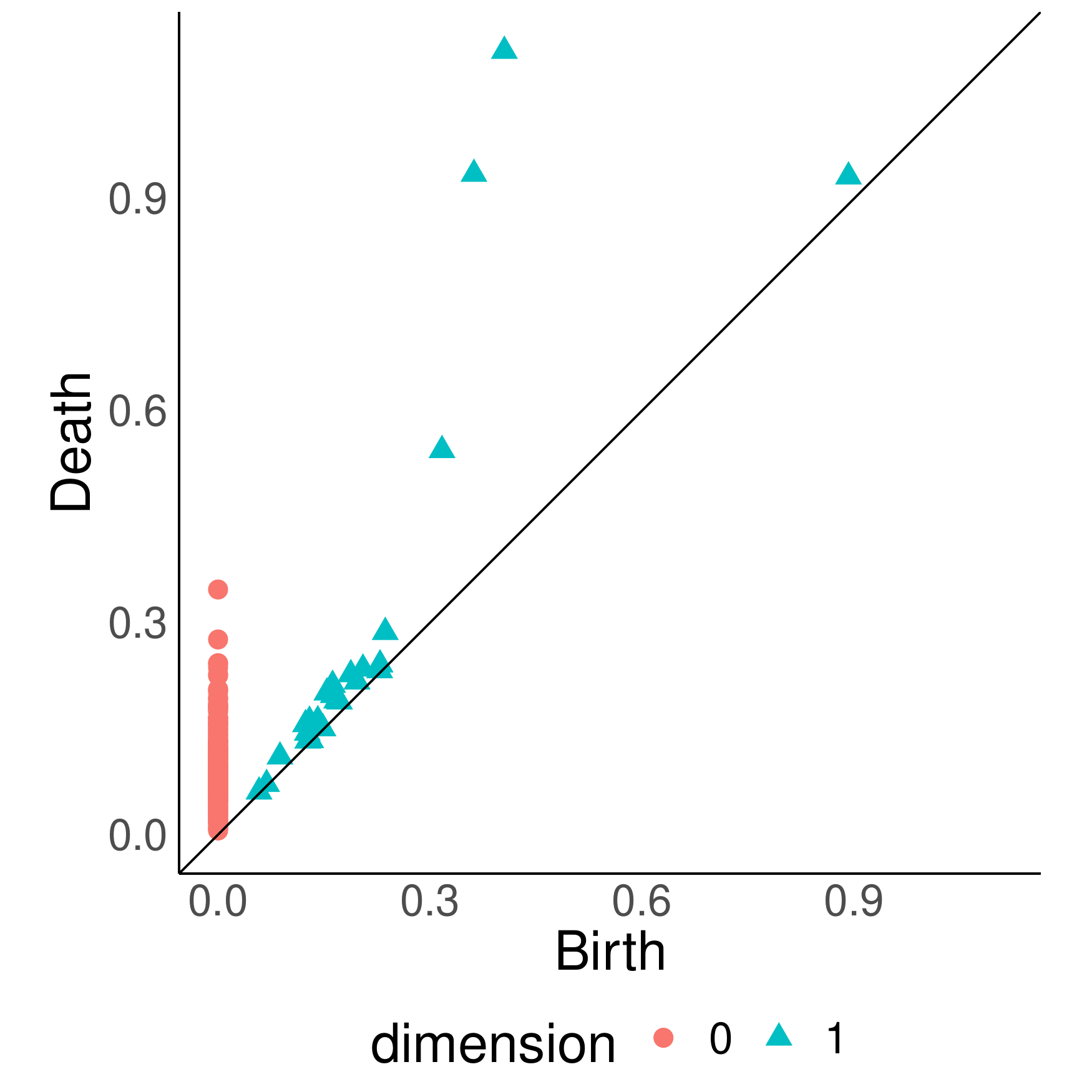}
\end{subfigure}
\caption{Representations of persistent homology via (a) barcode and (b) persistence diagram from Vietoris-Rips filtration of an intersecting circles example as shown in Figure \ref{fig:filtration}.}
\label{fig:PDbarcode}
\end{figure}

Given the persistent homology defined by a choice of complex, the information of topological features can be completely represented as a persistence diagram \citep{cohen-steiner_stability_2007} or a barcode \citep{collins_barcode_2004} as shown in Figure \ref{fig:PDbarcode}. A barcode is a multiset of intervals that connect $r$ values at which a feature first appears (birth) and disappears (death) for each dimension in persistent homology. The persistence diagram is obtained by mapping each interval's endpoints to $x$- and $y$-axis. The shorter the length of an interval is, the closer the birth and death are in that a feature represented as a short interval lies closer to the 45-degree reference line of the persistence diagram. An interesting observation is that the space of persistence diagrams is complete and separable metric space, allowing to define probability measures under the Wasserstein metric \citep{mileyko_probability_2011}. From now on, we may assume any persistence diagram $\sfD$ is composed of finitely many birth-death pairs, which can be justified by the fact that computationally we consider a point set of finite cardinality, and a common practice is to truncate all features beyond the maximum filtration value \citep{baas_persistence_2020}.

The persistence landscape \citep{bubenik_statistical_2015, baas_persistence_2020} is a functional representation of the persistence diagram in Banach space or even Hilbert space. One way to construct the persistence landscape uses birth-death pairs from a persistence diagram  $\sfD = \lbrace (b_i, d_i) \rbrace_{i\in I}$ for an index set $I$. For $b < d$, define a piecewise linear function$f (t\vert b,d) = \min (b+t, d-t)_+$ where $x_+$ denotes $\max(x, 0)$. The $k$-th persistence landscape function $\Lambda_k : \bbR \rightarrow \bbR$ is defined as 
\begin{equation*}
\Lambda_k (t) = \textrm{kmax} \lbrace f(t\vert b_i, d_i) \rbrace_{i\in I},
\end{equation*}
where $\textrm{kmax}$ denotes the $k$-th largest value. The persistence landscape $\Lambda : \bbN \times \bbR \rightarrow \bbR$ is a sequence of functions $\lbrace \Lambda_k  \rbrace_{k\in\bbN}$ where each function $\Lambda_k$ is again piecewise linear with slopes 1, 0, or -1. As a real-valued function on $\bbN\times\bbR$, we observe that the persistence landscape is nonnegative, integrable and piecewise linear in that it lies in a separable Banach space $L_p (\bbN \times \bbR)$ with norm
\begin{equation*}
\|\Lambda \|_p =  \left(\sum_{k=1}^\infty \|\Lambda_k\|_p^p\right)^{1/p} = \left(\sum_{k=1}^{\infty} \left\lbrack \int_{-\infty}^{\infty} \Lambda_k (t)^p dt \right\rbrack\right)^{1/p},
\end{equation*}
for $1\leq p < \infty$ with the product of the Lebesgue measure on $\bbR$ and the counting measure on $\bbN$. When $p=2$, the persistence landscape is a Hilbert space-valued object which ensures the validity of energy-based statistical inference that will be discussed in the following section.

We close this section by introducing some properties of the persistence landscape that make it an appealing object for inference \citep{bubenik_statistical_2015}. First, the mapping from a point cloud to a persistence landscape is non-expansive, and the mapping from persistence diagrams to persistence landscapes is invertible if diagrams are connected and arithmetically independent  \citep{baas_persistence_2020, betthauser_graded_2021}. The latter provides grounds for learning on the space of persistence landscapes as a proxy for that on the space of persistence diagrams. Second, the persistence landscape is stable. To describe, given two persistence diagrams $\sfD, \sfD'$ and their landscapes $\Lambda, \Lambda'$, 
\begin{equation*}
\| \Lambda - \Lambda' \|_\infty = \underset{(k,t) \in \bbN \times \bbR}{\sup} |\Lambda_k (t) - \Lambda'_k (t) | \leq d_B (\sfD, \sfD'),
\end{equation*}
where $d_B$ denotes the bottleneck distance \citep{cohen-steiner_stability_2007} between two persistence diagrams. Next, the law of large numbers and the central limit theorem, two fundamental building blocks of statistical inference, hold for persistence landscapes in a Banach space given finite second moment, i.e.~$\textrm{E} \| \Lambda \|^2 < \infty$. We refer to \cite{bubenik_statistical_2015} for a rigorous treatment of the law of large numbers and the central limit theorem for persistence landscapes and we assume $\textrm{E} \| \Lambda \|^2 < \infty$. 
Finally, the persistence landscape representation has computational benefits over other alternatives. For example, as a generalization of centroids, the \Frechet mean is a primary summary statistic in the analysis of persistence diagrams. However, computing the \Frechet mean is notoriously difficult with persistence diagrams, involving a combinatorial subproblem of high complexity $O(NM^3)$ for each iteration where $N$ is the number of sample diagrams and $M$ is the cardinality of a target diagram \citep{turner_frechet_2014}. In the machine learning community, a similar problem is known as the Wasserstein barycenter \citep{agueh_barycenters_2011, villani_optimal_2009}. However, complex operations with multiple persistence diagrams beyond the mean are still arduous, and the lack of computational dexterity prohibits further statistical analysis on multiple persistence diagrams. On the other hand, persistence landscape is a random object in some vector spaces in that standard computational pipelines are directly applicable.

\section{Learning with multiple latent space embeddings}\label{sec:main}

\begin{figure}[htbp]
\centering
\begin{tabular}{cc}
(a) & (b)  \\
\includegraphics[width=0.4\textwidth]{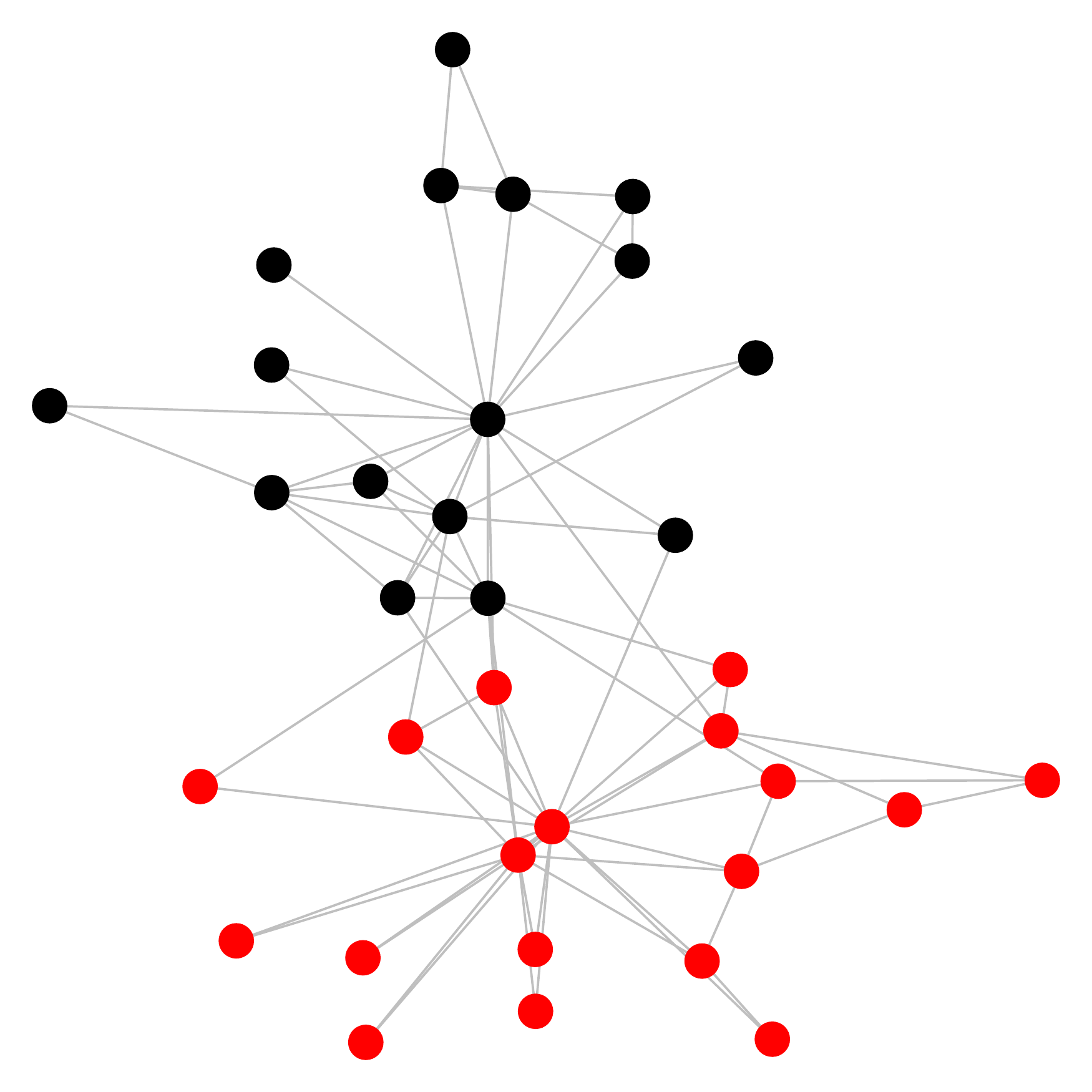} & 
\includegraphics[width=0.4\textwidth]{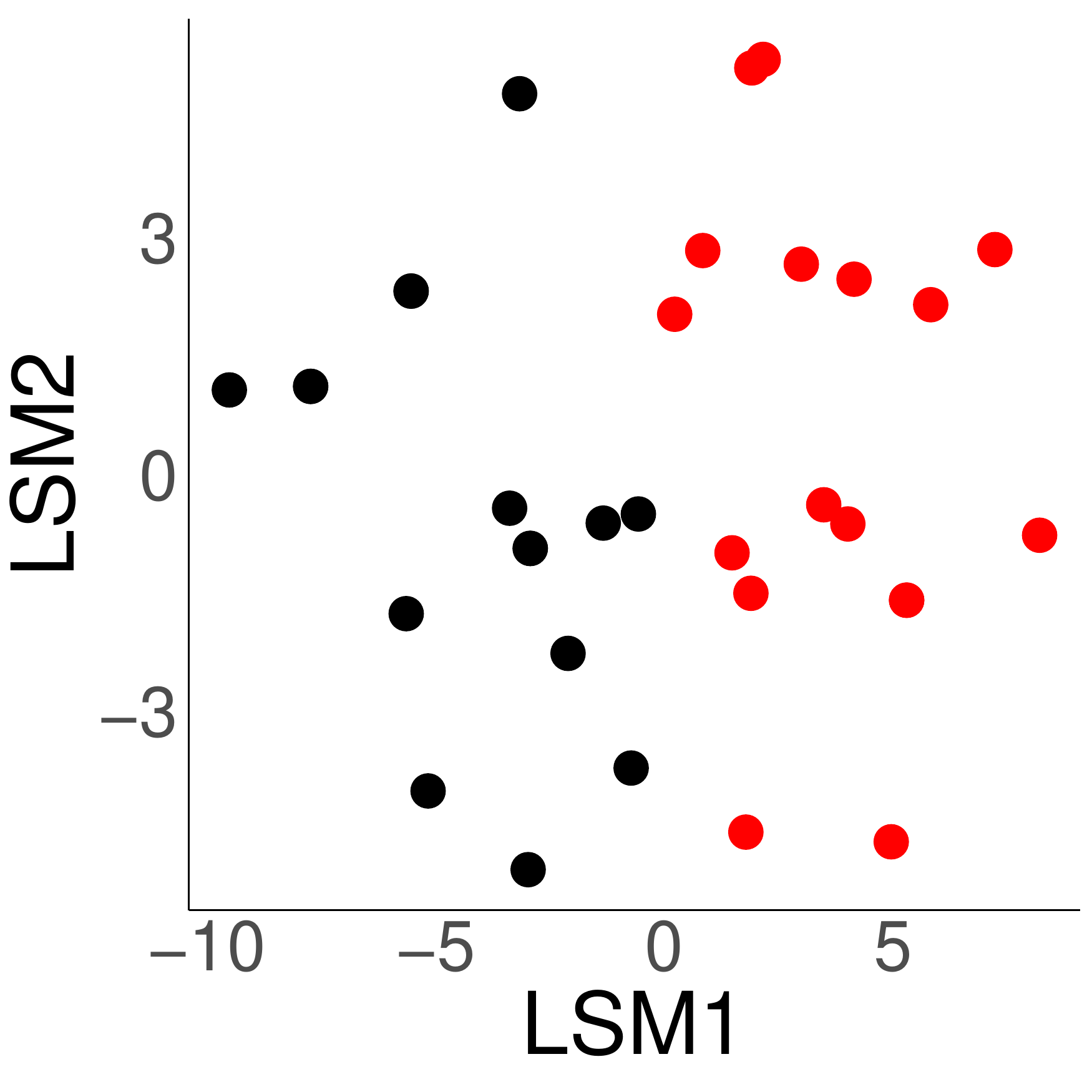}\\
(c) & (d)  \\
\includegraphics[width=0.4\textwidth]{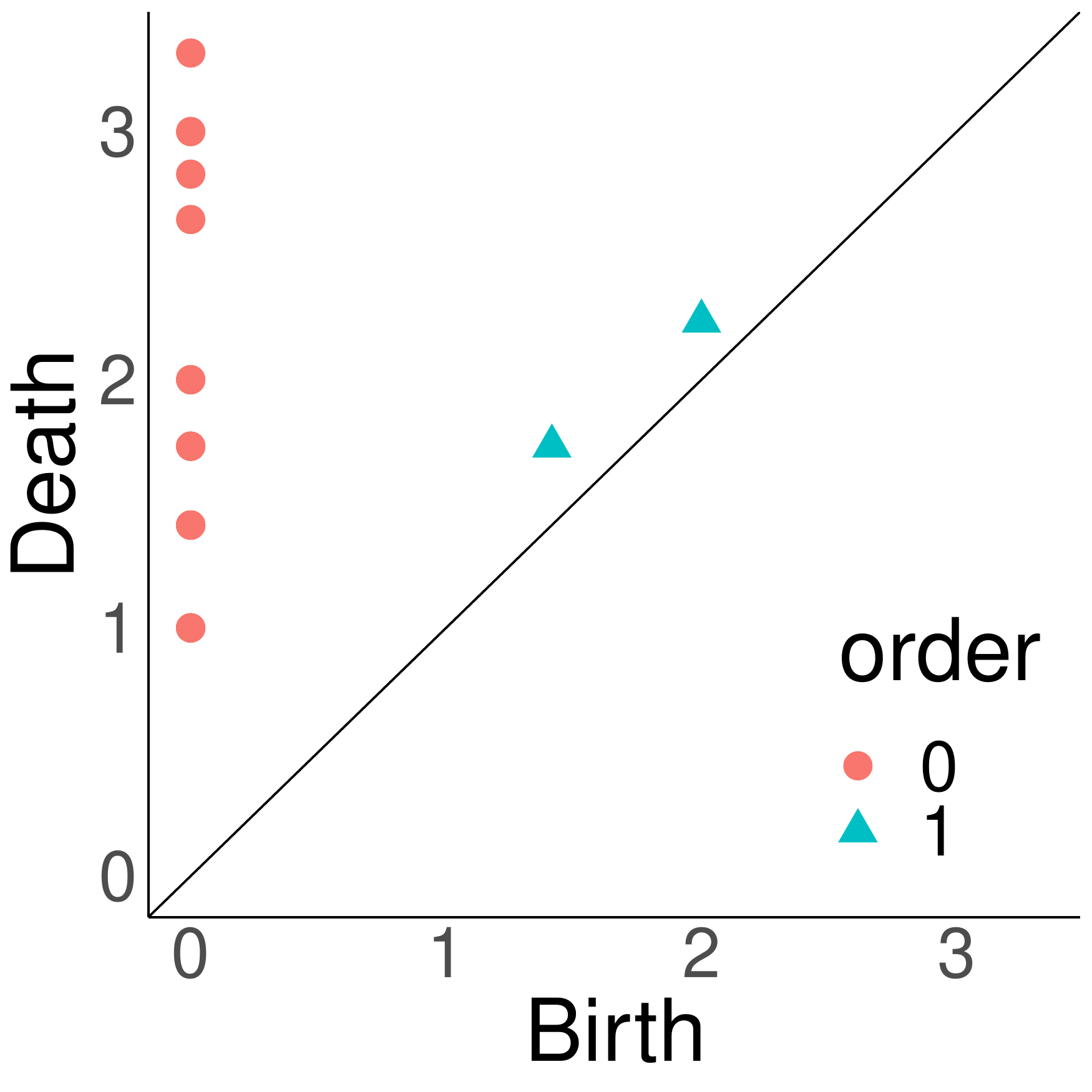} &
\includegraphics[width=0.4\textwidth]{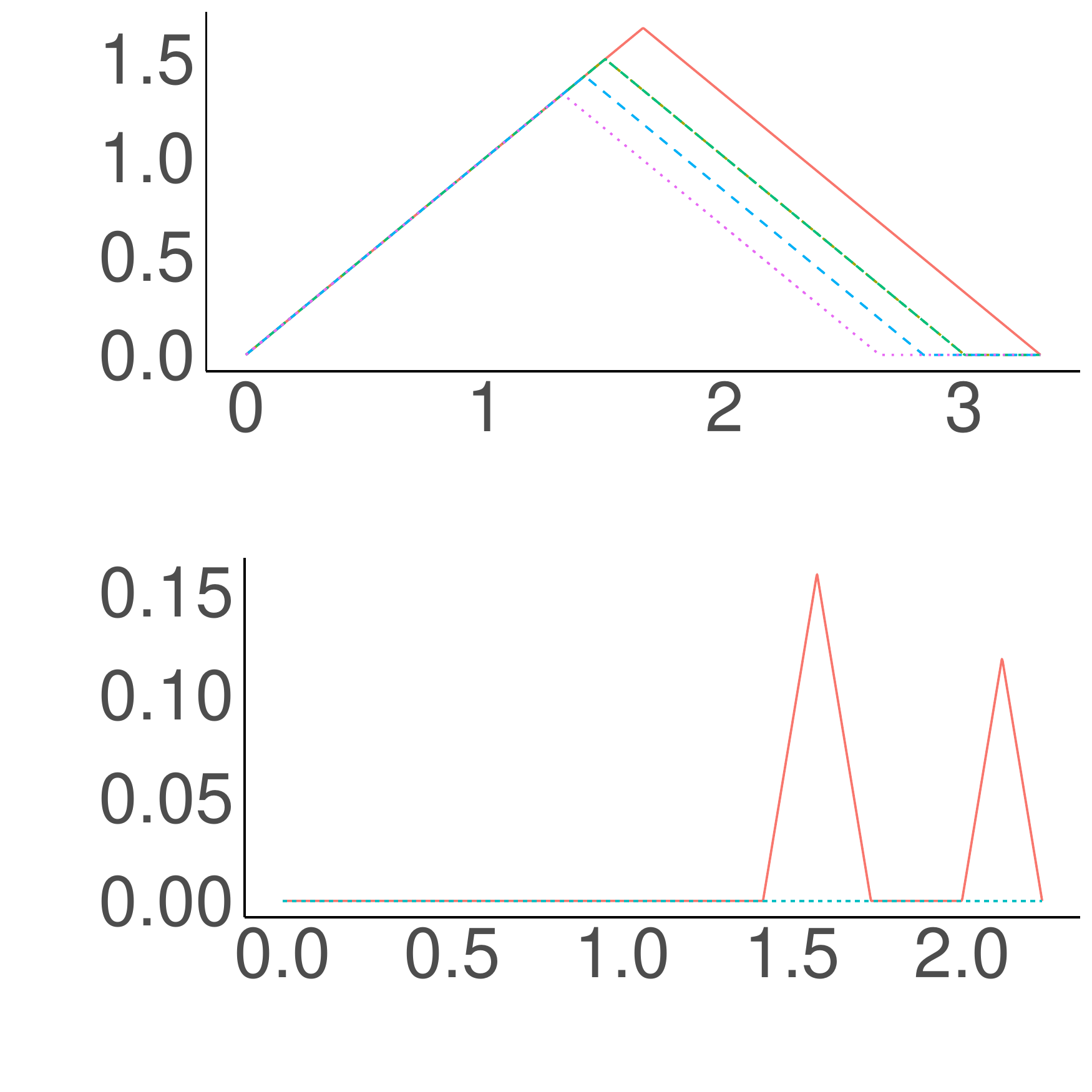}
\end{tabular}
\caption{
A diagram from a network to persistence landscape. Given a network (a), latent space model is fitted to find a low-dimensional embedding (b). Persistent homology of an embedding is represented by (c) persistence diagram, which is transformed into (d) persistence landscapes of order 0 (top) and 1 (bottom).}
\label{fig:flowchart}
\end{figure}

We present methods to learn with multiple networks of varying sizes with a class of latent space models. As mentioned before, we use persistence landscape as a stable summary statistic obtained as follows. For a given network $G$, a latent space model is fitted with an estimate of latent embedding $Z$, and its persistent homology is captured from a VR complex of the embedding  $\textsf{VR}_r (Z)$. The persistence landscape $\Lambda$ is obtained from representation in either a barcode or a diagram form with an order of interest. These steps play a role in finding a common representation for networks of varying sizes, which is summarized in Figure \ref{fig:flowchart}.

\subsection{Hypothesis testing}

We first consider the task of hypothesis testing to compare two or more samples of networks.   \cite{bubenik_statistical_2015} proposed a two-sample $z$-test using the central limit theorem for persistence landscapes. Although analytic properties of the test are well known, this approach is limited because the test procedure depends on the choice of functional that maps persistence landscapes to a scalar value. Furthermore, the null hypothesis of the $z$-test states that two sets of random variables in Banach space have the same mean, which does not account for higher-order moments. Here we propose an alternative approach for two- and multi-sample tests of equal distributions using energy statistics.

Energy statistics is a class of statistics based on distances between observations. In the Euclidean setting, the energy distance between two independent random variables $X$ and $Y$ is defined as
\begin{equation}\label{def:energy_distance}
\mathcal{E}(X,Y) = 2 \rmE [\delta(X,Y)]  - \rmE [\delta(X,X')] - \rmE[\delta(Y,Y')],
\end{equation}
where $X'$ and $Y'$ are independent and identical copies of $X$ and $Y$, respectively, and $\delta(\cdot,\cdot)$ is the standard Euclidean distance. An important fact about the energy distance is that the quantity $\mathcal{E}(X,Y) \geq 0$ and the equality holds if and only if $X$ and $Y$ are identically distributed random variables \citep{szekely_testing_2004}. This characteristic property leads to a two-sample test for equality of distributions for random variables in Euclidean space using an empirical estimate of $\mathcal{E}(X,Y)$. In particular, given independent random samples $X_1, \ldots, X_m$ and $Y_1,\ldots,Y_n$, the empirical energy distance between $X$ and $Y$ is computed as
\begin{equation}\label{def:energy_twosample_finite}
\mathcal{E}_{m,n}(X,Y) = \frac{2}{mn}\sum_{i=1}^m \sum_{j=1}^n \|X_i-Y_j\| - \frac{1}{m^2}\sum_{i=1}^m \sum_{i'=1}^m \|X_i - X_{i'}\| - \frac{1}{n^2} \sum_{j=1}^n \sum_{j'=1}^n \|Y_j - Y_{j'}\|.
\end{equation}
The null hypothesis of equal distributions is then rejected when the scaled energy test statistic 
\begin{equation*}
T_{m,n} = \frac{mn}{m+n} \mathcal{E}_{m,n}(X,Y).
\end{equation*}  
is larger than some threshold, which is known to be a consistent test against general alternatives~\citep{szekely_testing_2004}.

For general metric spaces, however, the energy distance $\mathcal{E}(X,Y)$ is neither necessarily non-negative nor a valid measure for testing equality of distributions \citep{klebanov_n-distances_2006}. In order to justify the use of the energy statistics in our setting, we build on several results from previous work. Let $(\bbX,\delta)$ be a metric space. We say that $(\bbX,\delta)$ has negative type if $\sum_{1 \leq i,j\leq n} \alpha_i \alpha_j \delta(x_i, x_j) \leq 0$
holds for all $n\geq 1$ with $x_1,\ldots,x_n \in \bbX$, and an arbitrary collection of coefficients $\alpha_1,\ldots,\alpha_n \in \bbR$ such that $\sum_{i=1}^n \alpha_i = 0$. A metric space $(\bbX, \delta)$ is called to have strong negative type if it has negative type and 
\begin{equation*}
\int \delta(x_1,x_2) d(\mu_1 - \mu_2)^2 (x_1,x_2) = 0
\end{equation*}
if and only if $\mu_1 = \mu_2$ where $\mu_1$ and $\mu_2$ are the Borel probability measures on $\bbX$ with finite first moments. The following proposition presents conditions under which the energy distance defined in a general metric space characterizes the equality of distributions as in the Euclidean space. This further validates our framework of energy-based learning for the persistence landscapes.

\begin{proposition}[Proposition 3 of \cite{szekely_energy_2017}]\label{prop:energy}
	Let $X,Y$ be independent random variables with the Borel probability measures $\mu,\nu$ on $(\bbX, \delta)$, respectively, and $X'$ and $Y'$ are i.i.d.~copies of $X$ and $Y$. The metric space $(\bbX, \delta)$ has negative type if and only if, for all $X, Y$, the following inequality holds
	\begin{equation*}
	2\mathrm{E} [\delta(X,Y)] - \mathrm{E} [\delta(X,X')] - \mathrm{E}[\delta(Y,Y')] \geq 0.
	\end{equation*}
	Furthermore, a necessary and sufficient for the claim that the equality holds if and only if $\mu=\nu$ is that the metric space has strong negative type.
\end{proposition}

\begin{proposition}\label{prop:energy_in_Hilbert}
	The space of persistence landscapes admits $\mathcal{E}(\mathbf{\Lambda}, \bfLambda') \geq 0$ and the equality holds if and only if distributions of $\bfLambda$ and $\bfLambda'$ are identical for $p=2$. 
\end{proposition}

\begin{proof}
	It is well known that $L_2 (\bbR)$ is a separable Hilbert space \citep{stein_functional_2011}. From a topological perspective, recall that a product of countably many separable spaces is separable \citep{willard_general_1970}. The product space is equivalently expressed as a direct sum of Hilbert spaces. If we denote $H_i = L_2 (\bbR)\textrm{ for all } i \in \bbN$, then 
	\begin{equation*}
	L_2 (\bbN \times \bbR) = \bigoplus_{i \in \bbN} H_i =  \left\lbrace h \in \prod_{i\in \bbN} H_i : \sum_{i \in \bbN}    \|h(i)\|^2 < \infty \right\rbrace,
	\end{equation*}
	with an inner product
	\begin{equation*}
	\langle g, h \rangle = \sum_{i \in \bbN} \langle g(i), h(i) \rangle,\quad \forall g,h \in \bigoplus_{i \in \bbN} H_i.
	\end{equation*}
	It was shown in \cite{conway_course_1997} that the Hilbert space direct sum is a vector space with a well-defined inner product, and every Cauchy sequence in $\bigoplus_{i \in \bbN} H_i$ converges in itself in that $L_2 (\bbN \times \bbR)$ is a Hilbert space.

	Theorem 3.16 of \cite{lyons_distance_2013} states that every separable Hilbert space has strong negative type. By Proposition \ref{prop:energy} and the strong negative type, $L_2 (\bbN \times \bbR)$ admits the energy distance with qualifications stated in the claim with the norm 
	\begin{equation*}	 
	\delta_2(\Lambda,\Lambda') := \| \Lambda -\Lambda' \|_2 = \left(
	\sum_{k=1}^\infty \left\lbrack\int_{-\infty}^{\infty}
	\left(	\Lambda_k (t) - \Lambda'_k(t)\right)^2 dt
	\right\rbrack
	\right)^{1/2}. 
	\end{equation*}\hfill $\square$
\end{proof}

We propose a modified version for the two-sample test of homogeneity \citep{szekely_testing_2004} based on Proposition \ref{prop:energy_in_Hilbert}. Let $\bfLambda_1 = \Lambda_{1,1}, \ldots, \Lambda_{1,n_1}$ and $\bfLambda_2 = \Lambda_{2,1}, \ldots, \Lambda_{2,n_2}$ be independent random samples for the Borel probability measures $\mu_1$ and $\mu_2$. The two-sample test statistic is then given by
\begin{equation}\label{eq:statistic_two_sample}
\begin{split}
T_{n_1,n_2} (\bfLambda_1, \bfLambda_2) &= \frac{n_1 n_2}{n_1 + n_2} 
\left( \frac{2}{n_1 n_2} \sum_{i=1}^{n_1} \sum_{j=1}^{n_2} \delta_2 (\Lambda_{1,i}, \Lambda_{2,j}) \right. \\
&\quad \left. - \frac{1}{n_1^2} \sum_{i,i'=1}^{n_1} \delta_2 (\Lambda_{1,i}, \Lambda_{1,i'})
- \frac{1}{n_2^2} \sum_{j,j'=1}^{n_2} \delta_2 (\Lambda_{2,j}, \Lambda_{2,j'}) \right),
\end{split}
\end{equation}
and we will simply write $T_{n_1,n_2} (\bfLambda_1, \bfLambda_2)$ as $T_{n_1,n_2}$ whenever it is clear from the context.

Since the null distribution of $T_{n_1,n_2}$ is unknown at least in finite-sample scenarios, we consider the permutation procedure to determine the significance of $T_{n_1,n_2}$~\citep{efron_introduction_1993-1}. To describe the permutation test, let $n = n_1 + n_2$ be the total size of the pooled sample~$\lbrace \tilde{\Lambda}_i \rbrace_{i=1}^n = \bfLambda_1 \cup \bfLambda_2$ and let $\boldsymbol{\sigma} = (\sigma(1),\ldots,\sigma(n))$ be a random vector uniformly distributed over the set of all possible permutations of $\{1,\ldots,n\}$ denoted by $\boldsymbol{\Sigma}_n$. Given $\boldsymbol{\sigma}_1,\ldots, \boldsymbol{\sigma}_B$, which are i.i.d.~copies of $\boldsymbol{\sigma}$, we compute the set of permuted test statistics $\{T_{n_1,n_2}^{(b)}\}_{b=1}^B$ where each $T_{n_1,n_2}^{(b)}$ is computed as in Equation~\eqref{eq:statistic_two_sample} based on $\bfLambda_1^{(b)} = \tilde{\Lambda}_{\boldsymbol{\sigma}_b(1)}, \ldots, \tilde{\Lambda}_{\boldsymbol{\sigma}_b(n_1)}$ and $\bfLambda_2^{(b)} = \tilde{\Lambda}_{\boldsymbol{\sigma}_b(n_1 + 1)}, \ldots, \tilde{\Lambda}_{\boldsymbol{\sigma}_b(n_1+n_2)}$. Then the corresponding permutation $p$-value is defined as 
\begin{equation}\label{eq:pvalue_twosample2}
	\hat{p}  = \frac{1}{B+1} \left( \sum_{b=1}^B I(T_{n_1,n_2} \leq T_{n_1,n_2}^{(b)}) + 1  \right),
\end{equation}
where $I(\cdot)$ denotes an indicator function. For the desired significance level $\alpha \in (0,1)$, the permutation test rejects the null hypothesis $H_0: \mu_1 = \mu_2$ when $\hat{p}$ is smaller than or equal to $\alpha$. Note that $T_{n_1,n_2},\{T_{n_1,n_2}^{(b)}\}_{b=1}^B$ are exchangeable under the null, which guarantees that the considered permutation test is level $\alpha$ due to Lemma 1 of \cite{romano2005exact}. In Appendix~\ref{sec: permutation p-value} of the Supplementary Material, we provide a slightly sharper result than Lemma 1 of \cite{romano2005exact} and its proof, which may be useful in other contexts as well.

The two-sample test can be easily extended for the multi-sample test of homogeneity, also called as  $k$-sample test. Let $\bfLambda_1,\ldots, \bfLambda_k$ be independent random samples of sizes $n_1, \ldots, n_k$ respectively, $\vec{n} = (n_1, \ldots, n_k)$ and $n = n_1 + \cdots + n_k$. The $k$-sample test statistic is defined by summing all $k(k-1)/2$ pairwise energy distances between two independent samples 
\begin{equation} \label{eq:statistic_multi_sample}
T_{n} = \sum_{i=1}^{k-1} \sum_{j=i+1}^k T_{n_i, n_j} (\bfLambda_i, \bfLambda_j),
\end{equation}
where $T_{n_i, n_j} (\bfLambda_i, \bfLambda_j)$ is given by Equation \eqref{eq:statistic_two_sample}. A permutation testing procedure for the $k$-sample test is given as follows; for each permutation $b = 1,2,\ldots,B$, randomly draw a permutation $\boldsymbol{\sigma} \in \boldsymbol{\Sigma}_n$ and generate partitioned samples $\bfLambda_i^{(b)},~i=1,\ldots,k$ of sizes $\vec{n}$ by the permutation $\boldsymbol{\sigma}$ without replacement from the pooled sample $\bfLambda_1 \cup \cdots \cup \bfLambda_k$. Let $T_n^{(b)}$ be the test statistic by the permuted samples $\bfLambda_1^{(b)}, \ldots, \bfLambda_k^{(b)}$ and we reject the null hypothesis $H_0 : \mu_1 = \cdots = \mu_k$ if the permutation $p$-value
\begin{equation*}
\hat{p} = \frac{1}{B+1} \left(\sum_{b=1}^B I(T_n \leq T_n^{(b)}) + 1 \right)
\end{equation*}
is smaller than or equal to to the significance level $\alpha$. By the same reason used for the two-sample case, $\hat{p}$ is a valid $p$-value under the null. We also note that consistency of the two- and $k$-sample tests against all fixed alternatives is shown in \cite{szekely_testing_2004} when $\lim_{n\rightarrow\infty} n_i /n \rightarrow \lambda_i \in (0,1)$ and $\rmE \| \bfLambda_i \| < \infty$ for $i=1,\ldots,k$. 

Distance components (DISCO) is another multi-sample test of homogeneity based on the energy distance \citep{rizzo_disco_2010-1}. DISCO is analogous to the analysis of variance (ANOVA) in the sense that the total dispersion is decomposed into the within- and between-sample dispersions that are measured by distances. For two samples $\bfLambda_1$ and $\bfLambda_2$ of sizes $n_1$ and $n_2$, respectively, let $d_\rho$-distance 
between two independent samples be defined as 
\begin{equation}\label{eq:dist_d_alpha}
d_\rho (\bfLambda_1, \bfLambda_2) := \frac{1}{n_1 n_2} \sum_{i=1}^{n_1} \sum_{j=1}^{n_2} \delta_2 (\Lambda_{1,i}, \Lambda_{2,j})^\rho,
\end{equation}
for $\rho \in (0,2)$. For the pooled sample $\bfLambda = \bfLambda_1 \cup \cdots \cup \bfLambda_k$ where $|\bfLambda_i| = n_i,~i=1,\ldots,k$ and $n = n_1+\cdots+n_k$, define the total dispersion of the observed random variables
\begin{equation}\label{eq:dispersion_total}
T_\rho = \frac{n}{2} d_\rho (\bfLambda, \bfLambda),
\end{equation}
which can be decomposed into the within- and between-sample energy statistics $W_\rho$ and $B_\rho$ 
\begin{gather}
W_\rho = \sum_{i=1}^k \frac{n_i}{2} d_\rho (\bfLambda_i, \bfLambda_i), \label{eq:dispersion_within}\\
B_\rho = \sum_{1 \leq i<j \leq k} \left\lbrack \frac{n_i n_j}{2n} \left( 2 d_\rho(\bfLambda_i, \bfLambda_j) - d_\rho (\bfLambda_i, \bfLambda_i) - d_\rho (\bfLambda_j, \bfLambda_j) \right) \right\rbrack,\label{eq:dispersion_between}
\end{gather}
where the decomposition $T_\rho = W_\rho + B_\rho$ holds and can be easily checked for $0 < \alpha \leq 2$ arithmetically. As pointed out before, this is only valid for $W_\rho, B_\rho \geq 0$ that is guaranteed by Proposition \ref{prop:energy_in_Hilbert}. Corollary 1 of \cite{rizzo_disco_2010-1} shows that $B_\rho = 0$ if and only if $\mu_1 = \cdots = \mu_k$ for $0 < \rho < 2$. When $\rho=2$, the equality holds if and only if $\rmE \mu_1 = \cdots \rmE \mu_k$, which entails similarity of the test to univariate and multivariate ANOVA for random variables in Euclidean space. The test procedure is identical to that of the $k$-sample test of homogeneity. A random permutation $\sigma_b \in \boldsymbol{\Sigma}_n$ is generated for $b=1,\ldots,B$ and  the statistic $B_{\rho}^{(b)}$ is computed correspondingly. The permutation $p$-value  is obtained as
\begin{equation}
\hat{p} = \frac{1}{B+1} \left(\sum_{b=1}^B I(B_\rho \leq B_\rho^{(b)}) + 1 \right).
\end{equation}
The null hypothesis of homogeneity is rejected if $\hat{p}$ is smaller than or equal to the desired significance level $\alpha$. The consistency of the DISCO test was proven in \cite{rizzo_disco_2010-1} against all fixed alternatives of finite second moments. We note that the proof is directly applicable to our setting with a separable Hilbert space as it does not involve any properties limited to random variables in finite-dimensional spaces.

\subsection{Clustering}

Unlike hypothesis testing where data are given with class labels, clustering is an unsupervised task of grouping a set of objects - persistence landscapes in our case. Since the persistence landscapes are elements of a separable Hilbert space, a class of algorithms from Functional Data Analysis (FDA) \citep{ramsay_functional_2005, wang_functional_2016} would be a first-hand choice although the basis expansion framework, one of the main pillars of FDA, may not be a feasible choice in the clustering task since it involves an arbitrary number of functions $\lambda_k (t), ~k=1,2,\ldots$ and proportionally increasing number of coefficients for a collection of bases over $k\in \bbN$. Instead, we introduce methods that only depend on pairwise dissimilarities for $n$ observed persistence landscapes $ \bfLambda = \lbrace \Lambda_1,  \ldots, \Lambda_n \rbrace$  that correspond to latent embeddings of networks. Throughout this section,  we denote by $\mathbf{S} = \lbrace S_1, S_2, \ldots, S_k \rbrace$ a partitioning of $\bfLambda$ into $k$ sets.

The first method is the $k$-medoids algorithm. While the $k$-medoids is similar to the $k$-means algorithm \citep{macqueen_methods_1967-1} that partitions the data by minimizing the distance from points to centers and assigning observations to the nearest centroid, the main difference is that cluster centroids, known as medoids, are chosen among actual observations. Thus there is no need to compute centers such as \Frechet mean in a general metric space and this characteristic makes the method applicable to data in an arbitrary space whenever the measure of pairwise dissimilarity is available. Furthermore, it has been empirically and theoretically shown that medoids are robust representations of the centroids \citep{kaufman_partitioning_1990, van_der_laan_new_2003}. 

The objective of the $k$-medoids is cast as 
\begin{equation}\label{eq:clustering_kmedoids}
\underset{\bfS}{\min} \sum_{i=1}^k \sum_{j=1}^n \delta_2 (\Lambda_j, \bar{\Lambda}_i)^2 \cdot I(\Lambda_j \in S_i)\quad \textrm{for}~~ \bar{\Lambda}_1,\ldots,\bar{\Lambda}_k \in \bfLambda,
\end{equation}
where the problem is known to be NP-hard and many heuristics for the problem \eqref{eq:clustering_kmedoids} have been proposed \citep{amato_faster_2019}. Among many candidates, we briefly mention the Partitioning Around Medoids (PAM) algorithm \citep{kaufman_partitioning_1990}. The PAM algorithm starts by randomly selecting $k$ data points as the medoids and each observation is assigned to the nearest medoid. Along the decreasing path of the objective \eqref{eq:clustering_kmedoids}, the algorithm iterates through all medoids. For each medoid and each non-medoidal observation, the cost change is computed for a configuration of swapping two points and the combination is recorded if the change is maximal. After one step of iteration is complete, perform the best swap among the recorded pairs of swaps. This process is repeated until the objective no longer improves.

Second, the $k$-groups algorithm \citep{li_k-groups_2017} is a generalization of the $k$-means algorithm using the framework of energy statistics. While $k$-means generates partitions based on differences of means, $k$-groups separates clusters in terms of energy distance to specify differences of distributions. Recall the decomposition of total dispersion $T_\rho$ into the within- and between-sample dispersions $W_\rho$ and $B_\rho$ respectively. The optimal clustering is a partition that distinguishes observations from different distributions in that it maximizes between-sample dispersion $B_\rho$. While the total dispersion being constant as it does not depend on a partition, the problem is equivalent to minimizing the within-sample dispersion $W_\rho$:
\begin{equation}\label{eq:clustering_kgroups}
\underset{\bfS}{\min}\ W_\rho(\bfS) = \underset{\bfS}{\min}\ \sum_{i=1}^k \frac{|S_i|}{2} d_\rho (S_i, S_i),
\end{equation}
where $d_\rho$-distance is given by Equation \eqref{eq:dist_d_alpha}. It can be easily shown that when observations lie in Euclidean space and $\rho=2$, $k$-groups and $k$-means share the same objective function. \cite{li_k-groups_2017} proposed a numerical procedure based on the idea of Hartigan and Wong's version of the $k$-means algorithm \citep{hartigan_algorithm_1979} that swaps single or multiple points at a time to minimize the objective function \eqref{eq:clustering_kmedoids} until convergence.

Spectral clustering is another class of methods for clustering which can be used in an arbitrary metric space \citep{von_luxburg_tutorial_2007-1}. Spectral clustering makes use of the spectrum information of the graph Laplacian matrix, which is a discrete analogue to the Laplacian operator. We refer interested readers to \cite{chung_spectral_1997-1} for a thorough introduction to the spectral graph theory and its applications thereof. 

We now turn to a practical description of the basic version of the algorithm. Let $A \in \bbR_{+}^{n\times n}$ be an affinity matrix where an entry $A(i,j)  \geq 0$ represents degree of similarity between observations $\Lambda_i$ and $\Lambda_j$. A popular choice of kernel $k(x,y) : \bbX \times \bbX \rightarrow \bbR_+$ on a metric space $(\bbX, \delta)$ to build an affinity matrix is Gaussian kernel $k_\sigma (x,y) = \exp (-\delta(x,y)^2 / \sigma^2) \in \lbrack 0, 1 \rbrack$ with a scale parameter $\sigma > 0$. This process is common in graph-based data analysis. In the classical graph theory, the presence of a binary edge between two nodes indicates proximity or relevance between the entries. When building an affinity matrix, one can easily observe that $k_\sigma (x,y) \rightarrow 1$ when the distance $\delta (x,y) $ approaches $0$. Similarly, the larger the distance between $x$ and $y$ is, the smaller the $k_\sigma (x,y)$ becomes, approaching 0. Therefore, an affinity matrix may be regarded as an approximation to the intrinsic geometry of a point set in $(\bbX, \delta)$. Given an affinity matrix $A:=A(i,j) = \exp(-\delta_2(\Lambda_i,\Lambda_j)^2 / \sigma^2)$, the graph Laplacian matrix  $L=D-A $ is constructed where $D(i,i) = \sum_{j=1}^n A(i,j)$ is a diagonal matrix. Let $v_1, \ldots, v_k$ be eigenvectors of $L$ that correspond to $k$ smallest eigenvalues so that $V = \lbrack v_1 | \cdots | v_k \rbrack \in \mathbb{R}^{n\times k}$. As the last step, the $k$-means or any other partitional algorithm is applied to rows of $V$. As a note, there exist other types of graph Laplacian matrix. The normalized cuts algorithm \citep{ng_spectral_2001} uses the symmetric normalized Laplacian $L = I - D^{-1/2}AD^{-1/2}$ whose eigenvalues $\lambda(L)$ lie in $[0,2]$. An equivalent characterization is the random-walk representation of the graph Laplacian $L = D^{-1}(D-A)$ \citep{shi_normalized_2000-1} while the other steps remain the same.

The choice of bandwidth parameter $\sigma$ plays a critical role in spectral clustering. As mentioned in the previous paragraph, the choice of Gaussian kernel puts a higher value close to 1 when two observations are very close. When $\sigma \rightarrow 0$, $k_\sigma (x,y)$ converges to 0, leading to sparsely connected graph representation of an underlying geometry of the data. On the other hand, $k_\sigma(x,y) \rightarrow 1$ as $\sigma \rightarrow \infty$ so that the induced topology converges to that of a complete graph. Furthermore, a single value of $\sigma$ across all data points only makes sense if all regions of the data manifold have similar degrees of density. To overcome such difficulties, many modifications have been proposed to find locally adaptive bandwidth parameter  \citep{zelnik-manor_self-tuning_2004, gu_improved_2009-1, zhang_spectral_2010-1, yang_spectral_2011-1}, apply post-processing such as neighborhood propagation of the affinity matrix \citep{li_constructing_2012-1}, and so on. Here we introduce one of the most simplest data-driven construction of the affinity matrix via nearest neighbor algorithm that was proposed in \cite{zelnik-manor_self-tuning_2004}. For $i=1,\ldots,n$, let $\sigma_i$ be the distance from $\Lambda_i$ to its $\tau$-th nearest neighbor. The locally-adaptive affinity matrix is constructed as
\begin{equation}\label{eq:clustering_zelnik}
A_{\tau}:=A_{\tau} (i,j) = k_{\sigma,\tau} (\Lambda_i, \Lambda_j) = \exp \left( - \frac{\delta_2 (\Lambda_i, \Lambda_j)^2}{\sigma_i \sigma_j}\right)\ \textrm{for} \ i,j=1,\ldots,n,
\end{equation}
and the $k$-means algorithms is applied to the row spaces of $k$ eigenvectors corresponding to smallest eigenvalues from the spectral decomposition of the normalized graph Laplacian $L_{\tau}=I-D_{\tau}^{-1/2}A_{\tau}D_{\tau}^{-1/2}$ where $D_{\tau} (i,i) = \sum_{j=1}^n A_{\tau}(i,j)$ is a diagonal degree matrix.

\section{Experiment}\label{sec:experiment}
\subsection{Simulation examples}

We consider two simulation scenarios with varying topological properties. For all simulations, we generate networks of the varying number of nodes $n \in [80, 120]$. Given a binary network, the latent space model of Equation \eqref{eq:lsm_binary} is fitted in $\bbX = \bbR^2$ via two-stage maximum likelihood estimation for the intercept $\alpha$ and embedding of nodes $\lbrace z_1, \ldots, z_n \rbrace \subset \bbR^2$ \citep{hoff_latent_2002-1}. We also summarize several topological descriptors of  sampled graphs, including average degree (AD), average shortest-path distance (ASD), betweenness centrality (BC), closeness centrality (CC), degree centrality (DC), density, diameter, modularity, and global transitivity \citep{newman_networks_2010}.

First, we perform experiments on networks sampled from the \Erdos-\Renyi (ER) model $G(n,p)$ where edges are given a fixed probability $p \in (0,1)$ of being present $(A_{ij} = 1)$ or absent $(A_{ij}=0)$ independently and identically \citep{erdos_random_1959-1}. The \Erdos-\Renyi model is central to random graph theory whose asymptotic  properties have long been studied \citep{newman_random_2001}. We consider 7 classes of ER models with varying edge probabilities $p \in \lbrace 0.01, 0.025, 0.05, 0.1, 0.15, 0.2, 0.25 \rbrace$, whose topological properties are shown in Figure \ref{fig:summary_ER} of the Supplementary Material. Two simulated graphs from $G(n,0.25)$ and $G(n,0.05)$ are presented in Figure \ref{fig:simulation1} of the Supplementary Material, which also contains visualization for a total of 100 networks, 50 from each class, via multidimensional scaling. This shows that two classes of networks are also distinguishable in terms of their persistent homology. 

We describe pairwise comparison procedures as follows. It starts by choosing edge probabilities $p_i$ and $p_j$ for $i < j=1,\ldots,7$ from $\lbrace 0.01, 0.025, 0.05, 0.1, 0.15, 0.2, 0.25 \rbrace$ and generating $m$ networks from each $G(n,p_i)$ and $G(n,p_j)$, denoting two sets of graphs as $G_i$ and $G_j$. We apply $k$-sample and DISCO tests given persistent homology reconstructed from latent space representation of networks to test whether two sets of networks are from equal distribution. The use of persistence landscapes of order 0 and 1 allows us to interpret whether two sets of networks are different with respect to the patterns of connectedness and holes. For both tests, we set the number of permutations $B=10^6$. 

Similarly, we apply three clustering algorithms with a fixed number of clusters $k=2$. Clustering accuracy is evaluated via Rand index \citep{rand_objective_1971-1}, which returns a numeric value in $[0,1]$ where the larger value indicates the higher coincidence of two clusterings up to permutation of labels. The experiment is repeated 100 times in all settings, and average $p$-values and Rand indices are reported.

\begin{figure}[htbp]
\centering
\begin{tabular}{c}
(a) \\ 
\begin{subfigure}[b]{\textwidth}
\centering
\includegraphics[width=.235\textwidth]{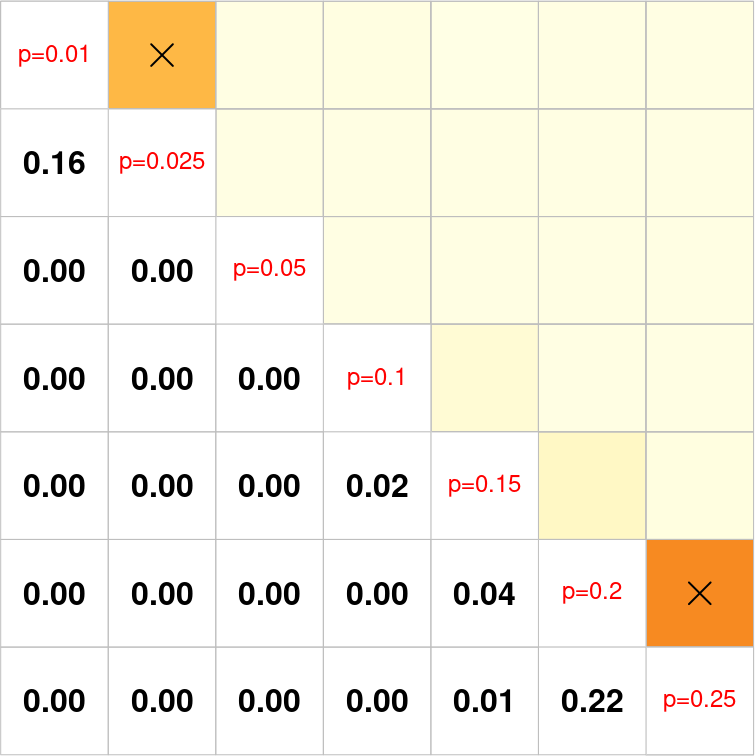}
\includegraphics[width=.235\textwidth]{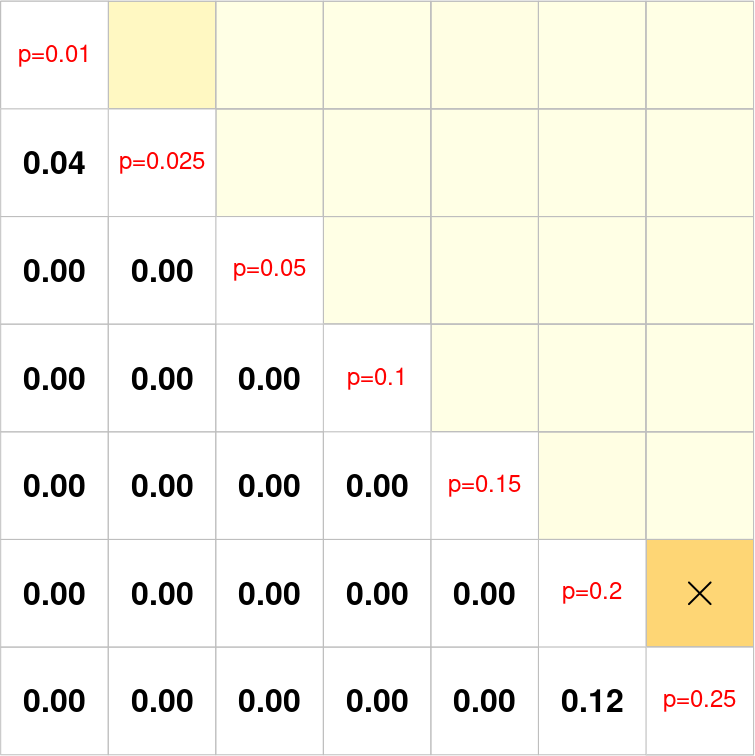}
\includegraphics[width=.235\textwidth]{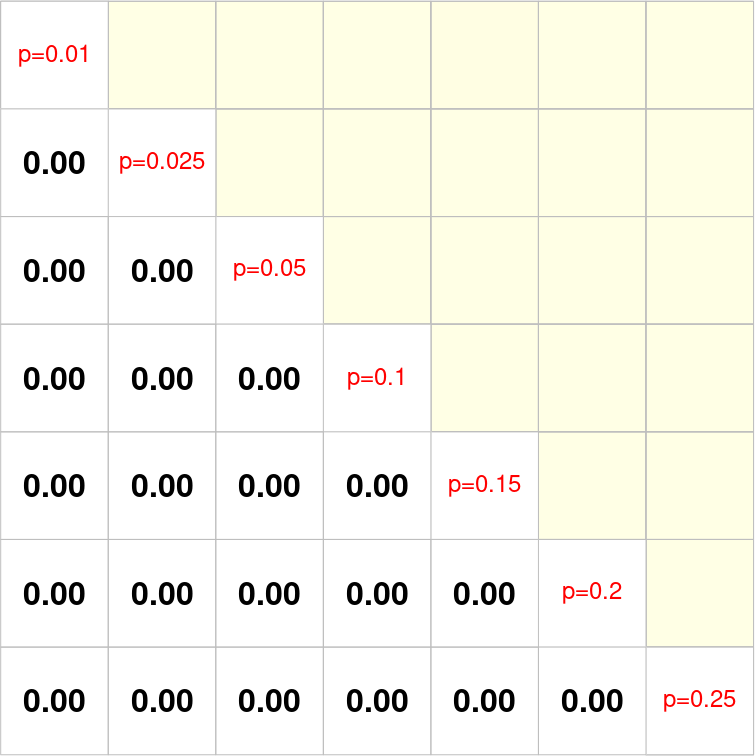}
\includegraphics[width=.235\textwidth]{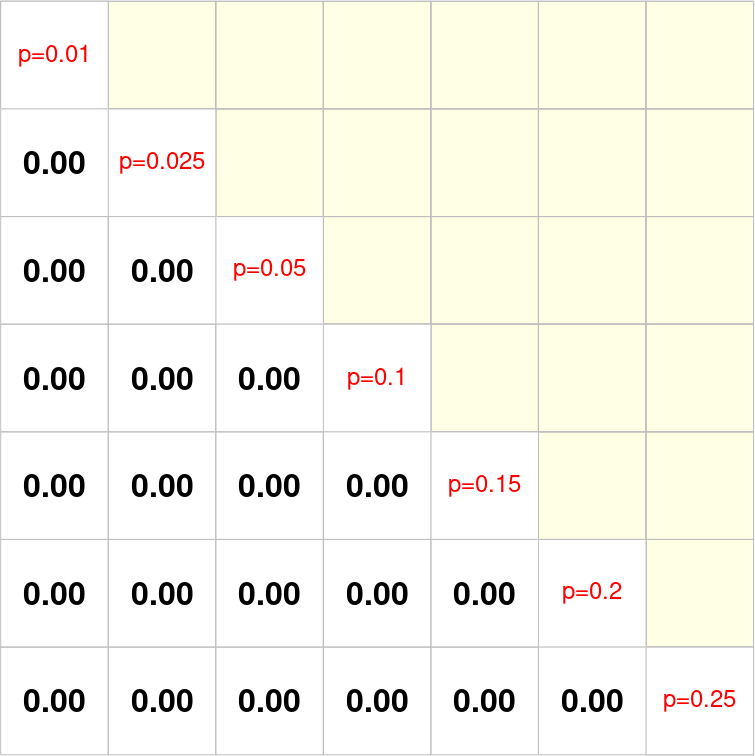}
\end{subfigure}\\[-1em]
(b) \\ 
\begin{subfigure}[b]{\textwidth}
\centering
\includegraphics[width=.235\textwidth]{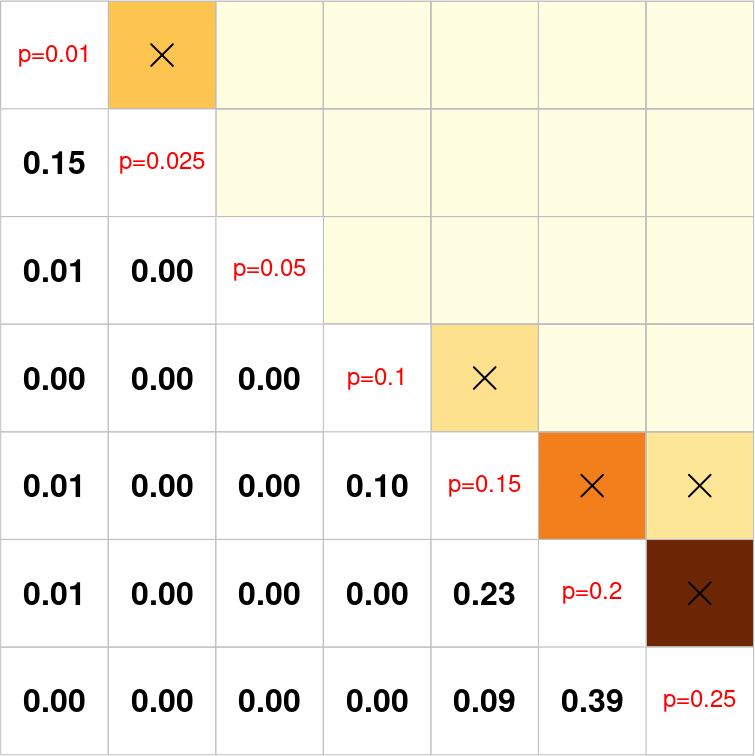}
\includegraphics[width=.235\textwidth]{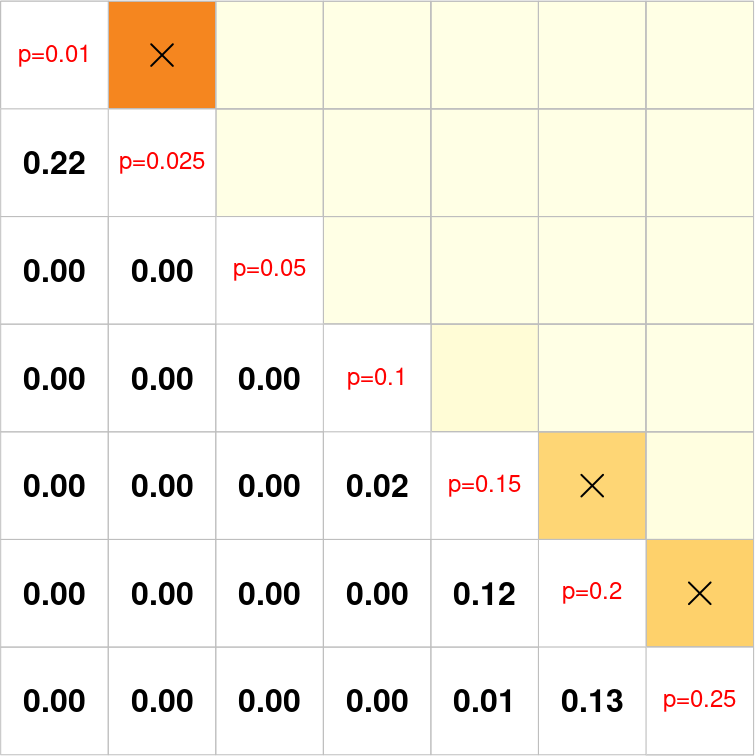}
\includegraphics[width=.235\textwidth]{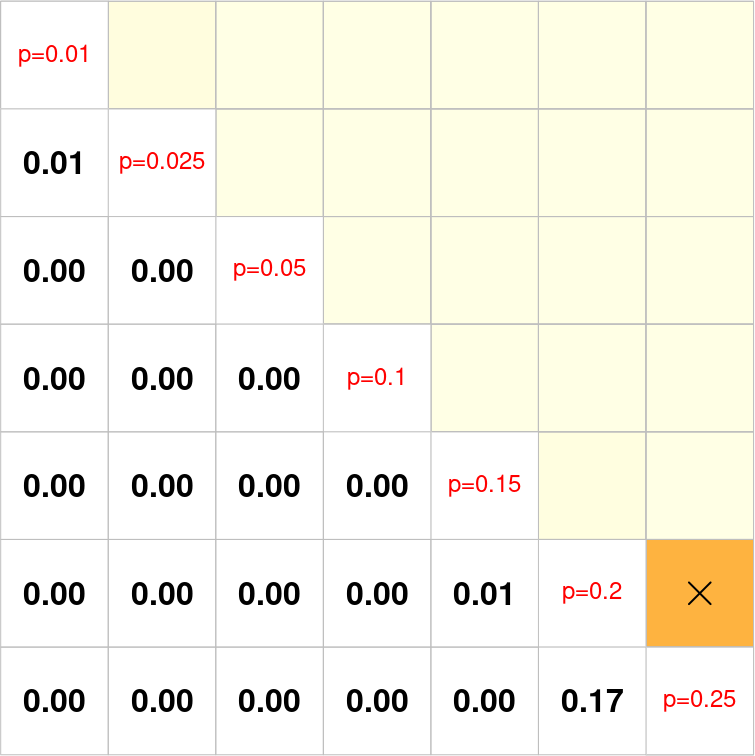}
\includegraphics[width=.235\textwidth]{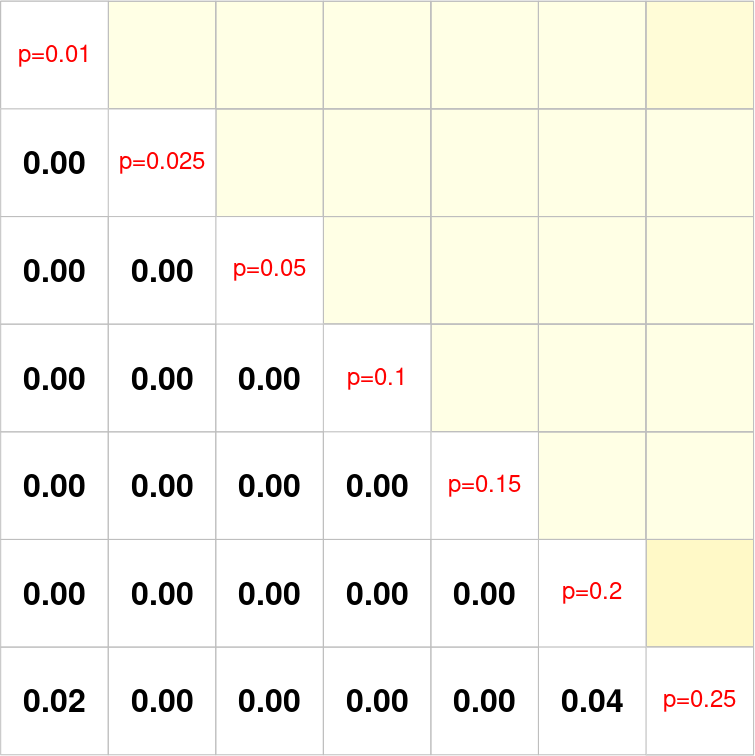}
\end{subfigure}\\[-1em]
(c) \\
\begin{subfigure}[b]{\textwidth}
\centering
\includegraphics[width=.235\textwidth]{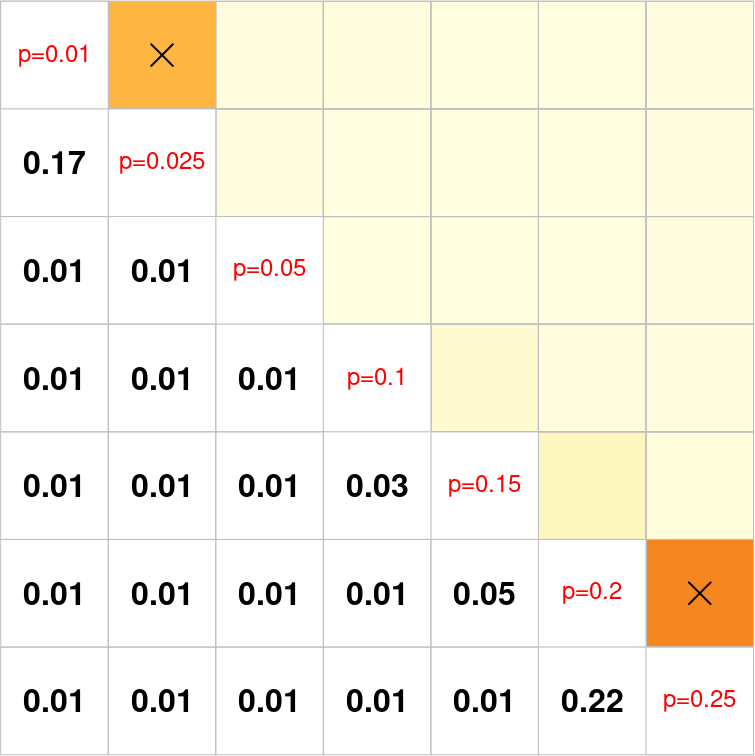}
\includegraphics[width=.235\textwidth]{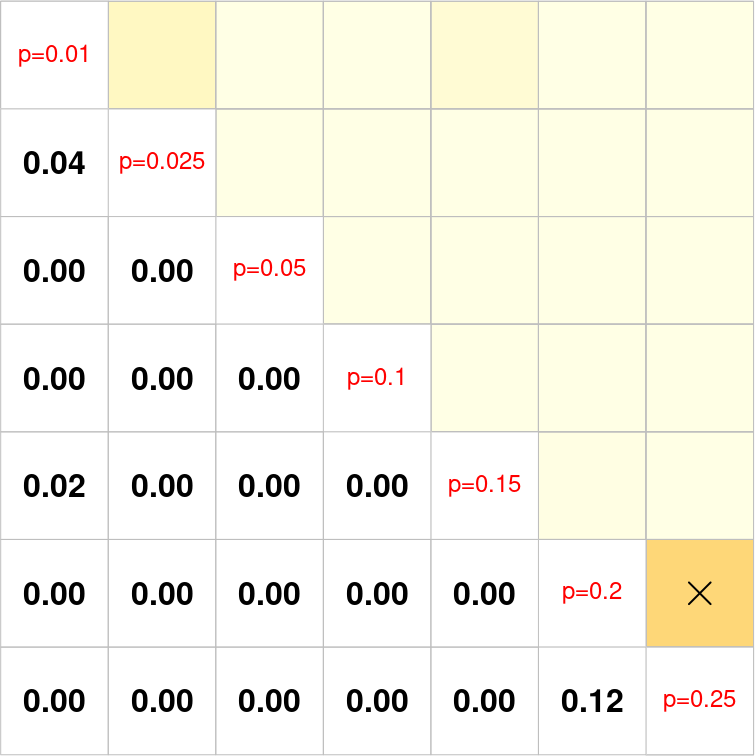}
\includegraphics[width=.235\textwidth]{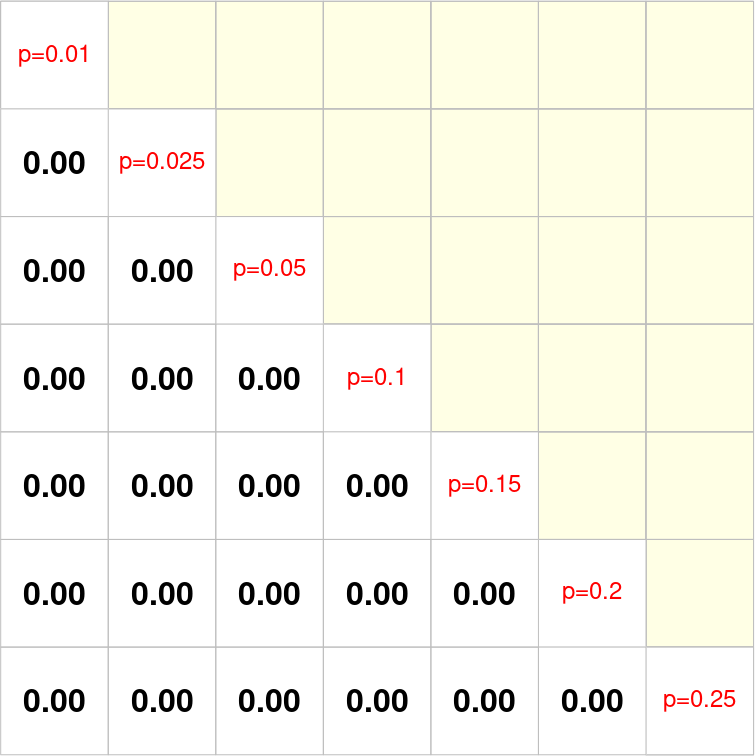}
\includegraphics[width=.235\textwidth]{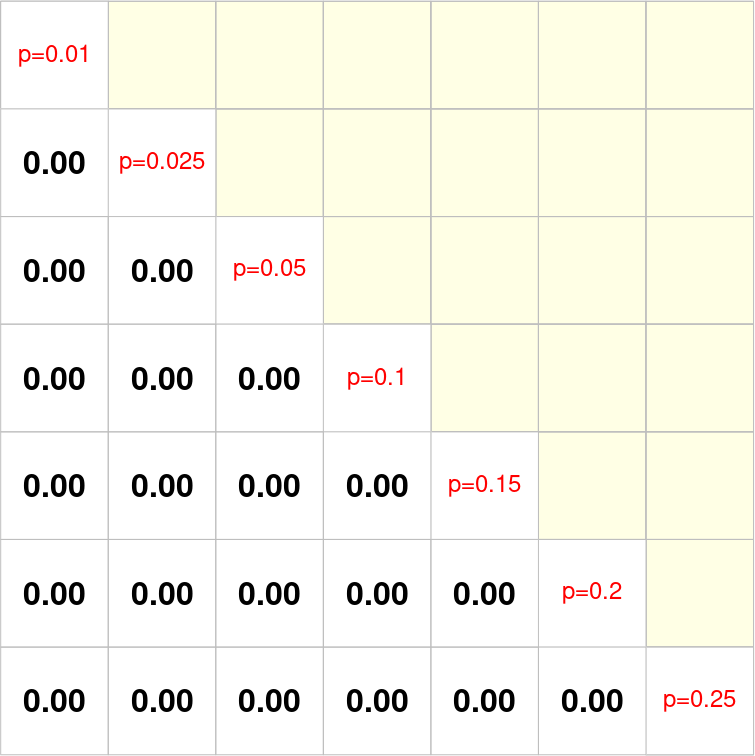}
\end{subfigure}\\[-1em]
(d) \\
\begin{subfigure}[b]{\textwidth}
\centering
\includegraphics[width=.235\textwidth]{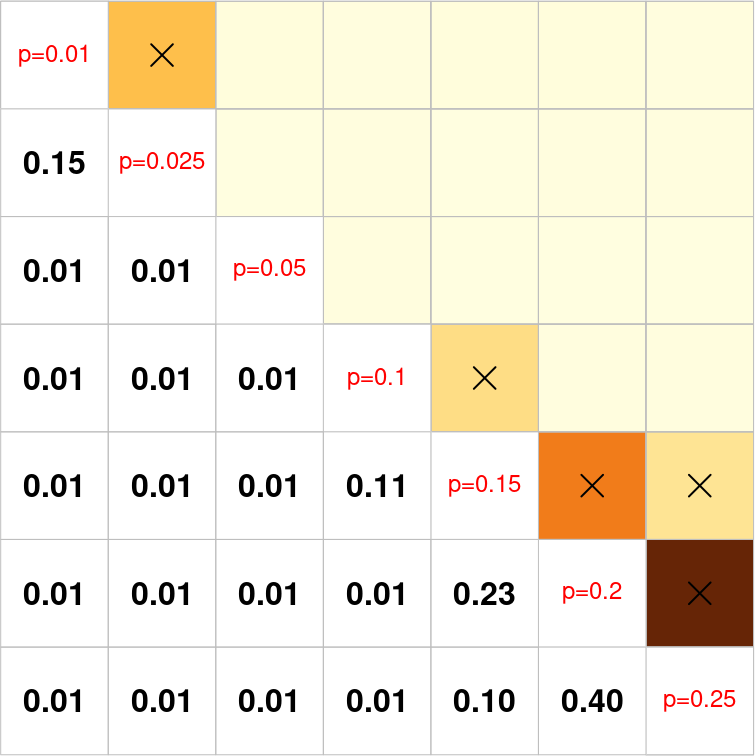}
\includegraphics[width=.235\textwidth]{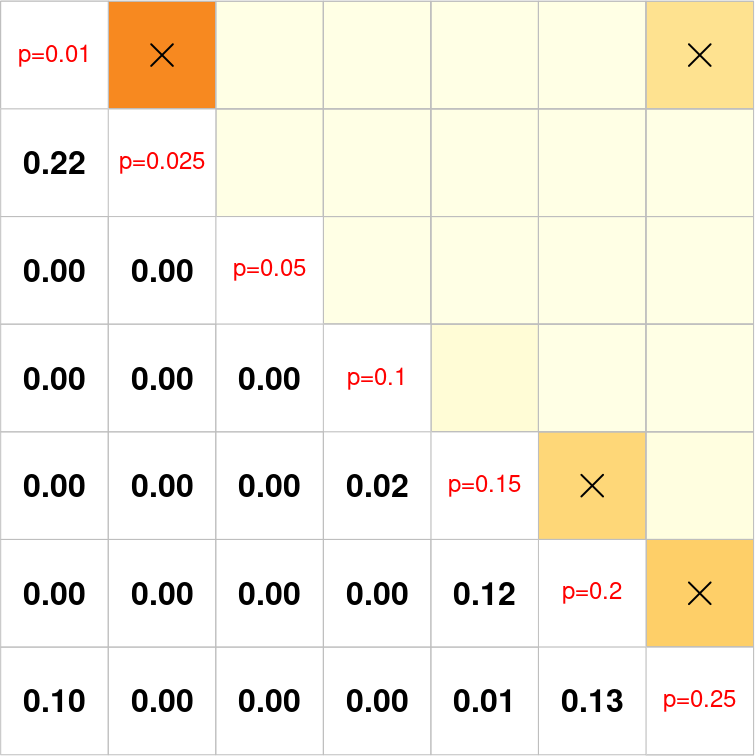}
\includegraphics[width=.235\textwidth]{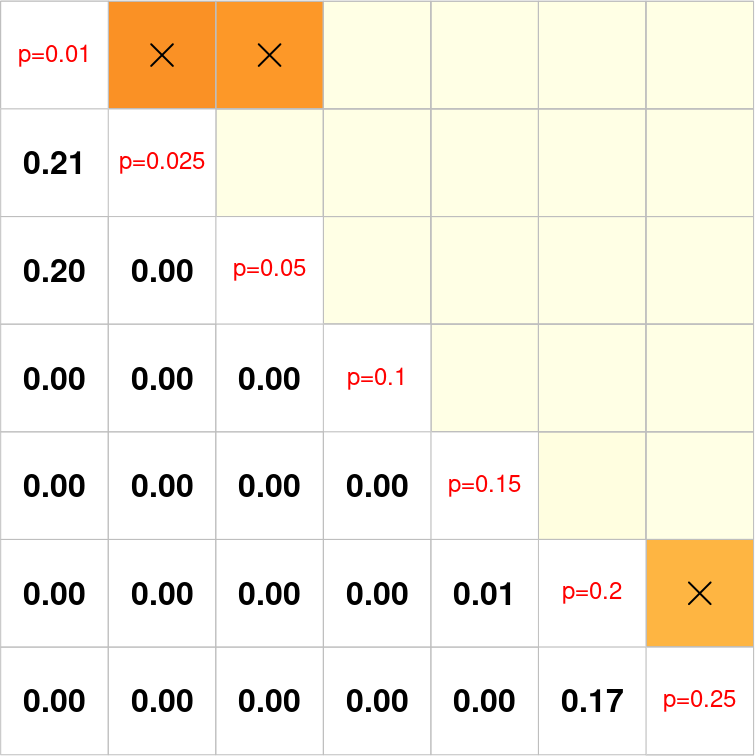}
\includegraphics[width=.235\textwidth]{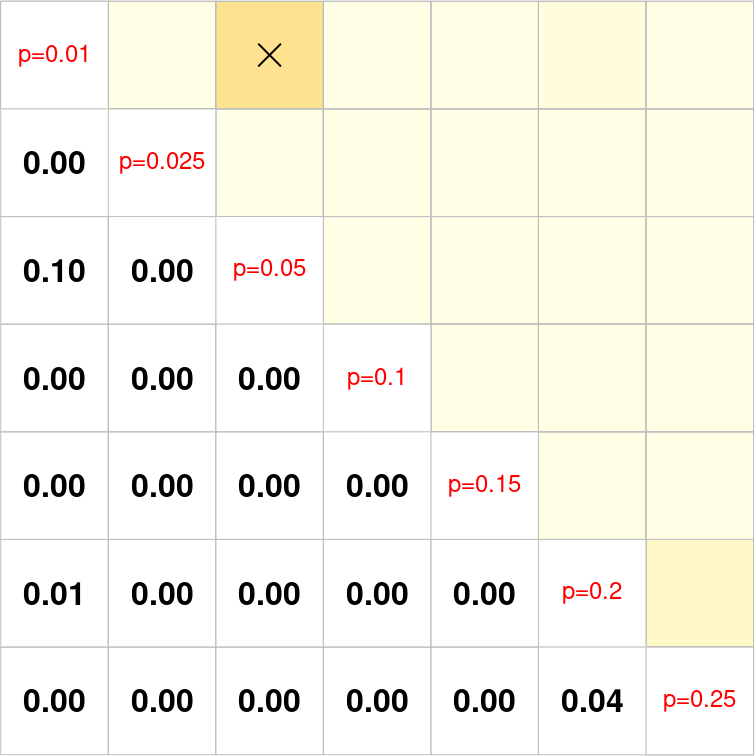}
\end{subfigure}
\end{tabular}
\caption{Pairwise comparison of ER models with hypothesis testing procedures; (a) $k$-sample test of order 0, (b) $k$-sample test of order 1, (c) DISCO of order 0, and (d) DISCO of order 1. For each row, results from different sample size are reported for $m=5,10,25,50$ from left to right. In each subplot, averages of 100 empirical $p$-values are presented in the lower triangular part. The darker the entries in the upper triangle are shaded, the larger the $p$-values are. 
}
\label{fig:results_ER_hypothesis}
\end{figure}

The results from pairwise hypothesis tests are summarized in Figure \ref{fig:results_ER_hypothesis}. One visible pattern is that non-significant $p$-values disappear as the sample size grows in both methods and orders. This phenomenon is expected in the sense that a small sample size does not fully characterize its generating law. Still, significant $p$-values were obtained in the small-sample regime when two model parameters $p_i$ and $p_j$ are different. When $m=5$, a consistent pattern was observed that two pairs $(p_i,p_j) = (0.01, 0.025)$ and $(0.1, 0.25)$ returned non-significant $p$-values. This pattern, however, fades as $m$ gets larger, which indicates that the proposed frameworks are indeed distinguishing two classes of networks well as expected. We also note that the order 1 shows more non-significant results than its order 0 counterparts across all settings. This aligns with what was observed in Figure \ref{fig:simulation1} of the Supplementary Material where the separation of two samples is less explicit in order 1 than order 0. 

We make similar reports on the results from cluster analysis in Figure \ref{fig:results_ER_cluster1} and \ref{fig:results_ER_cluster2} of the Supplementary Material. First, a similar pattern is observed that the larger sample size indicates higher clustering accuracy as shown in the hypothesis testing experiment. Comparison against the most sparse model $p=0.01$ shows poor results across all settings, which is suspected to stem from the fact that there exists no connected component in probability when $np < 1$ \citep{erdos_random_1959-1} and this lack of structure provides insufficient information for inferential algorithms. Second, spectral clustering shows superb performance against competing algorithms in most settings. As shown in Figure \ref{fig:simulation1} of the Supplementary Material, there is little guarantee that the pattern of separation be linear, leading to better performance of locally adaptive nonlinear methods like the spectral clustering algorithm.

Next, we consider networks having block structures. We model networks with community structure using the stochastic block model (SBM) where disjoint subsets of nodes are defined as communities \citep{holland_stochastic_1983-1}. In the literature of SBM, a community is conceptualized as a set of highly connected nodes and any pair of communities is weakly connected. We employ \Erdos-\Renyi models $G(n,p_{\textrm{high}})$ and $G(n,p_{\textrm{low}})$ to generate within- and between-community connections. In order to examine effectiveness of the multi-sample tests, we consider 5 classes of networks that have different numbers of communities $k_0=2,3,4,5,10$, whose topological properties and fitted intercept values are shown in Figure \ref{fig:summary_SBM} of the Supplementary Material. We note that compared to the ER models, centrality measures do not differ much across multiple models yet other descriptors show clearly distinctive patterns. 

Since sweeping over all possible combinations is not trivial, we opt to use 5 distinctive scenarios where each population consists of three types of networks $\lbrace 2,3,4\rbrace$, $\lbrace 2,3,5\rbrace$, $\lbrace 2,4,5\rbrace$, $\lbrace 3,4,5\rbrace$, and $\lbrace 2,5,10\rbrace $. Similar to the ER model case, an illustrated example is provided to demonstrate model networks and their distributions for the scenario of $\lbrace 2, 5, 10\rbrace$ in Figure \ref{fig:simulation2} of the Supplementary Material, showing heterogeneous characteristics of networks from SBMs by varying $k_0$.

For each scenario, $m$ networks are randomly drawn from one of three models. Probabilities for within- and between-cluster edges are set as $(p_{\textrm{high}}, p_{\textrm{low}}) = (0.8, 0.1)$ with varying number of network size in $n \in \lbrack 80, 120 \rbrace$. The choice of edge probabilities provides sufficient distinction between inter- and intra-community connections and contingent topological properties of a sampled network as shown in Figure \ref{fig:summary_SBM} of the Supplementary Material. Similar to the ER experiment, we apply $k$-sample and DISCO tests given persistent homology reconstructed from latent space representation of networks to test whether three types of networks are from equal distribution using $B=10^6$ permutations. Three clustering algorithms are applied with a fixed number of clusters $k=3$ and their accuracy is measured in terms of Rand index. Each experiment is repeated 100 times in all settings and average $p$-values and Rand indices are reported.

We first summarize results from multi-sample hypothesis tests in Table \ref{tab:results_SBM_hypothesis} of the Supplementary Material. It is easily observed that the small sample size of $m=5$ yields a larger $p$-value than the larger sample-size regime. Still, all settings returned significant results, which may be due to the fact that the comparison is performed on three sets of networks that are highly structured. This implies that the presence of differentiating structures benefits the test-based comparison even when an available sample size is small. 

For cluster analysis, we observe a distinct pattern for each topological dimension and summarize in Table \ref{tab:results_SBM_clustering} of the Supplementary Material. In most settings at order 0, $k$-medoids and $k$-groups algorithms both perform better than spectral clustering, while the opposite is observed when order is 1. Nevertheless, the minimal average Rand index for order 0 is 0.920, which indicates that all methods were successful to separate groups of networks from SBMs based on the connectedness. On the other hand, overall cluster performance for order 1 is not as impressive as that of order 1, which one may attribute to a weak analog of holes in the latent representation of networks to the context of SBMs.

\subsection{Real data analysis}

We apply the proposed topological framework to analyze the GEPS data that was introduced in Section \ref{sec:background}. The main objective is to test whether the latent dependence structures of innovative and regular schools are equally distributed per school level. As in NIRM, we used MCMC to estimate the model parameters.

\begin{figure}[htbp]
\centering
\begin{tabular}{cc}
(a) & (b) \\
\includegraphics[width=0.45\textwidth]{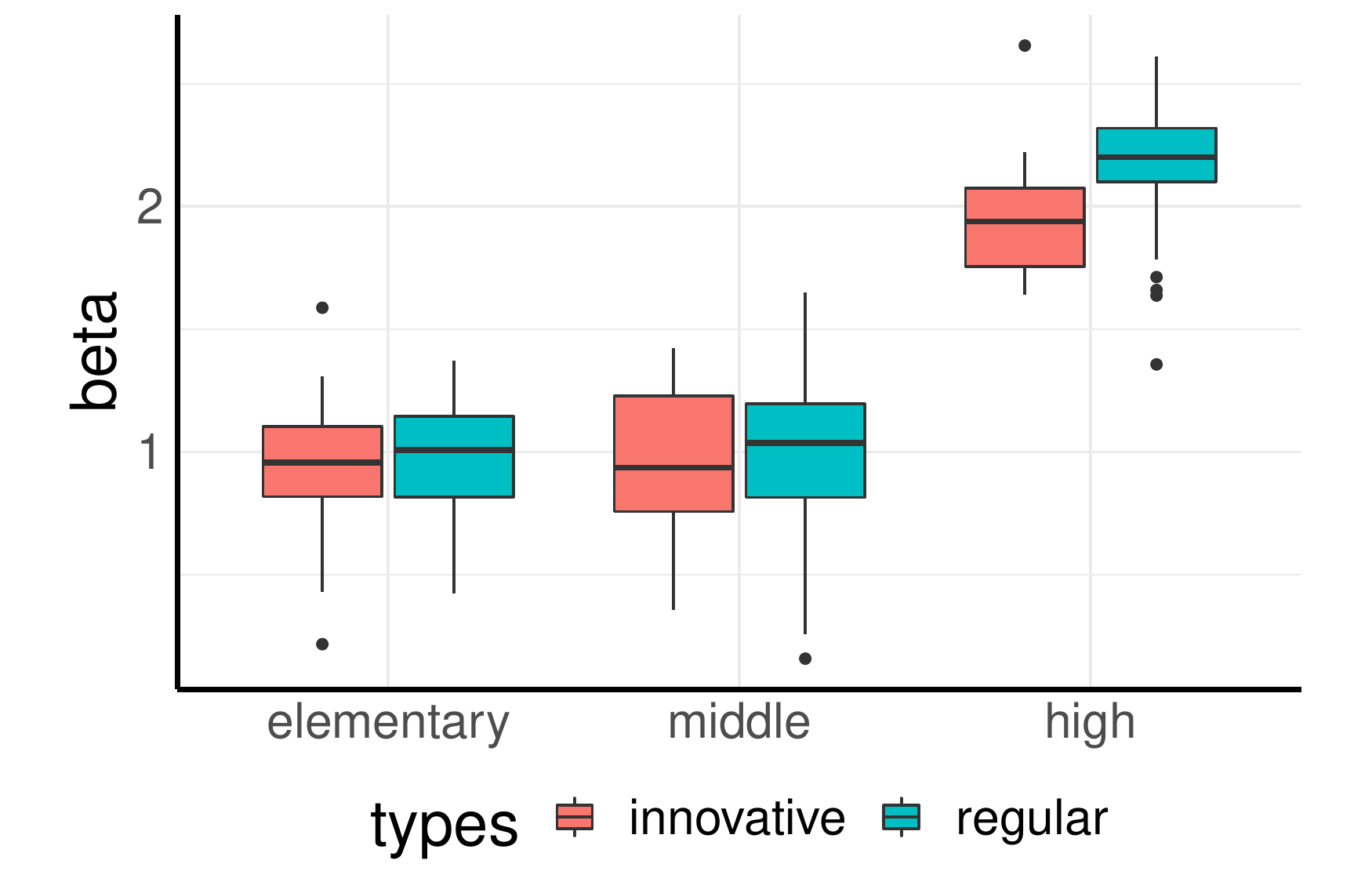} & 
\includegraphics[width=0.45\textwidth]{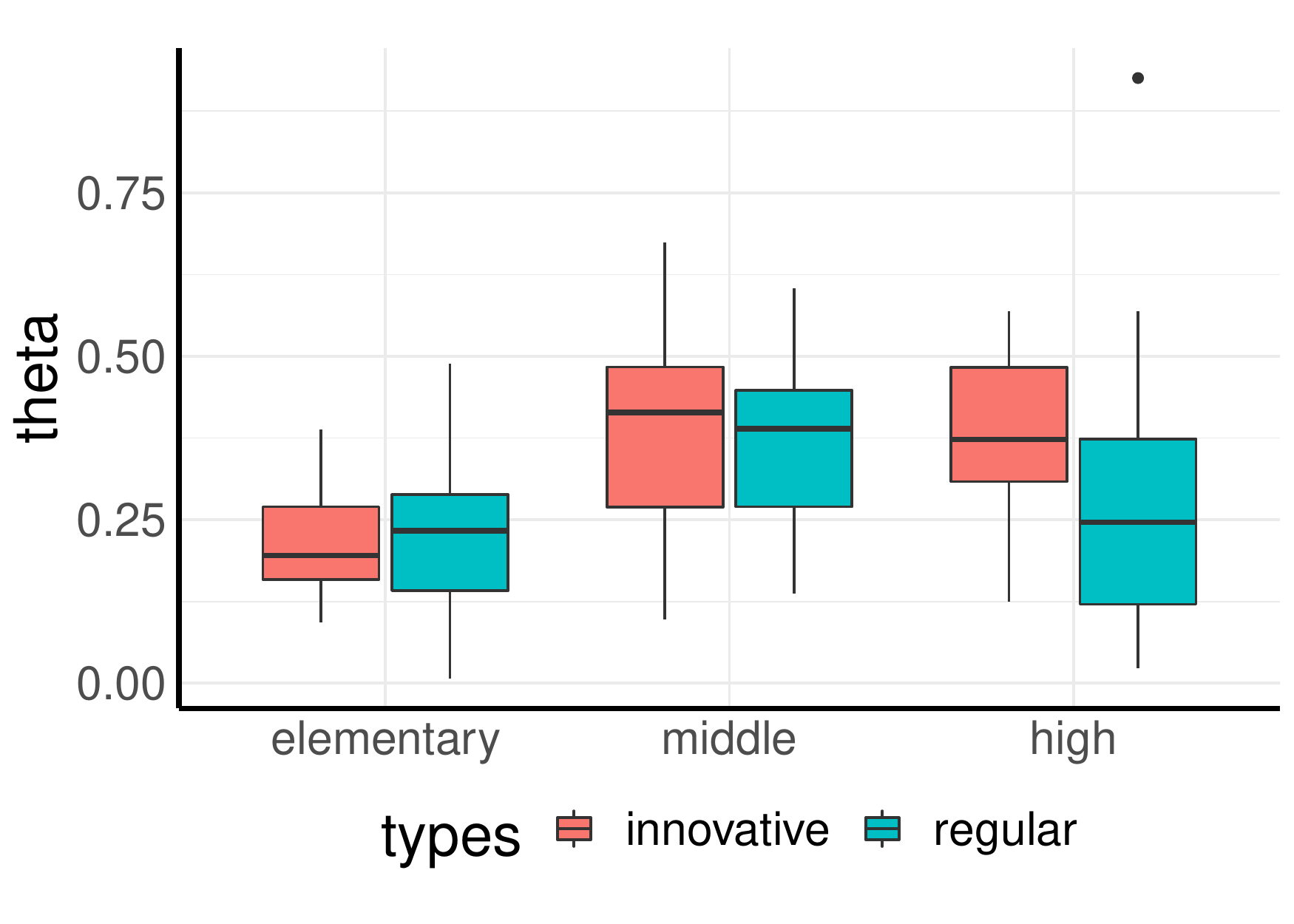}
\end{tabular}
\caption{Average MAP estimates of intercept parameters for (a) items $\beta$ and (b) individuals $\theta$ across three school levels. }
\label{fig:real_intercept}
\end{figure}

A latent representation for each school $[W,Z]$ is obtained by concatenating a pair of maximum a posteriori (MAP) estimates of items $W$ and individuals $Z$ locations. Figure \ref{fig:real_intercept} shows that average parameter estimates of items $\beta$ and individuals $\theta$ are not significantly different between innovation and regular schools. 
We test potential differences in dependence structure of $w$ and $z$  between the innovation and regular schools per school level using the $k$-sample test and DISCO under 
he null hypothesis and landscape orders $H_0 : \mu_{\textrm{innovative}} = \mu_{\textrm{regular}}$.  Table \ref{table:real_hypothesis} presents empirical $p$-values of the two tests. The results suggest that the dependence structure between the innovation and regular schools was indeed significantly different  at the middle-school level.

\begin{table}[htbp]
	\begin{center}
		\begin{tabular}{c|c|c|c|c|c|c}
		\hline
			\multirow{2}{*}{order} & \multicolumn{2}{c|}{elementary} & \multicolumn{2}{c|}{middle} & \multicolumn{2}{c}{high} \\
			\cline{2-7}
			& $k$-sample       & DISCO       &   $k$-sample       & DISCO           
			&  $k$-sample       & DISCO            \\ \hline 
			0 & 0.6861 & 0.6869 & 0.0049 & 0.0053 & 0.2249 & 0.2309\\
			1 & 0.2236 & 0.2268 & 0.0108 & 0.0117 & 0.3490 & 0.3529\\
			\hline
		\end{tabular}	
	\end{center}
	\caption{Empirical $p$-values of the two test procedures across three school levels and landscape orders 0 (connectedness) and 1 (holes).}
	\label{table:real_hypothesis}
\end{table}

\section{Conclusion}\label{sec:conclusion}

We have proposed a framework for analyzing multiple latent space embeddings based on TDA that overcomes the drawbacks of conventional approaches in the literature. Persistence landscape, a core tool of TDA, was adopted as an efficient representation with desirable theoretical properties. The theory of energy statistics provides two algorithms for multi-sample testing of equal distributions on top of recent results that guarantee theoretical validity to extend energy-based methods to persistence landscapes. Three algorithms for cluster analysis were also adopted to perform cluster analysis on the space of persistence landscapes. We demonstrated the effectiveness of our framework on two simulated sets of varying-size networks from \Erdos-\Renyi models and stochastic block models where our proposals were capable of distinguishing different sets of networks by both hypothesis testing and cluster analysis. Our proposal was applied to educational survey data and discovered that the newly adopted school system induced significant differences at the middle-school level.

We close this paper by discussing some of the potential issues and directions for future research. First is to design a pipeline to reflect certain types of fixed information. It is common in practice that a multitude of network data is derived on a shared set of nodes where nodal correspondence is of importance. Our real data analysis also raises a similar concern where the same questionnaire was used across all schools. While our topological approach provides unique advantages in comparing the shape of latent network representations, these scenarios necessitate a structured approach to take a fixed amount of partial information shared across networks into consideration. 
Another line of extension is to learn with empirical measures of networks rather than a single, static network. In our real data example, each network-compatible dataset was considered under a Bayesian context where a number of Markov chain Monte Carlo samples were drawn. We used MAP estimates according to our current proposal at the sacrifice of uncertainty engendered by a Bayesian framework. Therefore, it would be interesting to come up with an approach that can handle units of analysis represented by a collection of topological descriptors based on a solid theoretical framework.

\section*{Source code and data}

The \textsf{R} codes to replicate simulated examples in Section \ref{sec:experiment} can be found on GitHub at \url{https://github.com/kisungyou/papers}. The GEPS data is available upon the consent of the Gyeonggi Institute of Education in South Korea. We refer to English-version website of the institute at \url{https://www.gie.re.kr/eng/content/C0012-04.do} for interested readers. 

\section*{Acknowledgement}

This study was partially supported by the Yonsei University Research Fund 2019-22-0210 and by Basic Science Research Program through the National Research Foundation of Korea (NRF 2020R1A2C1A01009881). Correspondence should be addressed to Ick Hoon Jin, Department of Applied Statistics, Department of Statistics and Data Science, Yonsei University, Seoul. Republic of Korea. E-Mail: ijin@yonsei.ac.kr.

\bibliographystyle{dcu}
\bibliography{reference}

\newpage
\begin{center}
\LARGE	Supplementary material to ``Comparing multiple latent space embeddings using topological analysis''
\end{center}
\begin{appendices}
\section{Validity of permutation $p$-values} \label{sec: permutation p-value}

For a sequence of exchangeable random variables $X_1,\ldots,X_{n+1}$, Lemma 1 of \cite{romano2005exact} shows that
\begin{align*}
\mathbb{P} \biggl( \frac{1}{n+1} \bigg\{\sum_{i=1}^n I(X_{n+1} \leq X_i)  + 1 \bigg\} \leq \alpha \biggr) \leq \alpha, \quad \text{for all $\alpha \in [0,1]$.}
\end{align*}
In this section, we provide a slightly sharper result than Lemma 1 of \cite{romano2005exact}. In fact, \cite{romano2005exact} state their result without proof, which encourages us to present a full proof for completeness.

\begin{lemma}
	Suppose that $X_1,\ldots,X_n,X_{n+1}$ are exchangeable random variables. Then for any $\alpha \in [0,1]$, it holds that 
	\begin{align*}
	\mathbb{P} \biggl( \frac{1}{n+1} \bigg\{\sum_{i=1}^n I(X_{n+1} \leq X_i)  + 1 \bigg\} \leq \alpha \biggr) \leq \frac{\lfloor{(n+1)\alpha\rfloor}}{n+1} \leq \alpha,
	\end{align*}
	where $\lfloor{x\rfloor}$ is the largest integer smaller than or equal to $x$. Suppose further that $X_1,\ldots,X_{n+1}$ are all distinct with probability one. Then
	\begin{align*}
	\mathbb{P} \biggl( \frac{1}{n+1} \bigg\{\sum_{i=1}^n I(X_{n+1} \leq X_i)  + 1 \bigg\} \leq \alpha \biggr) = \frac{\lfloor{ (n+1)\alpha\rfloor}}{n+1}.
	\end{align*}
	\begin{proof}
		Let $X_{(1)} \leq X_{(2)} \leq \ldots \leq X_{(n+1)}$ be the order statistics of $X_1,\ldots,X_{n+1}$. Then we observe that
		\begin{align*}
		& \frac{1}{n+1} \bigg\{\sum_{i=1}^n I(X_{n+1} \leq X_i)  + 1 \bigg\} = \frac{1}{n+1} \sum_{i=1}^{n+1} I(X_{n+1} \leq X_i)  \leq \alpha \\[.5em]
		\overset{\text{iff}}{\Longleftrightarrow} ~ & \sum_{i=1}^{n+1} I(X_{n+1} \leq X_i)  \leq \lfloor{(n+1)\alpha\rfloor} \\[.5em]
		\overset{\text{iff}}{\Longleftrightarrow} ~  & \sum_{i=1}^{n+1} I(X_i < X_{n+1}) \geq (n+1) - \lfloor{(n+1)\alpha\rfloor}:= k \\[.5em]
		\overset{\text{iff}}{\Longleftrightarrow} ~ & X_{n+1} > X_{(k)}.
		\end{align*}
		Now by the exchangeability condition, we have
		\begin{align*}
		\mathbb{P}(X_{n+1} > X_{(k)}) ~=~& \mathbb{E}\bigg[ \frac{1}{n+1} \sum_{i=1}^{n+1} I(X_{i} > X_{(k)}) \bigg].
		\end{align*}
		On the other hand, by the definition of $X_{(k)}$, 
		\begin{align*}
		\frac{1}{n+1} \sum_{i=1}^{n+1} I(X_{i} > X_{(k)}) \leq \frac{n+1 - k}{n+1} = \frac{ \lfloor{(n+1)\alpha\rfloor} }{n+1}.
		\end{align*}
		Hence the first result follows. When $X_1,\ldots,X_{n+1}$ are all distinct, observe
		\begin{align*}
		\sum_{i=1}^{n+1} I(X_{i} > X_{(k)}) = n+1-k.
		\end{align*}
		Thus the second result follows as
		\begin{align*}
		\mathbb{E}\bigg[ \frac{1}{n+1} \sum_{i=1}^{n+1} I(X_{i} > X_{(k)}) \bigg] = \frac{n+1-k}{n+1} =  \frac{ \lfloor{(n+1)\alpha\rfloor} }{n+1}.
		\end{align*}
	\end{proof} 
\end{lemma}
\vfill

\newpage 
\section{Additional tables and figures}
\begin{figure}[htbp]
	\centering
	\includegraphics[width=\textwidth]{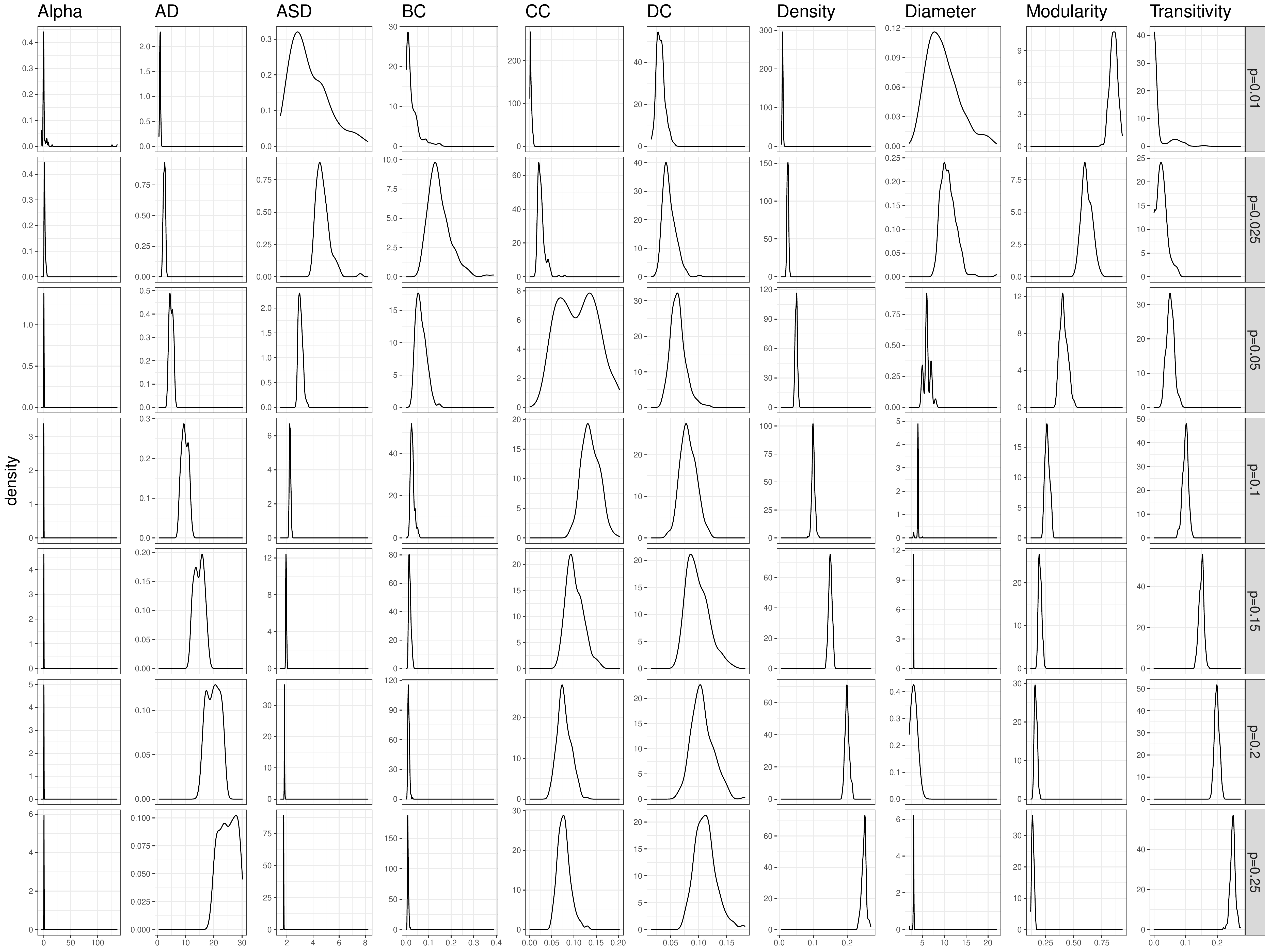}
	\caption{Summary of network characteristics sampled from the \Erdos-\Renyi (ER) models with varying edge probabilities. The first column represents the distribution of fitted intercept values $\alpha$, and the rest are those of topological descriptors, the latter of which indicates heterogeneity of topological properties across multiple ER models.}
	\label{fig:summary_ER}
\end{figure}

\begin{figure}[htbp]
	\centering
	\begin{tabular}{cc}
		(a) & (b) \\
		\includegraphics[width=0.4\textwidth]{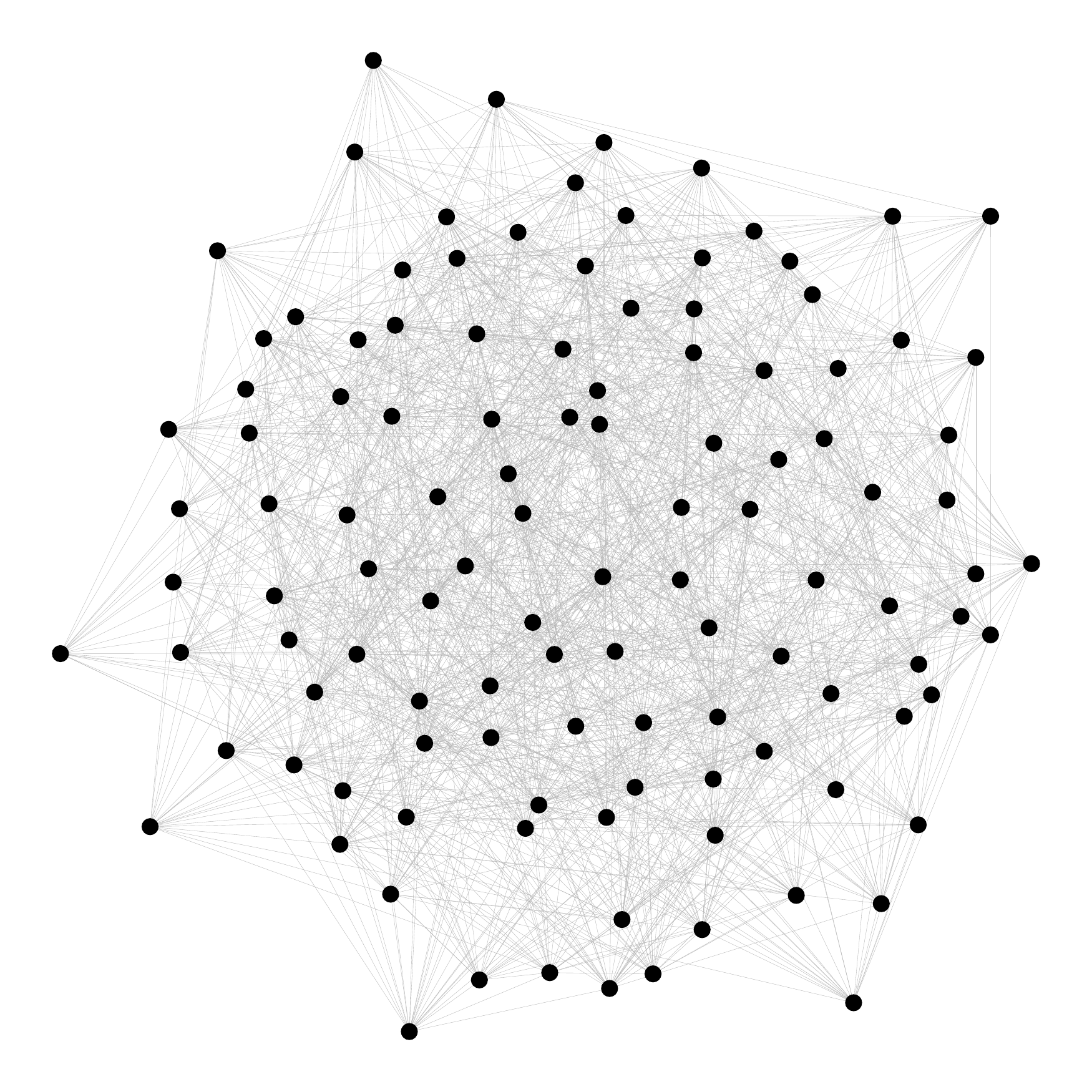} & 
		\includegraphics[width=0.4\textwidth]{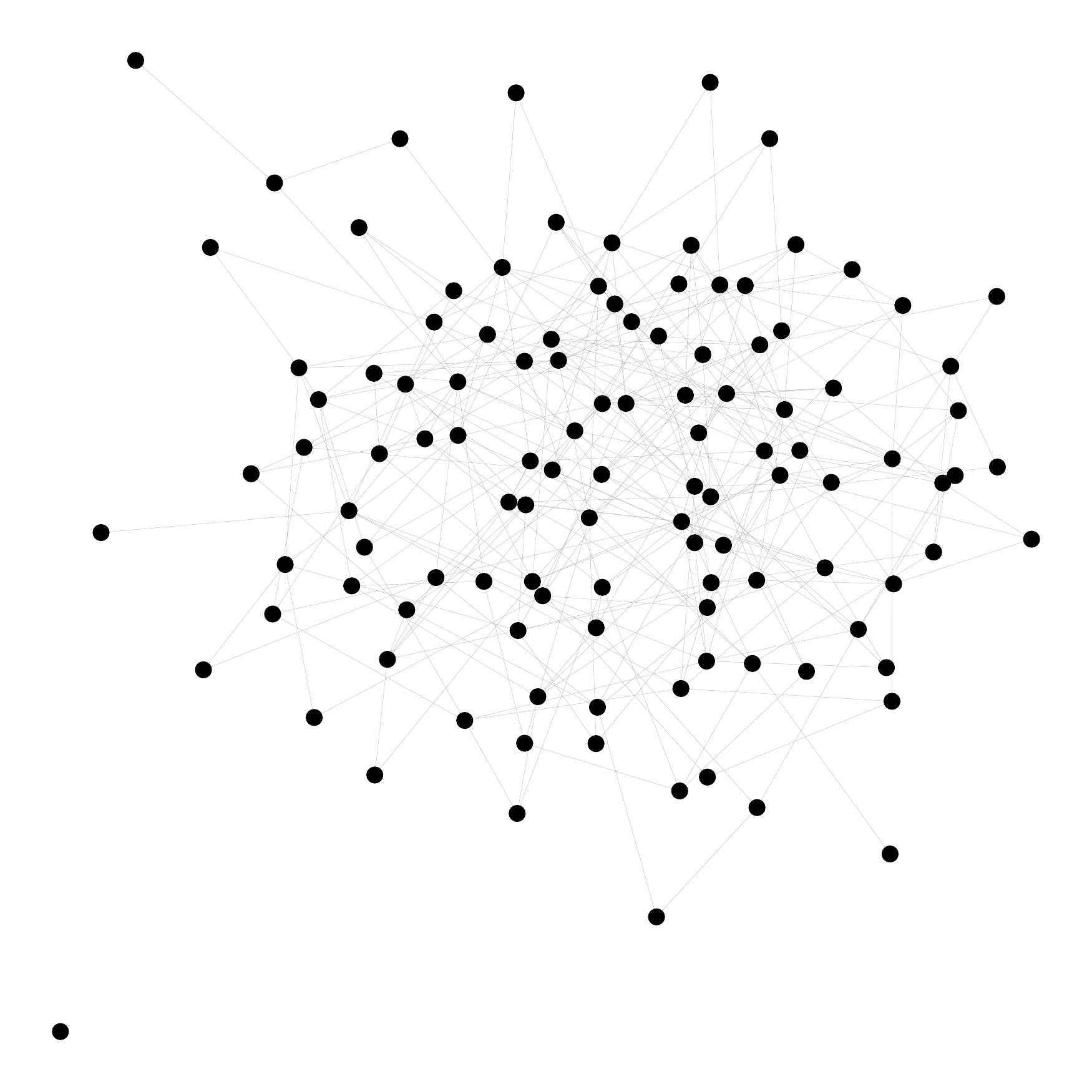}\\
		(c) & (d) \\
		\includegraphics[width=0.4\textwidth]{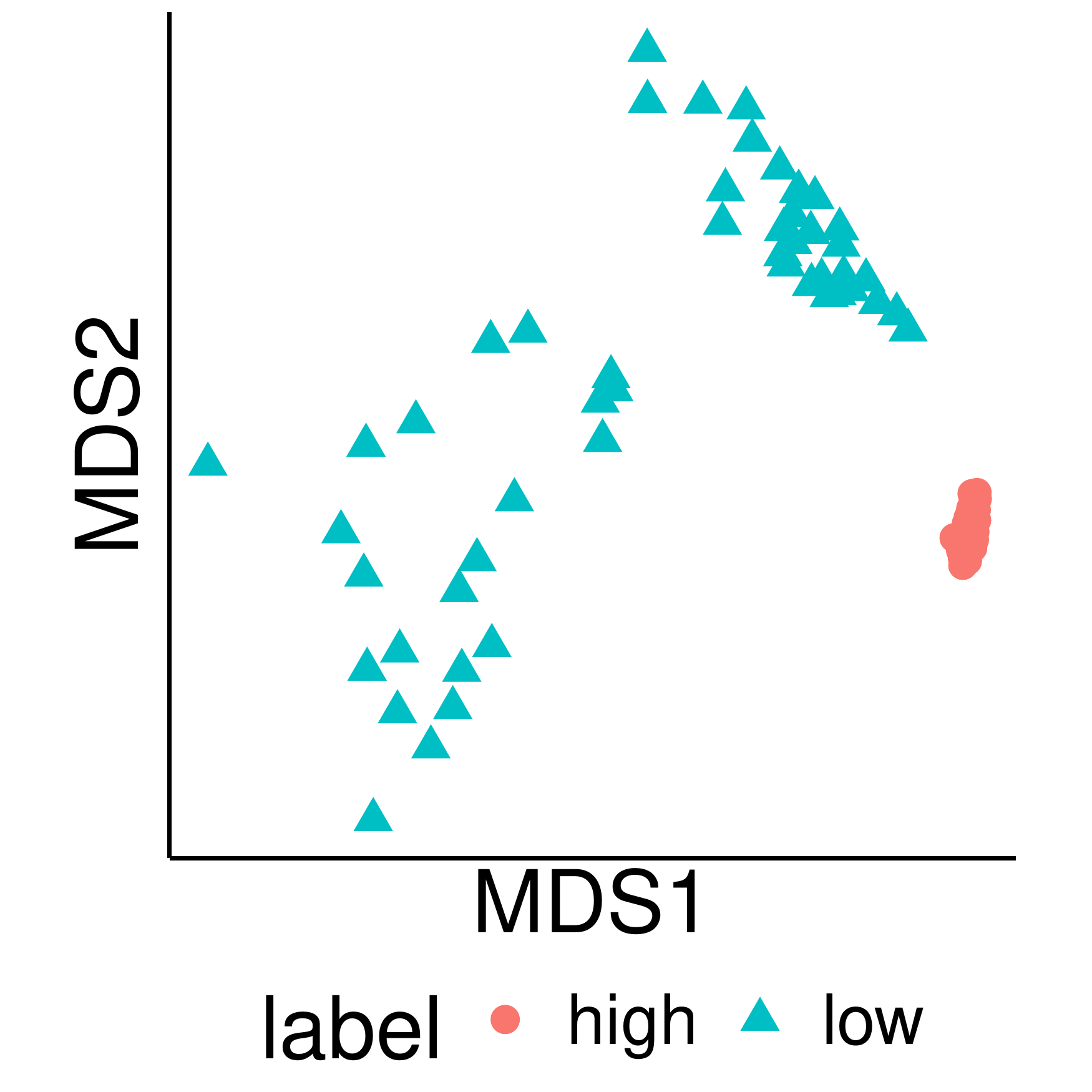} &
		\includegraphics[width=0.4\textwidth]{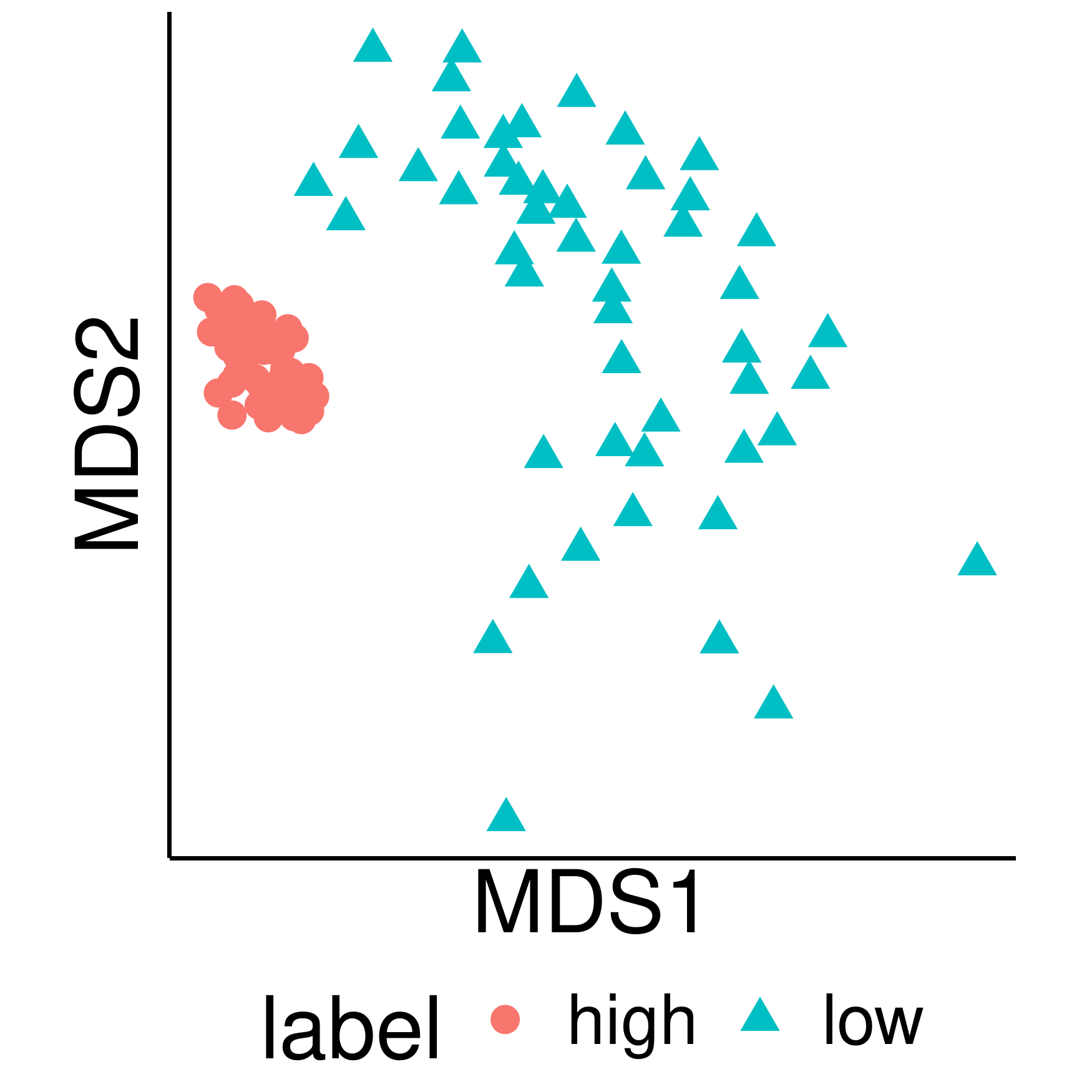}
	\end{tabular}
	\caption{Sample networks from (a) $G(n,p_{\textrm{high}} = 0.25)$ and (b) $G(n, p_{\textrm{low}} = 0.05)$. For each model, 50 networks are randomly generated and fitted using the latent space model on which persistence landscapes are computed. Distributions for a total of 100 persistence landscapes of (c) order 0 (connectedness) and (d) order 1 (holes) are visualized by multidimensional scaling.}
	\label{fig:simulation1}
\end{figure}

\begin{figure}[htbp]
	\centering
	\begin{tabular}{c}
		(a) \\ 
		\begin{subfigure}[b]{\textwidth}
			\centering
			\includegraphics[width=.24\textwidth]{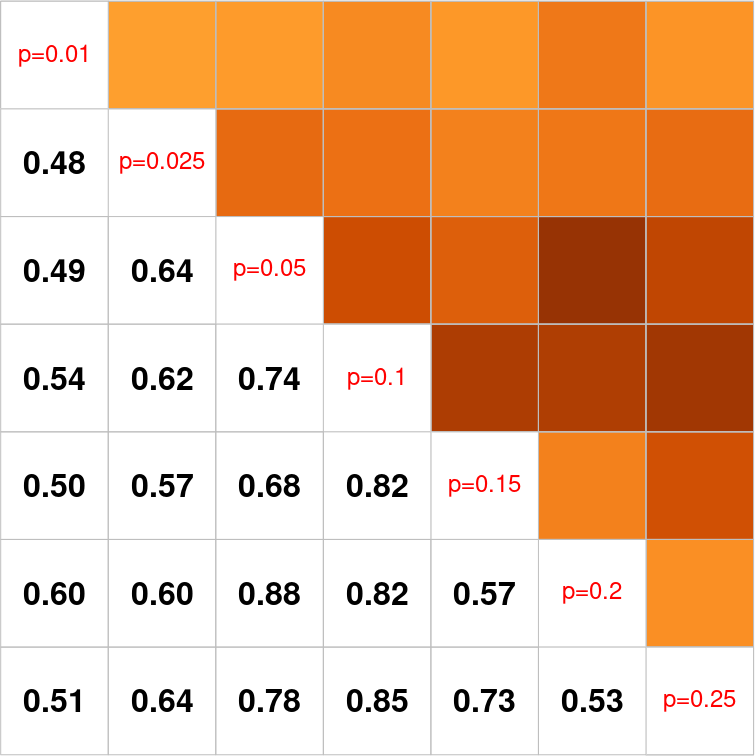}
			\includegraphics[width=.24\textwidth]{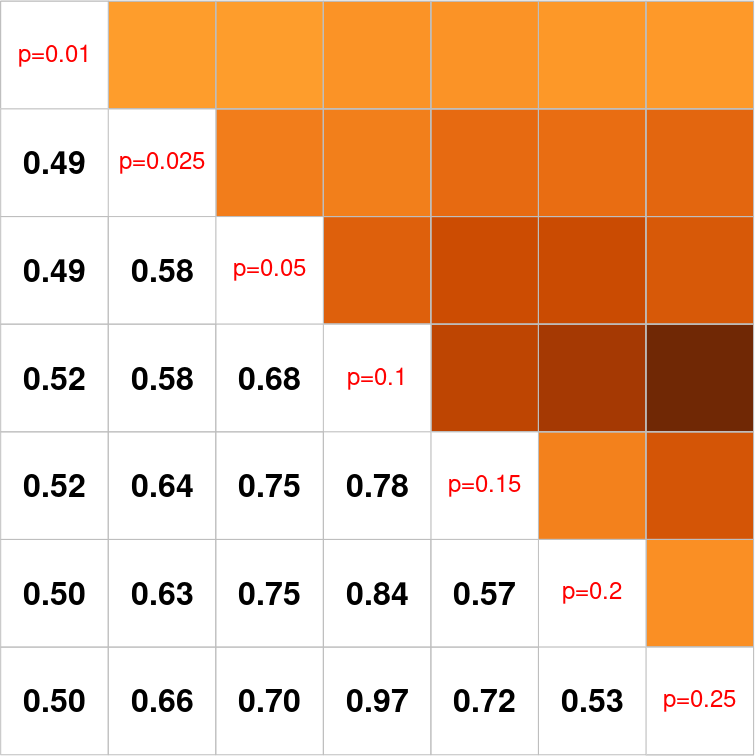}
			\includegraphics[width=.24\textwidth]{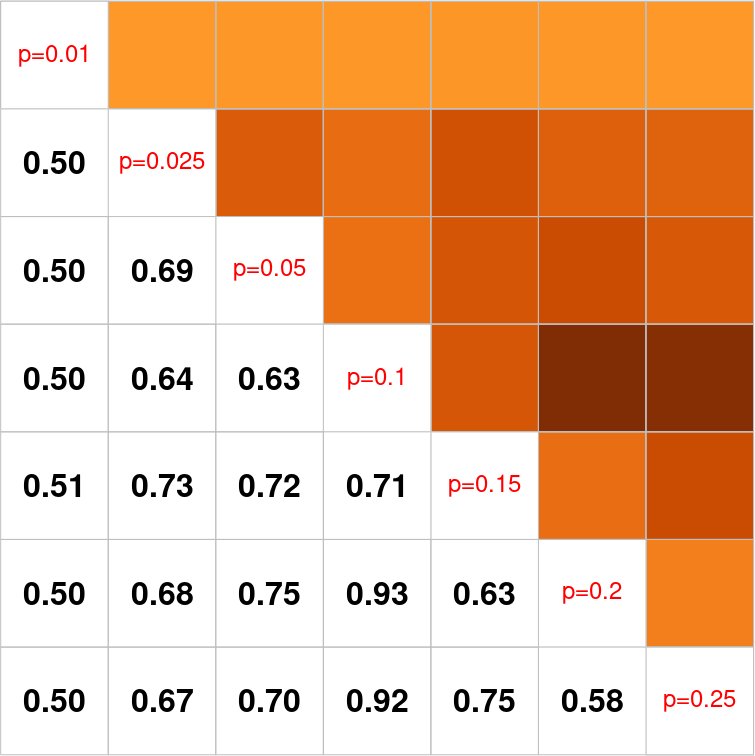}
			\includegraphics[width=.24\textwidth]{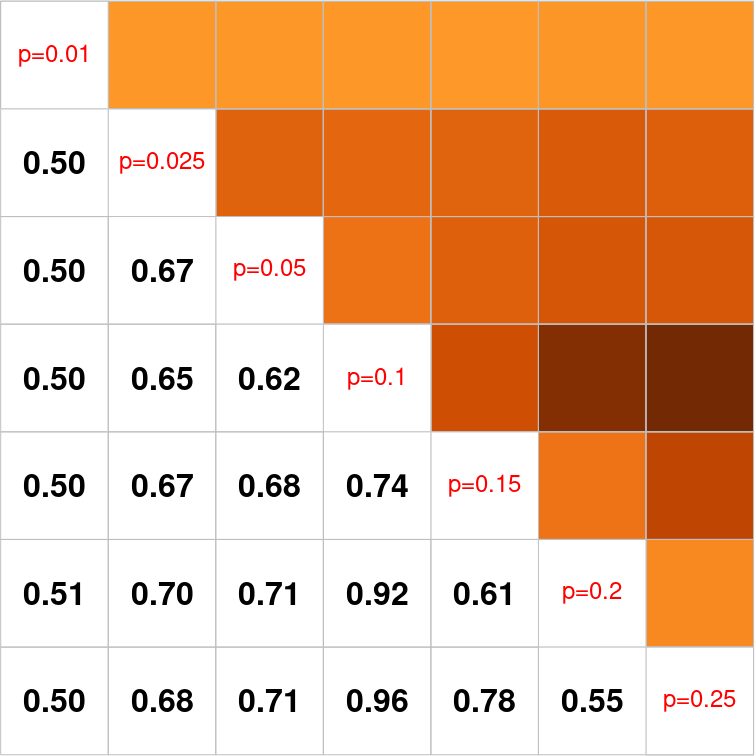}
		\end{subfigure}\\
		(b) \\ 
		\begin{subfigure}[b]{\textwidth}
			\centering
			\includegraphics[width=.24\textwidth]{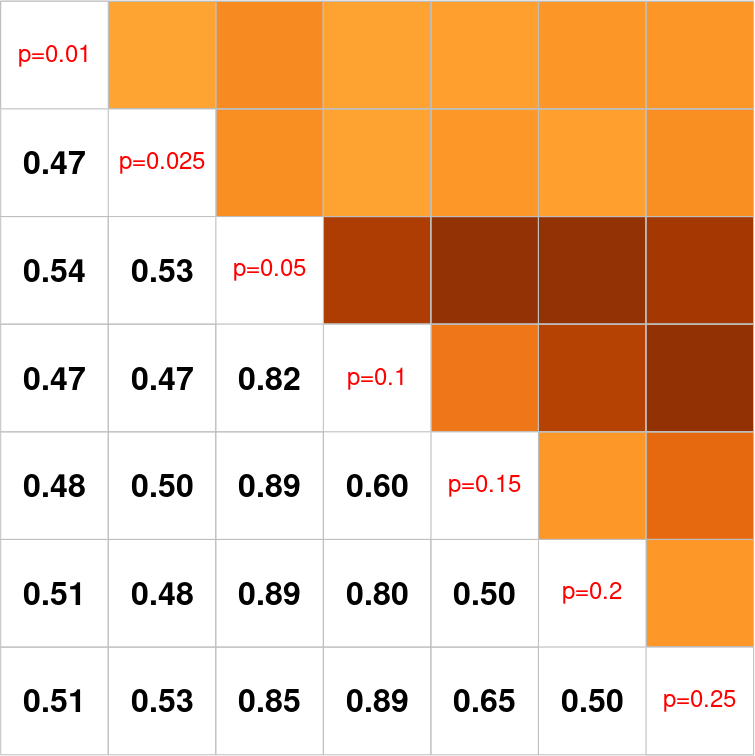}
			\includegraphics[width=.24\textwidth]{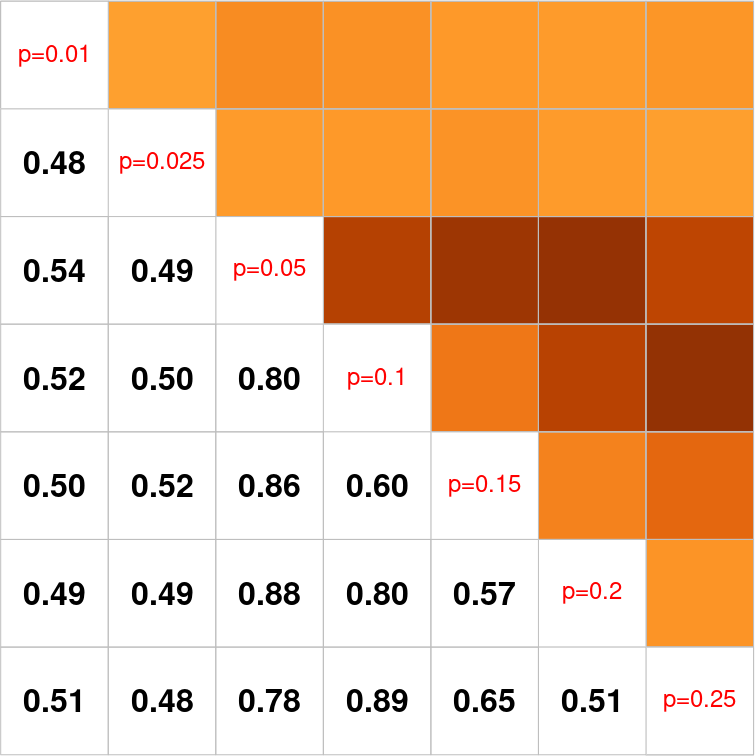}
			\includegraphics[width=.24\textwidth]{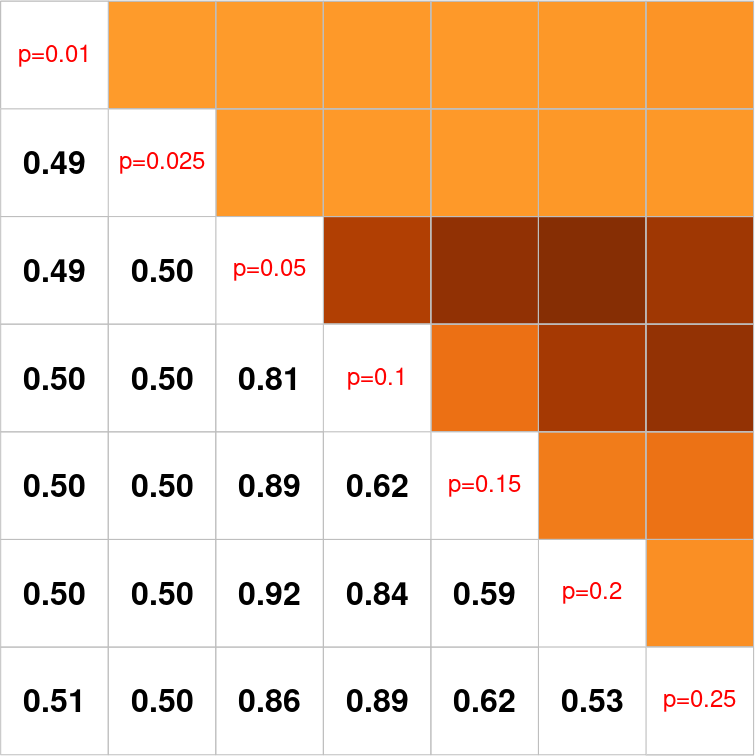}
			\includegraphics[width=.24\textwidth]{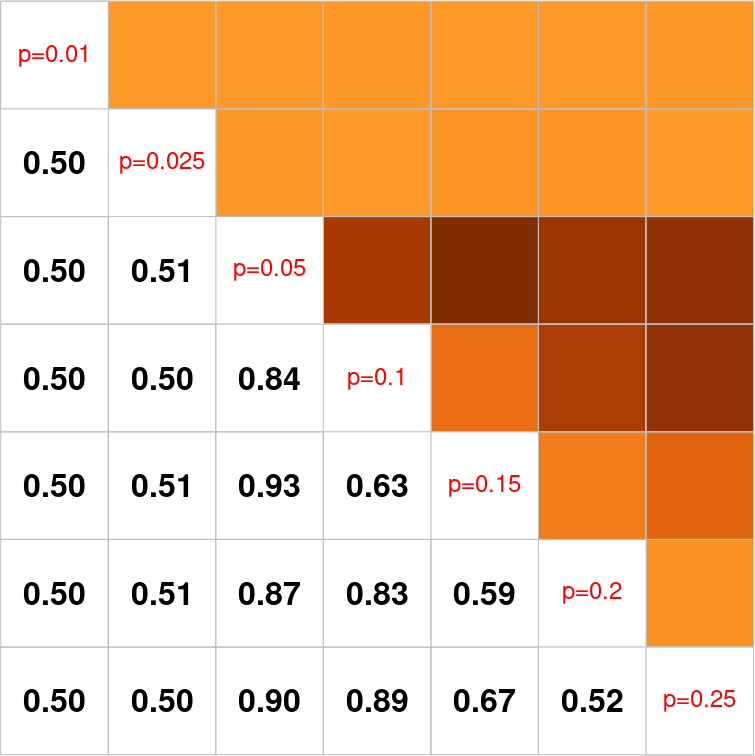}
		\end{subfigure}\\
		(c) \\ 
		\begin{subfigure}[b]{\textwidth}
			\centering
			\includegraphics[width=.24\textwidth]{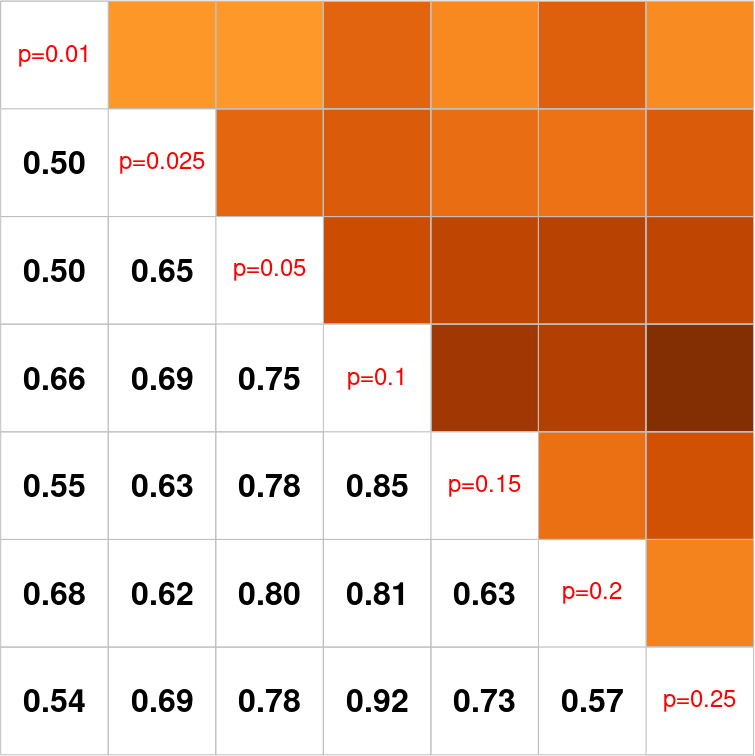}
			\includegraphics[width=.24\textwidth]{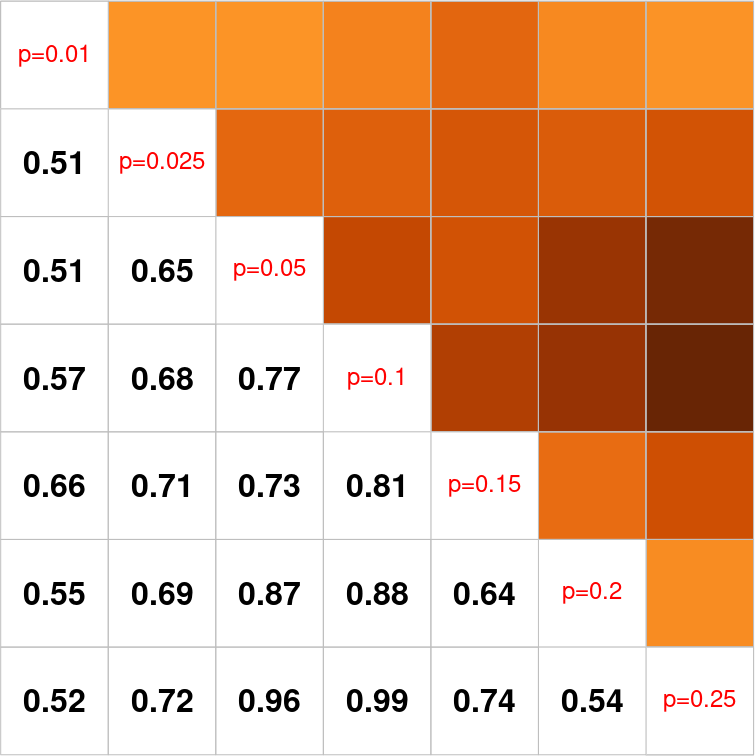}
			\includegraphics[width=.24\textwidth]{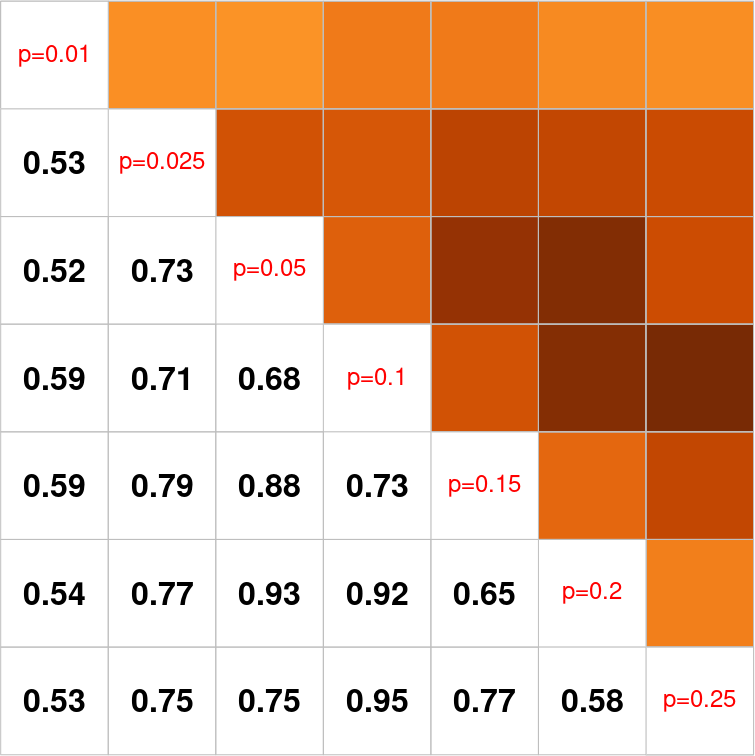}
			\includegraphics[width=.24\textwidth]{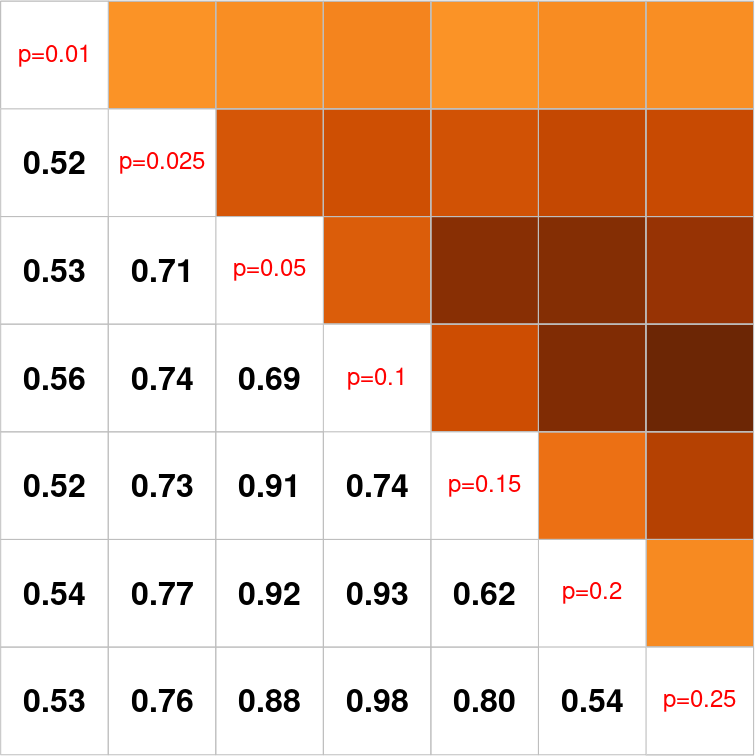}
		\end{subfigure}
	\end{tabular}
	\caption{Pairwise comparison of ER models with cluster analysis procedures with $k=2$; (a) $k$-medoids of order 0, (b) $k$-medoids of order 1, and (c) $k$-groups of order 0. For each row, results from different sample size are reported for $m=5,10,25,50$ from left to right. In each subplot, averages of 100 empirical Rand indices are presented in the lower triangular part. The darker entry means the larger value of an index, implying that it is more aligned with the ground-truth label of networks.}
	\label{fig:results_ER_cluster1}
\end{figure}

\begin{figure}[htbp]
	\centering
	\begin{tabular}{c}
		(d) \\ 
		\begin{subfigure}[b]{\textwidth}
			\centering
			\includegraphics[width=.24\textwidth]{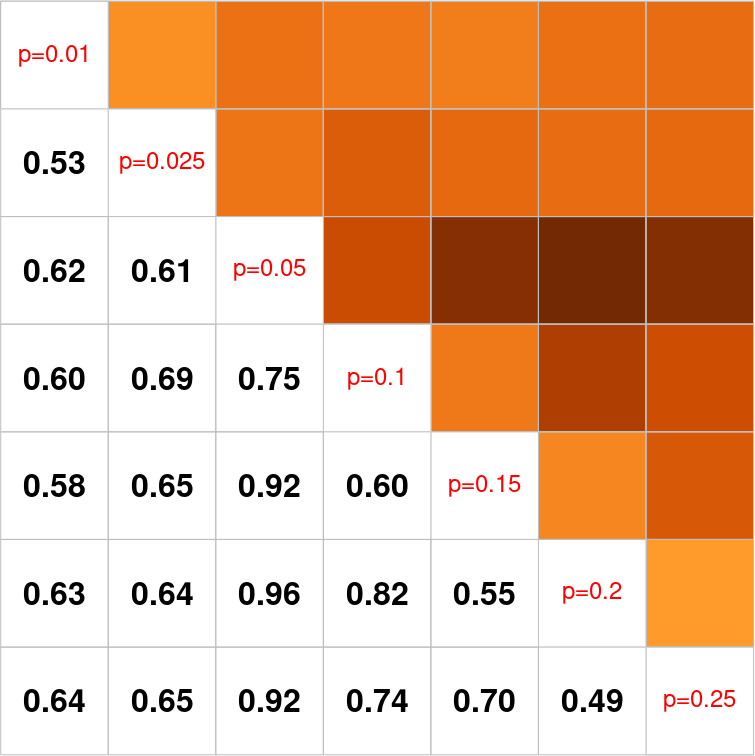}
			\includegraphics[width=.24\textwidth]{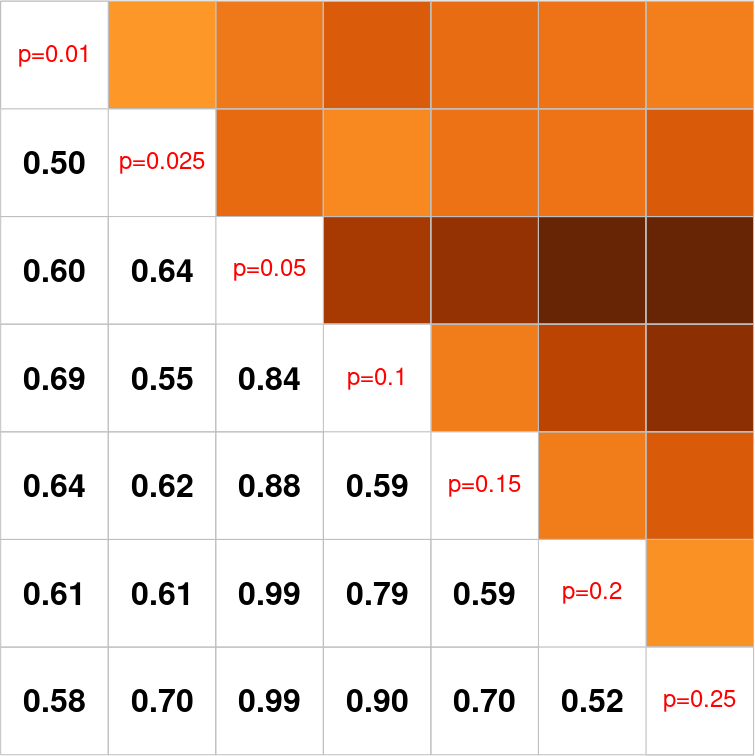}
			\includegraphics[width=.24\textwidth]{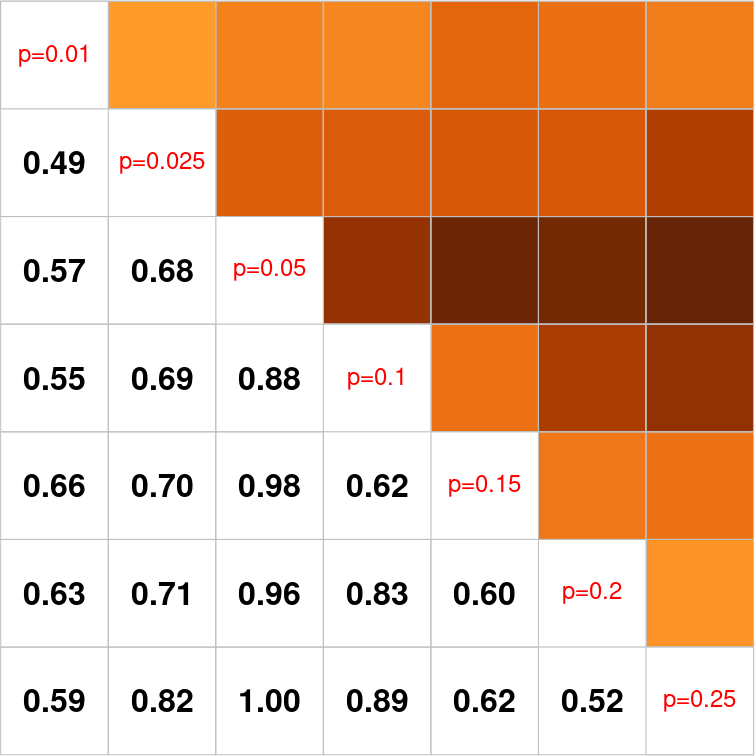}
			\includegraphics[width=.24\textwidth]{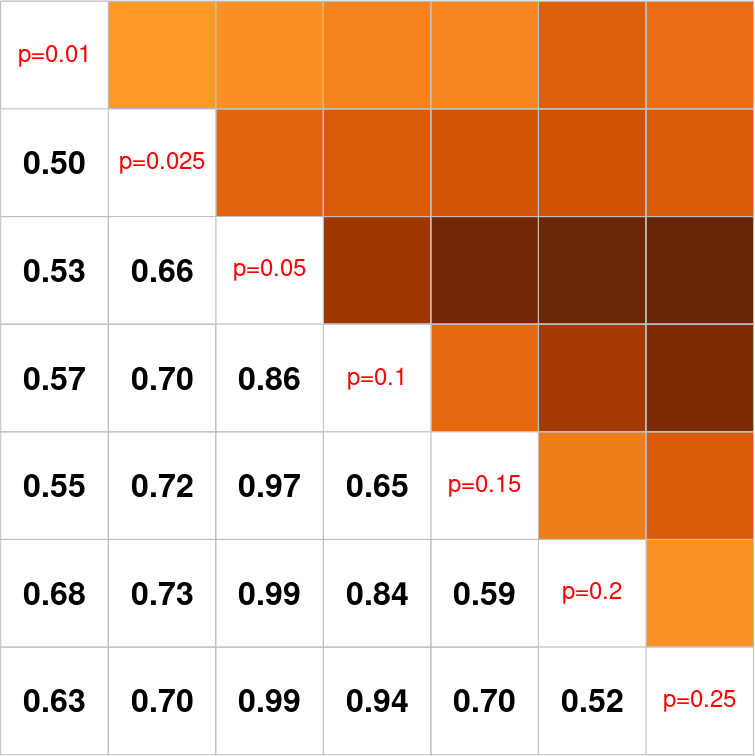}
		\end{subfigure}\\
		(e) \\ 
		\begin{subfigure}[b]{\textwidth}
			\centering
			\includegraphics[width=.24\textwidth]{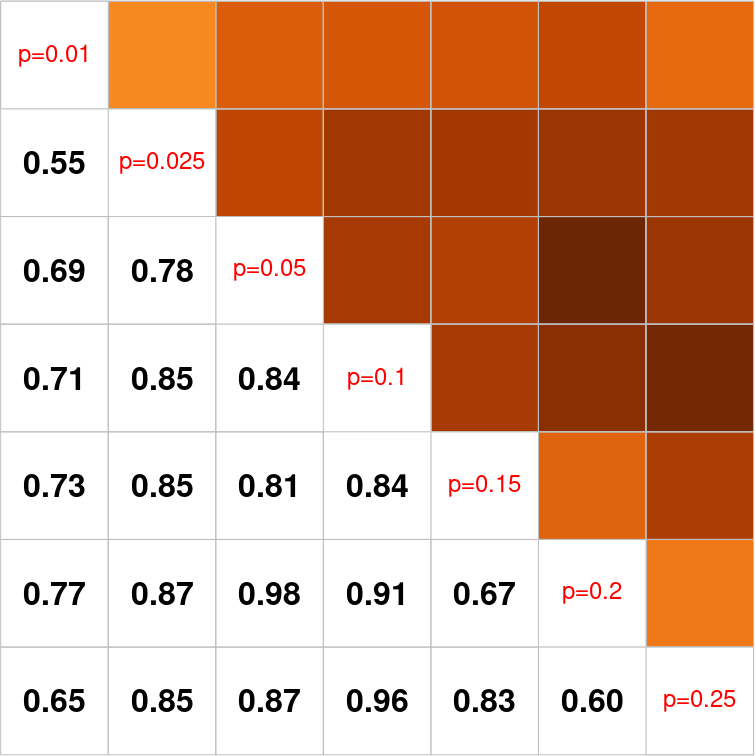}
			\includegraphics[width=.24\textwidth]{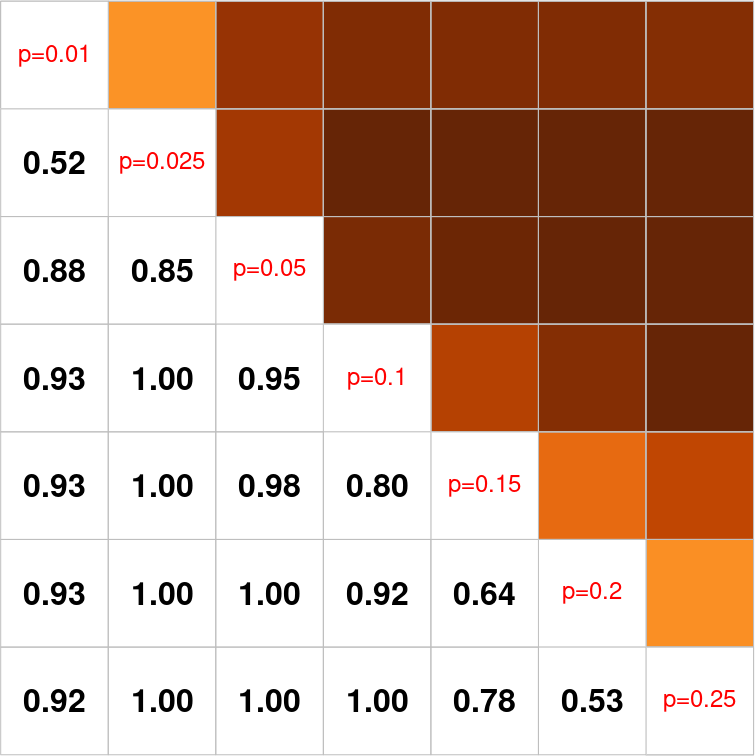}
			\includegraphics[width=.24\textwidth]{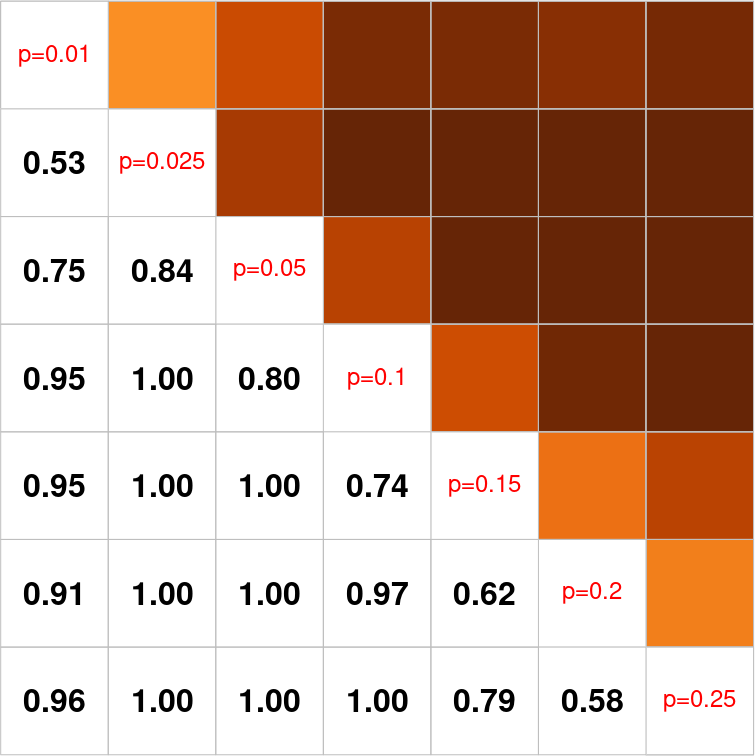}
			\includegraphics[width=.24\textwidth]{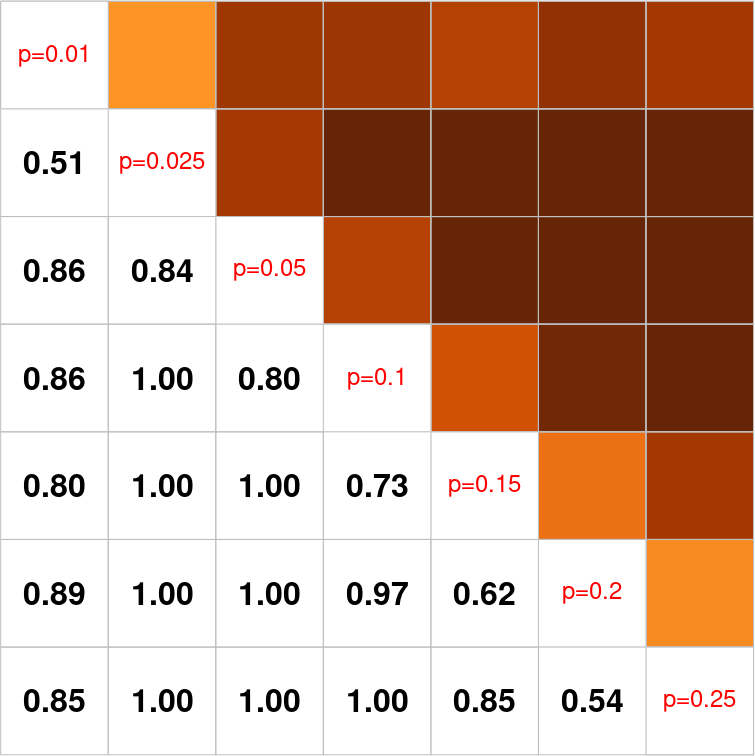}
		\end{subfigure}\\
		(f) \\ 
		\begin{subfigure}[b]{\textwidth}
			\centering
			\includegraphics[width=.24\textwidth]{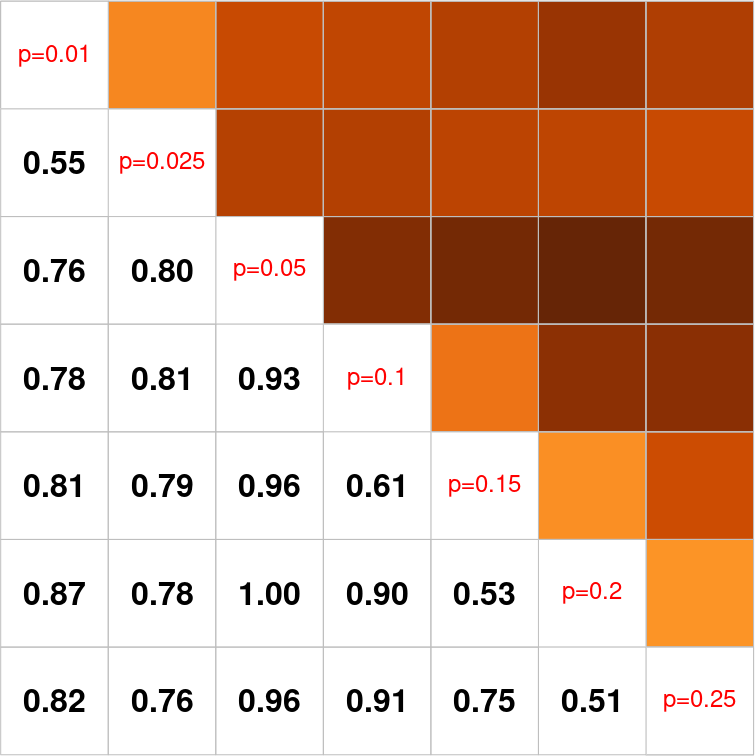}
			\includegraphics[width=.24\textwidth]{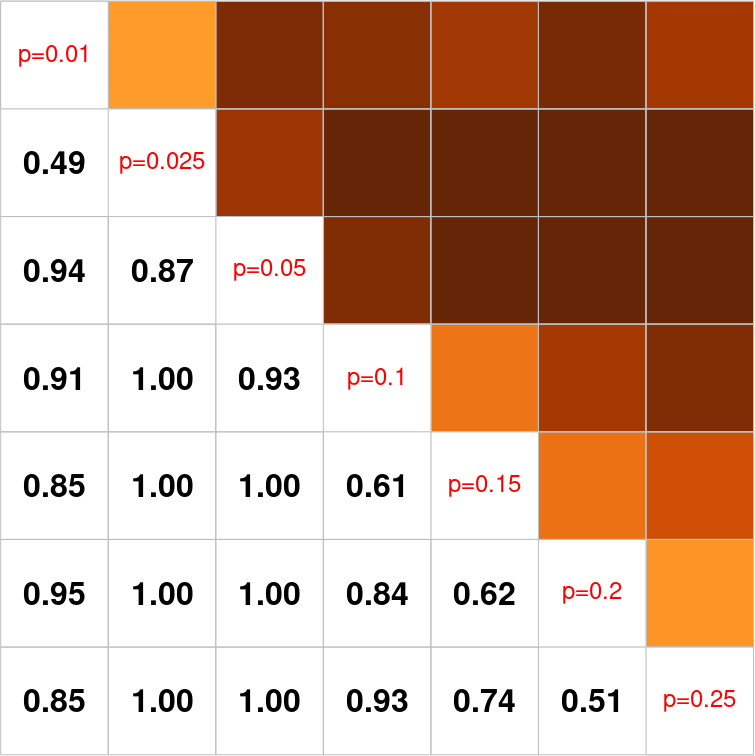}
			\includegraphics[width=.24\textwidth]{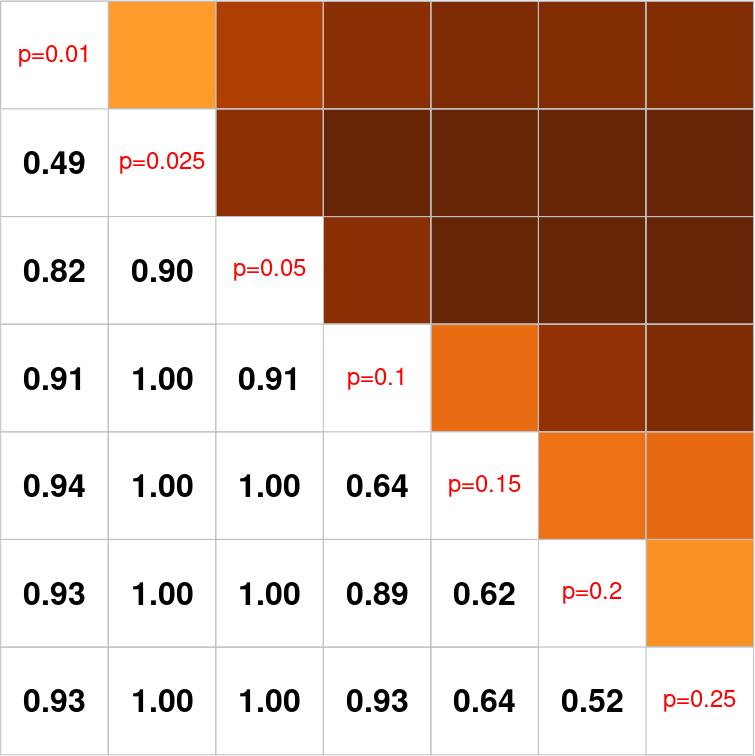}
			\includegraphics[width=.24\textwidth]{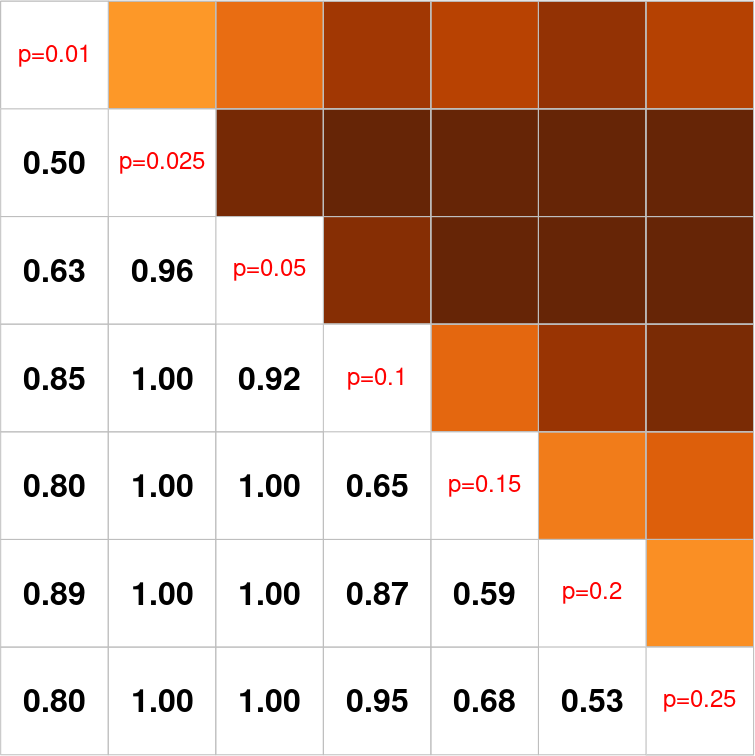}
		\end{subfigure}\\
	\end{tabular}
	\caption{Pairwise comparison of ER models with cluster analysis procedures with $k=2$; (d) $k$-groups test of order 1, (e) spectral clustering of order 0, and (f) spectral clustering of order 1. For each row, results from different sample size are reported for $m=5,10,25,50$ from left to right. In each subplot, averages of 100 empirical Rand indices are presented in the lower triangular part. The darker entry means the larger value of an index, implying that it is more aligned with the ground-truth label of networks.}
	\label{fig:results_ER_cluster2}
\end{figure}

\begin{figure}[htbp]
	\centering
	\includegraphics[width=\textwidth]{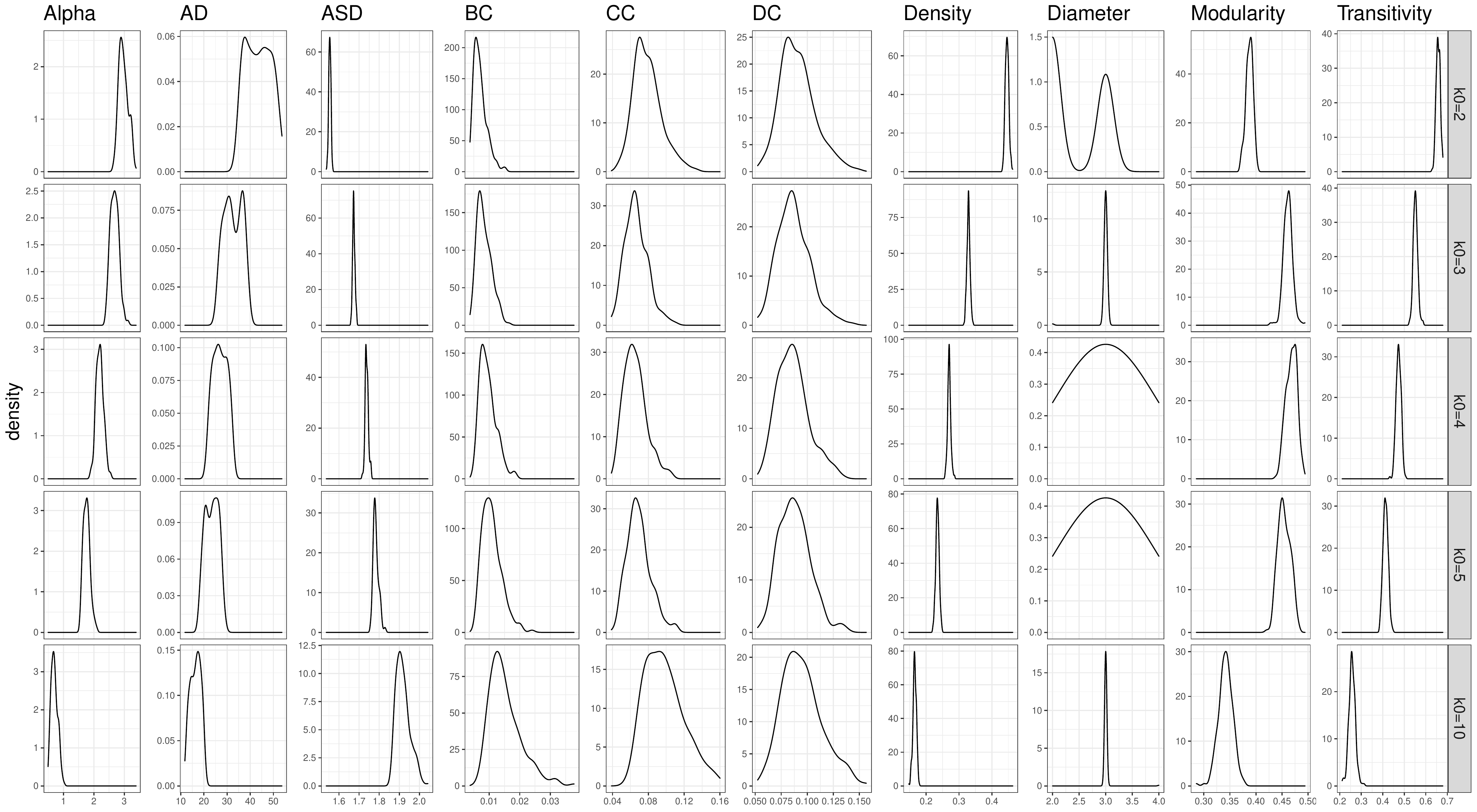}
	\caption{Summary of network characteristics sampled from the SBM models with varying number of communities. The first column represents distribution of fitted intercept values $\alpha$ and the rest are those of topological descriptors, the latter of which indicates heterogeneity of topological properties across multiple SBM models.}
	\label{fig:summary_SBM}
\end{figure}

\begin{figure}[htbp]
	\begin{tabular}{cc}
		\multicolumn{2}{c}{(a) \hspace*{4cm} (b) \hspace*{4cm} (c)}\\
		\multicolumn{2}{c}{
			\includegraphics[width=0.3\textwidth]{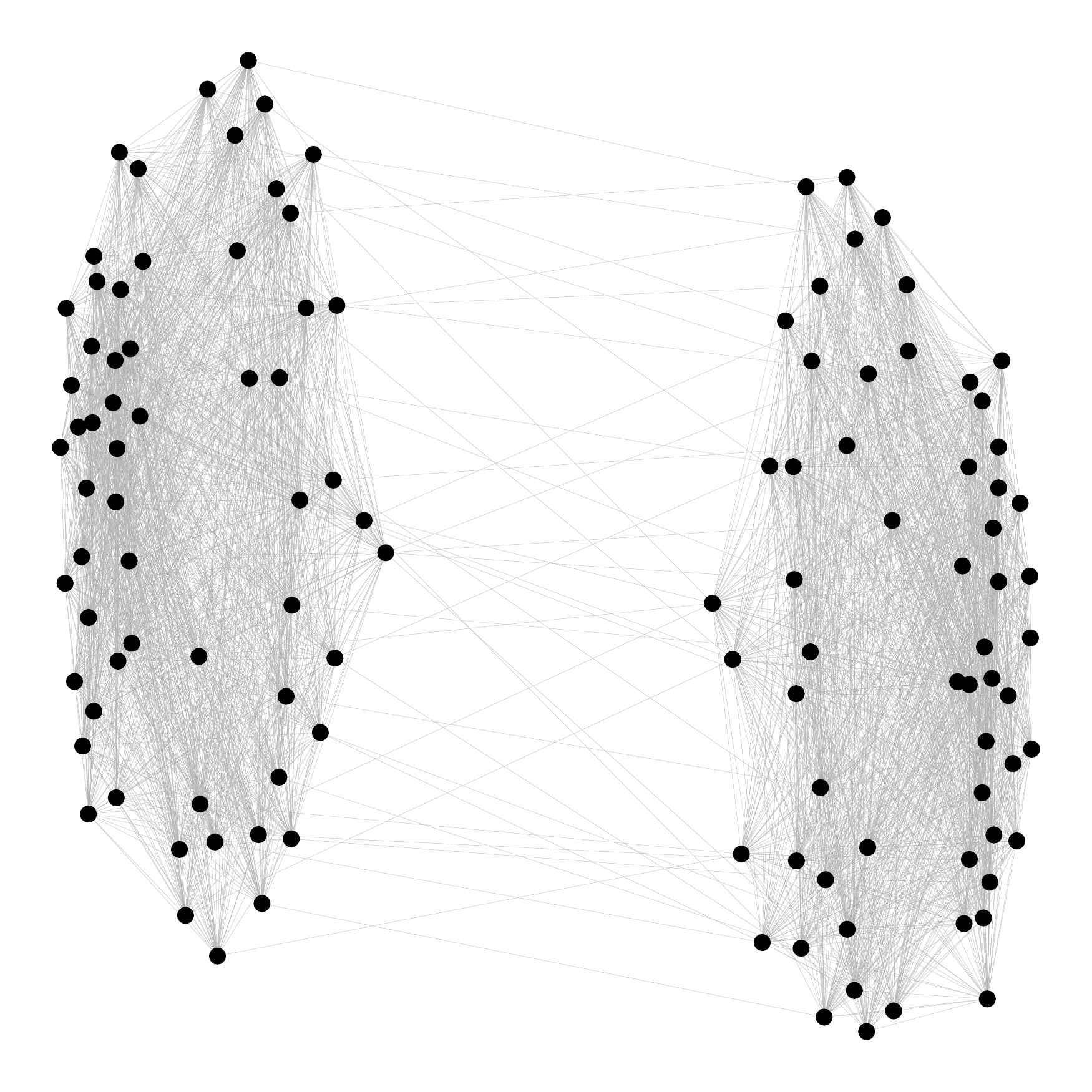}
			\includegraphics[width=0.3\textwidth]{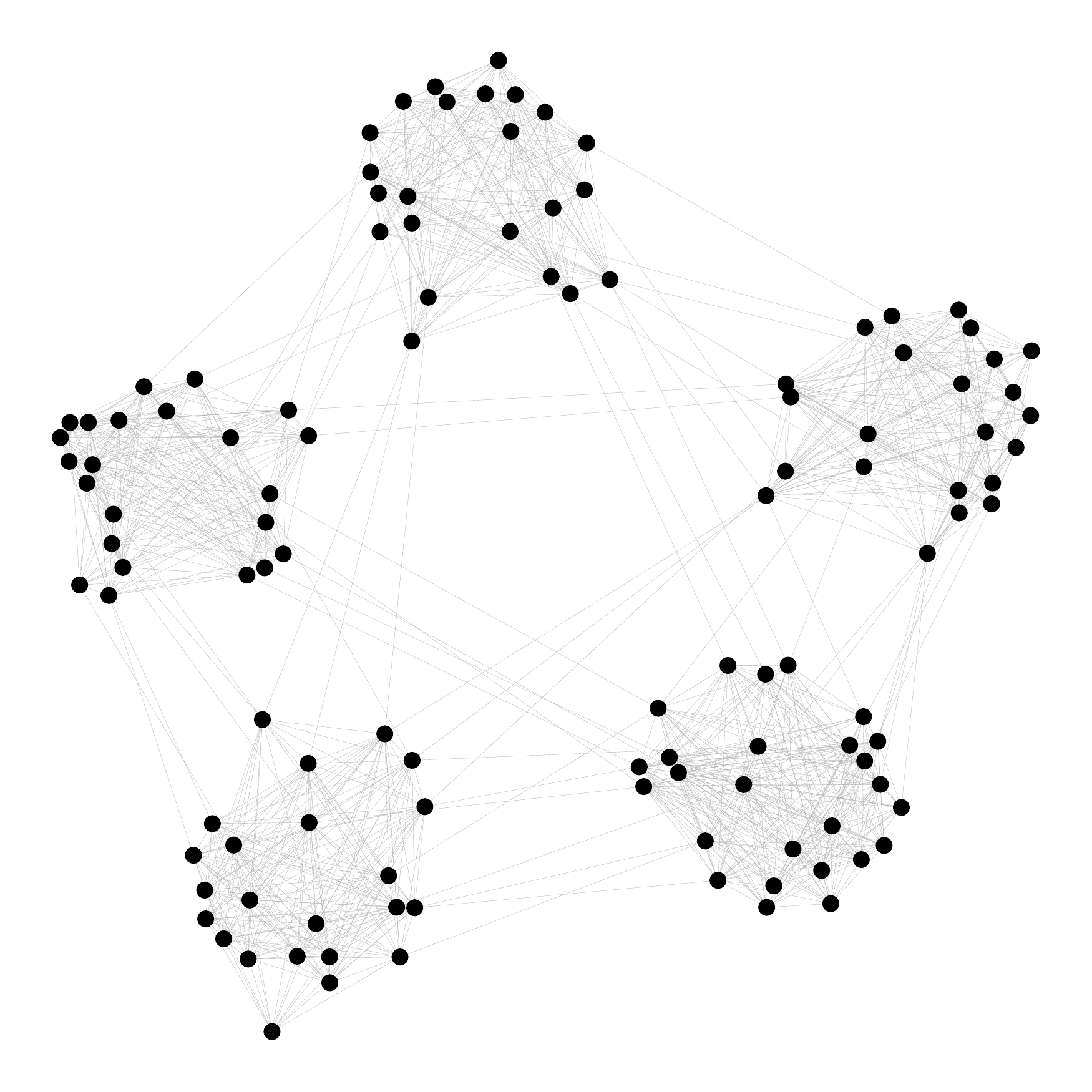}
			\includegraphics[width=0.3\textwidth]{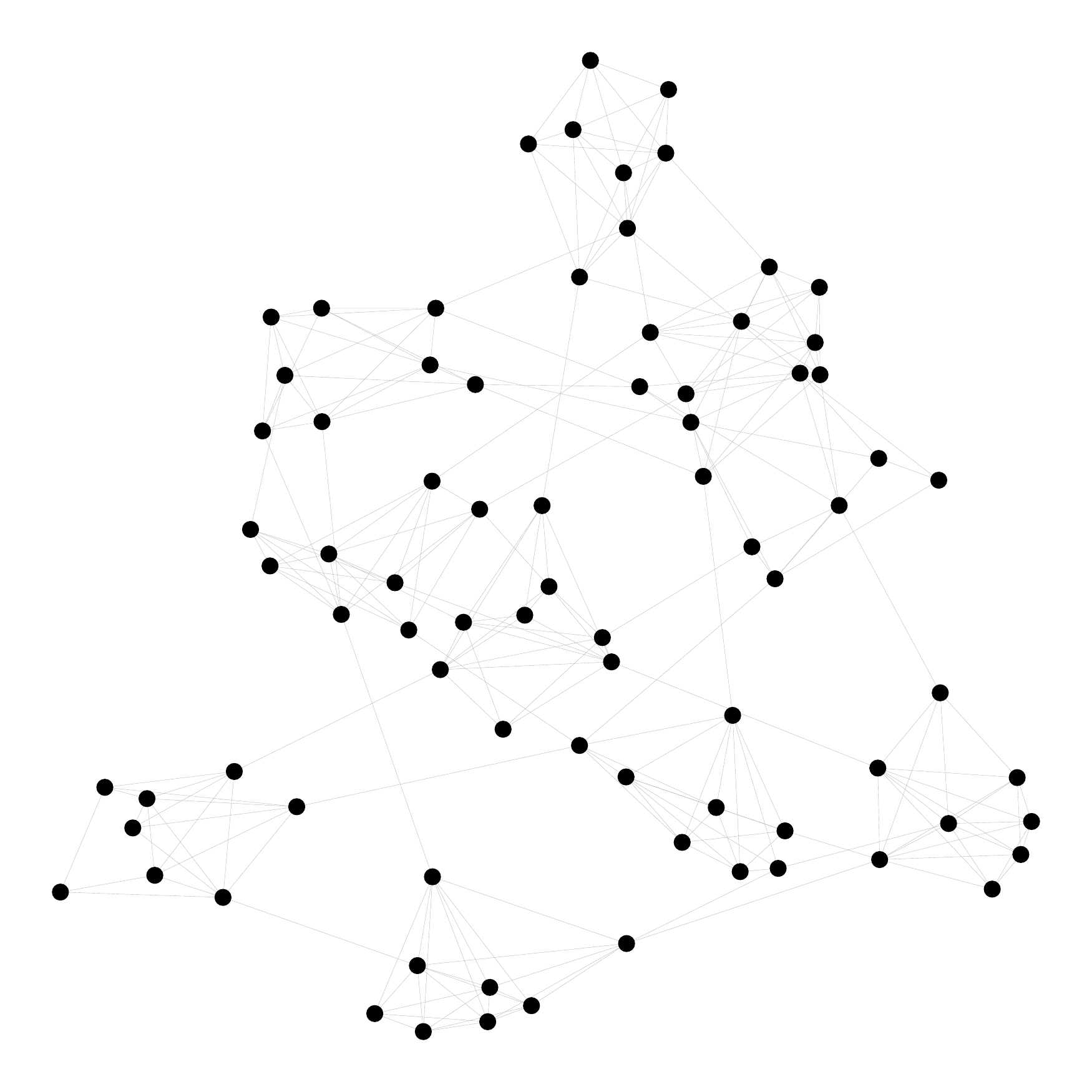}
		}\\
		(d) & (e) \\
		\includegraphics[width=0.45\textwidth]{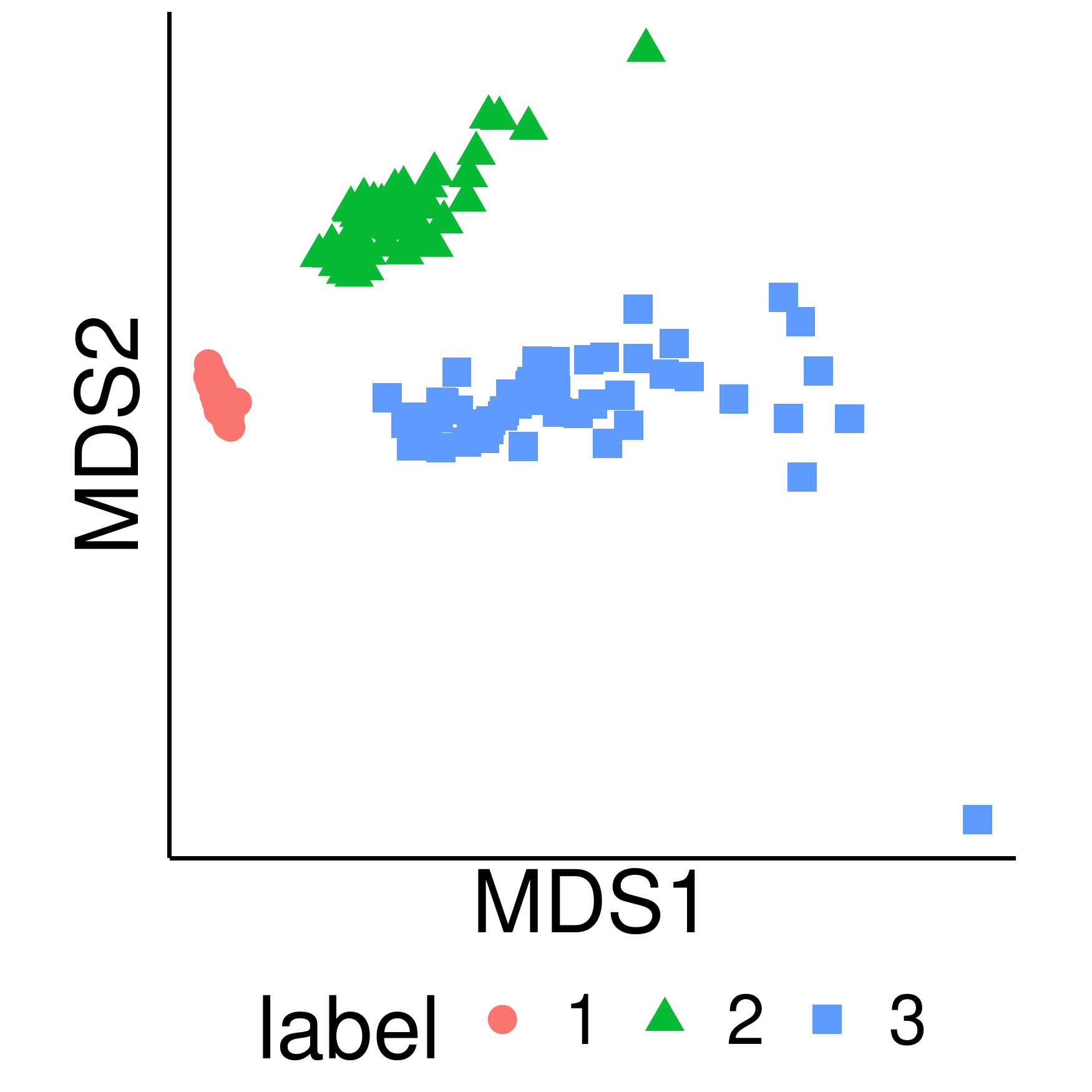} & 
		\includegraphics[width=0.45\linewidth]{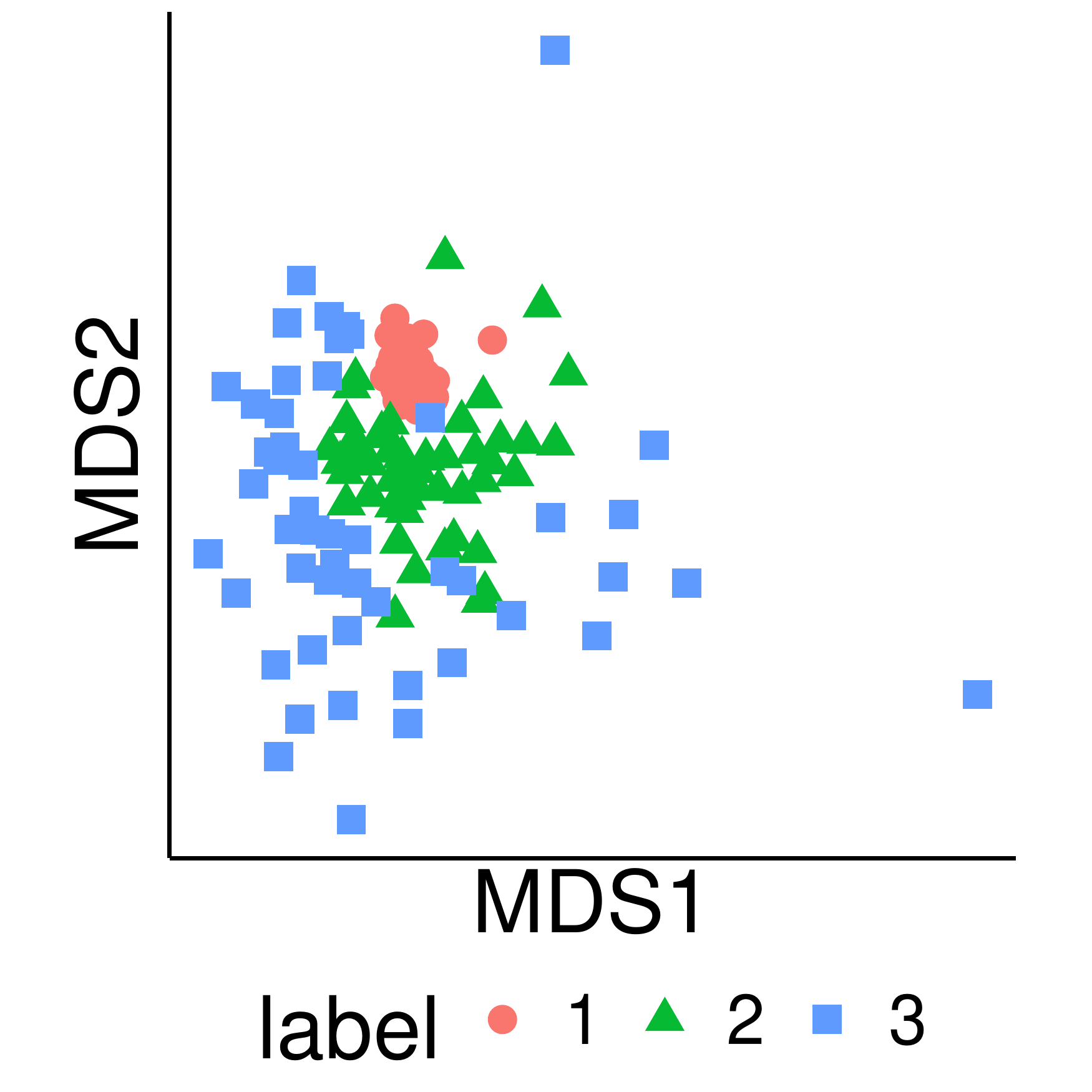}
	\end{tabular}
	\caption{Sample networks from stochastic block models with (a) 2 (label 1), (b) 5 (label 2), and (c) 10 (label 3) communities. For each model, 50 networks are randomly generated and fitted using the latent space model on which persistence landscapes are computed. Distributions of persistence landscapes for (d) order 0 (connectedness) and (e) order 1 (holes) are visualized by multidimensional scaling.}
	\label{fig:simulation2}
\end{figure}

\begin{table}[htbp]
	\centering
	\begin{tabular}{c}
		(a) \\ 
		\begin{subtable}{\textwidth}
			\centering
			\begin{tabular}{ccccccccc}
				\hline
				\multirow{2}{*}{scenario} & \multicolumn{4}{c}{order 0} & \multicolumn{4}{c}{order 1} \\ \cline{2-9}
				& $m=5$     & $m=10$ & $m=25$ & $m=50$ & $m=5$     & $m=10$ & $m=25$ & $m=50$ \\ \hline
				$\lbrace 2,3,4\rbrace$ & 1.50e-06 & 1.00e-06 & 1.00e-06 & 1.00e-06 & 1.18e-02 & 3.70e-06 & 1.00e-06 & 1.00e-06 \\ 
				$\lbrace 2,3,5\rbrace$ & 1.10e-06 & 1.00e-06 & 1.00e-06 & 1.00e-06 & 8.19e-03 & 1.00e-06 & 1.00e-06 & 1.00e-06 \\ 
				$\lbrace 2,4,5\rbrace$ & 1.30e-06 & 1.00e-06 & 1.00e-06 & 1.00e-06 & 7.21e-04 & 1.00e-06 & 1.00e-06 & 1.00e-06 \\
				$\lbrace 3,4,5\rbrace$ & 1.40e-06 & 1.00e-06 & 1.00e-06 & 1.00e-06 & 3.46e-04 & 1.00e-06 & 1.00e-06 & 1.00e-06 \\ 
				$\lbrace 2,5,10\rbrace$& 1.10e-06 & 1.00e-06 & 1.00e-06 & 1.00e-06 & 1.00e-06 & 1.00e-06 & 1.00e-06 & 1.00e-06 \\ 
				\hline
			\end{tabular}
		\end{subtable}
		\\ (b) \\
		\begin{subtable}{\textwidth}
			\centering
			\begin{tabular}{ccccccccc}
				\hline
				\multirow{2}{*}{scenario} & \multicolumn{4}{c}{order 0} & \multicolumn{4}{c}{order 1} \\ \cline{2-9}
				& $m=5$     & $m=10$ & $m=25$ & $m=50$ & $m=5$     & $m=10$ & $m=25$ & $m=50$ \\ \hline
				$\lbrace 2,3,4\rbrace$ & 4.00e-06 & 1.00e-06 & 1.00e-06 & 1.00e-06 & 1.17e-02 & 2.60e-05 & 1.00e-06 & 1.00e-06 \\ 
				$\lbrace 2,3,5\rbrace$ & 6.00e-06 & 1.00e-06 & 1.00e-06 & 1.00e-06 & 8.32e-03 & 1.00e-06 & 1.00e-06 & 1.00e-06 \\ 
				$\lbrace 2,4,5\rbrace$ & 1.00e-06 & 1.00e-06 & 1.00e-06 & 1.00e-06 & 7.11e-04 & 1.00e-06 & 1.00e-06 & 1.00e-06 \\ 
				$\lbrace 3,4,5\rbrace$ & 1.00e-06 & 1.00e-06 & 1.00e-06 & 1.00e-06 & 3.59e-04 & 1.00e-06 & 1.00e-06 & 1.00e-06 \\ 
				$\lbrace 2,5,10\rbrace$& 9.00e-06 & 1.00e-06 & 1.00e-06 & 1.00e-06 & 1.00e-06 & 1.00e-06 & 1.00e-06 & 1.00e-06 \\ 
				\hline
			\end{tabular}
		\end{subtable}
	\end{tabular}
	\caption{Average empirical $p$-values for multi-sample hypothesis testing procedures on SBM model scenarios; (a) $k$-sample test and (b) DISCO.}
	\label{tab:results_SBM_hypothesis}
\end{table}

\begin{table}[htbp]
	\centering
	\resizebox{15cm}{!}{
		\begin{tabular}{c}
			(a) \\ 
			\begin{subtable}{\textwidth}
				\centering
				\begin{tabular}{ccccccccc}
					\hline
					\multirow{2}{*}{scenario} & \multicolumn{4}{c}{order 0} & \multicolumn{4}{c}{order 1} \\ \cline{2-9}
					& $m=5$     & $m=10$ & $m=25$ & $m=50$ & $m=5$     & $m=10$ & $m=25$ & $m=50$ \\ \hline
					$\lbrace 2,3,4\rbrace$ & 1.000 & 0.987 & 0.998 & 0.996 & 0.574 & 0.566 & 0.566 & 0.588 \\ 
					$\lbrace 2,3,5\rbrace$ & 1.000 & 1.000 & 0.998 & 0.997 & 0.508 & 0.588 & 0.585 & 0.557 \\ 
					$\lbrace 2,4,5\rbrace$ & 1.000 & 0.974 & 0.960 & 0.978 & 0.700 & 0.626 & 0.671 & 0.641 \\ 
					$\lbrace 3,4,5\rbrace$ & 0.956 & 0.971 & 0.981 & 0.964 & 0.671 & 0.589 & 0.626 & 0.599 \\ 
					$\lbrace 2,5,10\rbrace$& 0.983 & 1.000 & 1.000 & 0.997 & 0.595 & 0.611 & 0.628 & 0.625 \\ 
					\hline
				\end{tabular}
			\end{subtable}
			\\ (b) \\
			\begin{subtable}{\textwidth}
				\centering
				\begin{tabular}{ccccccccc}
					\hline
					\multirow{2}{*}{scenario} & \multicolumn{4}{c}{order 0} & \multicolumn{4}{c}{order 1} \\ \cline{2-9}
					& $m=5$     & $m=10$ & $m=25$ & $m=50$ & $m=5$     & $m=10$ & $m=25$ & $m=50$ \\ \hline
					$\lbrace 2,3,4\rbrace$ & 0.983 & 0.991 & 0.998 & 0.998 &0.610 & 0.596 & 0.619 & 0.624 \\ 
					$\lbrace 2,3,5\rbrace$ & 1.000 & 1.000 & 0.998 & 0.997 &0.669 & 0.640 & 0.635 & 0.598 \\ 
					$\lbrace 2,4,5\rbrace$ & 0.960 & 0.969 & 0.966 & 0.978 &0.732 & 0.669 & 0.699 & 0.668 \\ 
					$\lbrace 3,4,5\rbrace$ & 0.929 & 0.971 & 0.981 & 0.962 &0.660 & 0.625 & 0.639 & 0.629 \\ 
					$\lbrace 2,5,10\rbrace$& 1.000 & 1.000 & 1.000 & 0.999 &0.653 & 0.740 & 0.751 & 0.775 \\ 
					\hline
				\end{tabular}
			\end{subtable}
			\\ (c) \\
			\begin{subtable}{\textwidth}
				\centering
				\begin{tabular}{ccccccccc}
					\hline
					\multirow{2}{*}{scenario} & \multicolumn{4}{c}{order 0} & \multicolumn{4}{c}{order 1} \\ \cline{2-9}
					& $m=5$     & $m=10$ & $m=25$ & $m=50$ & $m=5$     & $m=10$ & $m=25$ & $m=50$ \\ \hline
					$\lbrace 2,3,4\rbrace$ & 0.943 & 0.971 & 0.920 & 0.953 & 0.725 & 0.655 & 0.658 & 0.649 \\ 
					$\lbrace 2,3,5\rbrace$ & 0.970 & 0.944 & 0.927 & 0.947 & 0.705 & 0.736 & 0.673 & 0.638 \\ 
					$\lbrace 2,4,5\rbrace$ & 0.970 & 0.949 & 0.950 & 0.965 & 0.787 & 0.769 & 0.747 & 0.715 \\ 
					$\lbrace 3,4,5\rbrace$ & 0.955 & 0.904 & 0.934 & 0.972 & 0.708 & 0.674 & 0.662 & 0.653 \\ 
					$\lbrace 2,5,10\rbrace$& 0.972 & 1.000 & 0.976 & 0.977 & 0.832 & 0.774 & 0.757 & 0.694 \\ 
					\hline
				\end{tabular}
			\end{subtable}
		\end{tabular}
	}
	\caption{Average Rand indices for estimated community structure from cluster analysis algorithms with a fixed number of cluster $k=3$ in SBM model scenarios; (a) $k$-medoids, (b) $k$-groups, and (c) spectral clustering.}
	\label{tab:results_SBM_clustering}
\end{table}
\end{appendices}

\end{document}